\documentclass{aa}  

\usepackage{lscape}
\usepackage{graphicx}
\usepackage{txfonts}
\usepackage{amsmath}
\usepackage{multirow}
\usepackage{adjustbox}
\usepackage{tabularx}
\usepackage{amsmath}
\usepackage{bm}
\usepackage{array}
\usepackage{xcolor}
\usepackage{pifont}
\usepackage{graphicx}
\usepackage{txfonts}
\usepackage{lipsum}
\usepackage{subcaption}         
\usepackage{lscape}            
\usepackage{placeins} 

\usepackage{booktabs}
\usepackage{siunitx}

\def\nic{\textit{NICER}\xspace}

\def\ros{\textit{ROSAT}\xspace}

\def\srgero{{{\it SRG}/eROSITA}\xspace}

\def\swix{\textit{Swift}-XRT\xspace}
\def\swiu{\textit{Swift}-UVOT\xspace}
\def\swi{\textit{Swift}\xspace}
\def\swib{\textit{Swift}-BAT\xspace}

\def\xmm{\textit{XMM-Newton}\xspace}

\begin{document}

   \title{Searching for supermassive black holes binaries within SRG/eROSITA-De I:}
      \subtitle{Properties of the X-ray selected candidates.}

   \author{Dusán Tubín-Arenas\thanks{\email{dtubin@aip.de}}
          \inst{1,2}
          \and
          Mirko Krumpe\inst{1}
          \and
          David Homan\inst{1}
          \and
          Alex Markowitz\inst{3}
          \and
          Meredith Powell\inst{1}
          \and
          Georg Lamer\inst{1}
          \and
          Tanya Urrutia\inst{1}
          \and
          Axel Schwope\inst{1}
          \and
          Hartmut Winkler\inst{5}
          \and
          Sabina Bahic\inst{1,2}
          \and
          Johannes Buchner\inst{4}
          \and
          Carolina Andonie\inst{4}
          \and
          Mara Salvato\inst{4}
          \and
          Andrea Merloni\inst{4}
          \and
          Jan Kurpas\inst{1,2}
          \and
          Stefano Ciroi\inst{6}
          \and
          Francesco di Mille\inst{7}
          \and
          Avinash Chaturvedi\inst{1}
          \and
          Arne Rau\inst{4}
          \and
          Zsofi Igo\inst{4}
          \and
          Iuliia Grotova\inst{4}
          \and
          Zhu Liu\inst{8,4}
          \and 
          Kirpal Nandra\inst{4}
          }

   \institute{Leibniz-Institut f\"ur Astrophysik Potsdam (AIP), An der Sternwarte 16, 14482 Potsdam, Germany
         \and
    Potsdam University, Institute for Physics and Astronomy, Karl-Liebknecht-Straße 24/25, 14476 Potsdam, Germany 
    \and
    Nicolaus Copernicus Astronomical Center, Polish Academy of Sciences, ul. Bartycka 18, 00-716 Warszawa, Poland
    \and
    Max-Planck-Institut f\"ur extraterrestrische Physik, Gießenbachstraße 1, 85748 Garching, Germany
    \and
    Dept. Physics, University of Johannesburg, PO Box 524, 2006 Auckland Park, Johannesburg, South Africa
    \and
     Department of Physics and Astronomy, University of Padova, Via F. Marzolo 8, I-35131 Padova, Italy
     \and
     Las Campanas Observatory – Carnegie Institution for Science, Colina el Pino, Casilla 601, La Serena, Chile
     \and
     Centre for Astrophysics Research, University of Hertfordshire, College Lane, Hatfield AL10 9AB, UK
     }        

   \date{Received ...; accepted ...}

% \abstract{}{}{}{}{} 
% 5 {} token are mandatory
 
  \abstract
  % context heading (optional)
  % {} leave it empty if necessary  
    {Supermassive black hole binaries (SMBHBs) separated by (sub)-pc scales represent one of the latest stages of hierarchical galaxy assembly. However, many of these objects are hidden behind large columns of gas and dust at the center of galaxies and are difficult to detect. In these systems, accretion is expected to take the form of two individual accretion disks around the individual black holes, fed by a larger circumbinary disk. The X-ray and UV emission in these systems are predicted to vary regularly on timescales comparable to that of the orbital period of the binary.}
    {This is the first of a series of papers in which we search for and characterize SMBHB candidates based on quasi-periodic light curves from the soft X-ray instrument eROSITA on board the Spectrum-Roentgen-Gamma (SRG) observatory and extensive X-ray follow-up.}
    {We searched the multi-epoch \srgero all-sky surveys for extragalactic sources that show an `up-down-up-down' or `down-up-down-up' profile (from scan to scan) in their 0.2 – 2.3 keV flux light curves. We selected sources where the `up' and `down' flux levels are different by at least $3\sigma$. The `down' states are also allowed to correspond to non-detections. We excluded stellar objects, blazars, and radio-loud AGN via Gaia DR3 parallaxes and proper motions and by visually inspecting images from the Legacy Survey DR10 and the SIMBAD database.
    }
    {We compiled a sample of 16 sources that are suitable for X-ray follow-up campaigns given their brightness and significant variability between bright and faint \srgero flux levels. We triggered extensive \swix and \nic monitoring campaigns on the best SMBHB candidates to confirm or discard their tentative periodicities. Optical spectroscopic observations confirmed the nuclear and extragalactic nature of 15/16 objects and enabled single-epoch SMBH mass measurements and BPT classifications of the dominant ionization in the host galaxy. Our most promising candidate, eRASSt J0530-4125, shows X-ray quasi-periodic variability with a typical time scale of one year in the observed frame. By stacking the X-ray observations of each source in our sample, we find that 14/15 sources can be modeled by a power law with a photon index ranging from $\Gamma\sim1.8-2.8$. 
    Based on our selection, we estimate an optimistic upper limit on the fraction of SMBHB candidates to be $\sim 0.05$ per galaxy. We emphasize that further observational evidence is needed to confirm the SMBHB nature of our sources.} 
    {}
\authorrunning{D. Tubín-Arenas et al.}

   \maketitle
%
%-------------------------------------------------------------------

\section{Introduction}

The formation of supermassive black hole (SMBH) binaries is thought to be a direct consequence of the $\Lambda$CDM cosmological framework where galaxies hierarchically grow through frequent mergers \citep[see e.g.,][]{white+78,white+91}, thus linking large-scale structure formation and gravitational wave physics. 

During a major galaxy merger event, two (or more) SMBHs may be active and embedded in the merging system separated by kilo-parsec scales. This is the so-called dual active galactic nuclei (AGN) phase. Detailed studies of dual AGNs reveal a vast set of astrophysical processes between the active SMBHs and the host galaxies, such as high-velocity outflows \citep{treister+18}, extended AGN ionization in the host galaxy \citep{tubin+21}, or even triple SMBH systems \citep{kollatschny+20}. Statistically, the dual AGN fraction, luminosity, and obscuration all increase with decreasing AGN separation \citep[see e.g.,][]{koss+12,koss+18,ricci+17,derosa+23}.

During the course of the merger, dynamical friction \citep[][]{Chandrasekhar+43} will efficiently remove angular momentum from the SMBH orbits and form an inspiraling binary system at pc separations \citep[][]{Begelman+1980}. At these spatial scales, the SMBHs become gravitationally bound in a SMBH binary (SMBHB). The projected distances between the SMBHB are small and are virtually impossible to resolve for most of the telescope facilities, especially for non-local galaxies.

Since the dynamical friction becomes inefficient due to the low stellar density at pc-scales, the SMBHB can potentially remain orbiting for times longer than the Hubble time. This is known as the final parsec problem \citep{Milosavljevic_2003} and highlights the need for external effects to bring the SMBHs to milli-pc scales. Non-spherical stellar potentials of galaxies \citep[see e.g.,][]{Vasiliev+14,mirza+17}, injection of gas into the system \citep[][]{escala+05,lodato+09} or scattering with stars and massive black holes \citep[see e.g.,][]{ryu+2018} are all processes that could remove angular momentum from the binary.

The recent detection of the gravitational-wave background (GWB) using pulsar timing arrays \citep[PTA;][]{epta+23gw,parkes+23,chinesepta+23} suggests the presence of a population of massive binaries (with masses ranging between $10^{8}-10^{10}\; \rm M_{\odot}$) that are close enough (0.001–0.01 pc separation) to produce gravitational waves in nano-hertz frequencies across the universe \citep[][]{nanograv+23smbhb,epta+24}. The GWB results and the expected population of SMBHBs suggest that there is a mechanism that removes angular momentum from these SMBHBs. This brings them closer until gravitational wave production efficiently removes angular momentum from the binary, eventually leading to mergers on timescales shorter than the Hubble time.

However, the biggest challenge in confirming the existence of SMBHBs and characterizing their demographics is detecting the electromagnetic counterparts of SMBHBs. These objects are usually separated by (sub)-pc scales to be spatially resolved by current facilities and likely obscured at the centers of galaxies. A well-characterized population of SMBHBs could constrain the physics and timescales associated with the gravitationally bounded SMBHs.

The presence of gas surrounding the binary provides a way to detect SMBHBs and their electromagnetic features. Hydrodynamic simulations predict that each SMBH at sub-pc separation accretes material from a circumbinary disk \citep[][]{noble+12,farris+14,farris+15} via accretion inflows, forming smaller and independent mini-accretion disks around each SMBHs \citep[][]{Hayasaki+08,dAscoli+18,bowen+18}. The presence of the two SMBHs will create cavities in the circumbinary disk \citep[][]{dorazio+13}, creating a deficit of optical-UV radiation compared to single-SMBH accretion disks \citep[see, e.g.,][]{gultekin+12,Roedig_2014,wang+23}. X-ray and hard UV emission are expected to originate mainly from the inner mini-disks around each SMBH \citep[][]{sesana+12,dAscoli+18}, while optical and infrared emission come from the circumbinary disk \citep[see e.g.,][]{tanaka+12}. 

Comprehensive reviews on the search and observational evidence of SMBHBs have been reported by \cite{derosa+18}, \cite{bogdanovic+22}, and \cite{DOrazio+23}. \cite{DOrazio+23} also compile a recent and complete summary of the multi-wavelength observational efforts to search for SMBHBs and the known SMBHB candidates. In the following paragraphs, we describe the most relevant multi-wavelength and multi-technique efforts to search for SMBHBs.

Extragalactic sources with periodic light curves have long been discussed and used as a first step to identify close SMBHB candidates. X-ray and UV light curves are expected to vary regularly with periods comparable to that of the binary period \citep[][]{bowen+18,tang+18,bowen+19,kelley+19,Westernacher-Schneider+22} since the mini-disks are periodically fed by gas inflow from the circumbinary disk \citep[][]{Hayasaki+08,farris+14}. Doppler beaming \citep[][]{dorazio+15,charisi+18} or gravitational lensing \citep[][]{haiman+17,dorazio+18} can also produce periodical features in the light curves of SMBHBs. Further evidence for the binary scenario comes from detecting double-peaked or ripples in the Fe K$\alpha$ emission line profile in the X-rays \citep[see, e.g.,][]{sesana+12,McKernan+2013,McKernan+15,2020CoSka..50..219J}. 

However, periodic behavior observed in X-ray light curves can have several origins. One of the most common producers of spurious periodicities is the `normal' AGN variability. Accretion onto single SMBHs produces highly variable X-ray emission with a stochastic behavior \citep[known as red noise, see, e.g.,][]{Press+1978,Vaughan+03}, and this stochastic behavior can mimic strictly- and quasi-periodic variability for a few cycles \citep[][]{Krishnan+21,witt+22}. 
This implies that tests for periodicities can easily yield false positives \citep[][]{Press+1978,Krishnan+21} even if no periodic signals are inherently present in the data. In the presence of real periodicity, deconvolving the red noise and the periodic signal is difficult and can be hampered by the quality and cadence of the data. 

Regular monitoring campaigns spanning several years at high energies are only available in exceptional cases. For example, \cite{Serafinelli+20} and \cite{liu+20} carried out independent systematic searches for periodic variable AGN in hard X-ray (14--195 keV) sources within the 105-month \swib survey. \cite{Serafinelli+20} found two SMBHB candidates with tentative periodicities, namely MCG+11-11-032 and Mrk 915. On the other hand, \cite{liu+20} did not find evidence for periodic AGNs in a slightly larger \swib sample that includes the previous two sources. 

MCG+11-11-032 is a SMBHB candidate initially selected as a dual AGN candidate based on its double-peaked optical emission lines. Its X-ray emission hints at a period of 26 months with a double-peaked Fe K$\alpha$ line (shifted by $\Delta E = 0.4\pm0.2\rm \; keV$) that has been claimed at 4$\sigma$ confidence \citep[][]{Severgnini+18}. Mrk~915, on the other hand, does not show a double-peaked Fe K$\alpha$ line but has a tentative X-ray periodicity of 35 months \citep[][]{Serafinelli+20}. In these objects, the distances between the SMBHBs would be 5 and 6.5 milli-pc, respectively. Another claimed SMBHB candidate is OJ 287 \citep[][]{Sillanpaa+88,dey+19}. Extensive multi-wavelength monitoring observations, including \swix campaigns, have been implemented on OJ 287 to claim the presence of an eccentric secondary SMBH orbit with a mass ratio of $q\simeq 0.01$ that crosses the accretion disk of the main SMBH twice every $\sim$12 years \citep[see e.g,][for the project MOMO: multi-wavelength Observations and Modeling of OJ 287]{Komossaxray+21,Komossa+21}.

Periodicities in other wavelengths have also been proposed as tracers of SMBHBs. Optical photometric variability studies performed on survey data provide a vast initial sample of  $\sim150$ periodic SMBHB candidates \citep[see e.g.,][]{graham+15,charisi+16,liu+16}. Most of these candidates are have timescales in the regime where the robust separation of red noise from periodicities is difficult, as stochastic accretion effects can lead to spurious signals given a limited observation duration or gappy data sampling \citep[][]{Vaughan+16,Liu_2018,Krishnan+21}. Recently, \cite{luo+2025} conducted the first systematic search for SMBHB candidates with periodic light curves from Wide-field Infrared Survey Explorer (WISE) IR data. However, their mock simulation of the parent sample indicates that stochastic variability can produce a similar number of periodic sources. Periodicity observed in radio-loud sources has been attributed to the presence of a SMBHB in the center of the galaxy but could also be due to other single AGN processes \citep[][]{holgado+18} such as jet precession \citep[][]{caproni+13,liska+18} or intrinsic variability of the accretion flow in the accretion disc \citep[][]{king+13}. The presence of SMBHBs have also been invoked to explain periodicities of a few cycles that have been found in radio data of individual blazars \citep[see e.g.,][]{ren+21,oneill+22,Kiehlmann2024,Parra2024}.

Spectroscopic features like line asymmetries, double-peaks, velocity shifts, or line profile variability on broad emission lines such as H$\beta$ or H$\alpha$ have been proposed as evidence of orbital motion and a consequence of a black hole binary in the center of AGNs \citep[][]{Tsalmantza+2011,Eracleous+2012,shen+13,liu+14,runnoe+17,doan+20,2025AdSpR..75.1441J}. One example of this is the Seyfert galaxy NGC 4151. Radial velocity curves show flux variations in the H$\alpha$ light curve that suggest orbital motion consistent with the signature of a sub-pc SMBHB system with an orbital period of about $\sim5700$ days \citep{Bon_2012}.
However, in some sources, radial velocity shifts of the broad-line components are found to be inconsistent with the expected orbital motion of a SMBHB \citep[][]{Wang_2017}. \cite{Eracleous+09} and \cite{Liu_2016} find evidence that double-peaked line profiles can also be produced by single (non-binary) AGN broad line regions(BLR). 
Detailed modeling and extensive time-resolved spectroscopic studies are still needed to understand the spectral features expected for SMBHBs \citep[see e.g.,][for a detailed discussion on the hypothesis and challenges using broad emission lines as tracers of SMBHBs]{POPOVIC201274,derosa+18}.

Periodic X-ray emission seems to be one of the most promising ways to identify SMBHB candidates. The X-ray photons can penetrate obscuring gas and dust, and in SMBHBs are mainly emitted in regions very close to the orbiting SMBHs. Therefore, the X-ray emission is expected to vary regularly with periods comparable to that of the binary period \citep[e.g.,][]{Hayasaki+2007,MacFadyen+2008,dorazio+13,farris+14}. However, the X-ray coverage of the sky and the regular monitoring of X-ray sources have been limited. The launch of the extended ROentgen Survey with an Imaging Telescope Array \citep[eROSITA;][]{merloni+12,predehl+21} has opened a window to the search and discovery of peculiar AGNs in one of the largest and most sensitive X-ray catalogs to date. eROSITA is the soft X-ray instrument on board the {\it Spectrum-Roentgen-Gamma} \citep[SRG;][]{sunyaev+21} observatory and provides an unprecedented opportunity for regular, all-sky X-ray monitoring thanks to its scanning strategy. Therefore, eROSITA enables extensive time-domain studies of the X-ray sky. The survey has already been used to compile a catalog of highly variable extragalactic transients between the first and second eROSITA sky scans \citep{grotova+25}.

This work is the first in a series of papers that aim to find and confirm SMBHBs based on data from one of the largest X-ray all-sky catalogs ever obtained. Here, we exploit the \srgero all-sky data to identify extragalactic objects that show quasi-periodic X-ray light curves as a first attempt to compile a list of SMBHB candidates. We obtain additional extensive X-ray follow-up, optical spectra, and complementary multi-wavelength data to constrain the nature of the selected sources. 
According to the current theories, a confirmation of the SMBHB scenario is based on the detection of i) persistent quasi-periodic X-ray signatures for multiple periods and ii) the confirmation of double-peaked Fe K$\alpha$ lines that shift in energy with the orbital period of the SMBHB. We note that there is other interesting proposed evidence for SMBHB, such as the periodic ripples in the line profile, which consist of a pair of dips in the broad line blue-ward and red-ward from the line centroid energy \citep[see e.g.,][]{Jovanovic+2014,McKernan+15,2020CoSka..50..219J}. However, high signal-to-noise X-ray observations are needed to observe these features \citep[see e.g.,][]{McKernan+15}.
 Since our selection is mainly based on quasi-periodic behavior that can also be produced by stochastic accretion, a detailed assessment of the red noise in our sample selection will be made in the second paper of the series (Tubín-Arenas et al. in prep.). Upcoming papers will also present longer and deeper X-ray follow-up data obtained after the time of writing this paper. 

The paper is structured as follows: Sect. 2 describes the \srgero scanning pattern and how we search for quasi-periodic signals within the eROSITA data. In Sect. 3, we present the additional infrared, optical, UV, and X-ray data for the selected objects.
Since we continuously monitor the sources in X-rays, this paper is based on all X-ray data obtained until Dec. 31st, 2024. In Sect. 4, we analyze the multi-wavelength data from follow-up observations. In Sect. 5, we present the optical and X-ray properties of the selected sample, including a detailed discussion of the two most well-monitored SMBHB candidates. We discuss the nature of the variability from different physical perspectives and constrain their abundance in Sect. 6. Finally, we present our conclusions in Sect. 7. 

Throughout this paper, we assume a $\Lambda$CDM cosmology with $h_{0}=0.7$,
$\Omega_{m}=0.27$ and $\Omega_{\Lambda}=0.73$ \citep{2009ApJS..180..225H}. Unless explicitly stated otherwise, we report all errors corresponding to 68\% confidence (1$\sigma$) intervals. 
 
\begin{figure*}[ht]
\centering
\includegraphics[width=\textwidth]{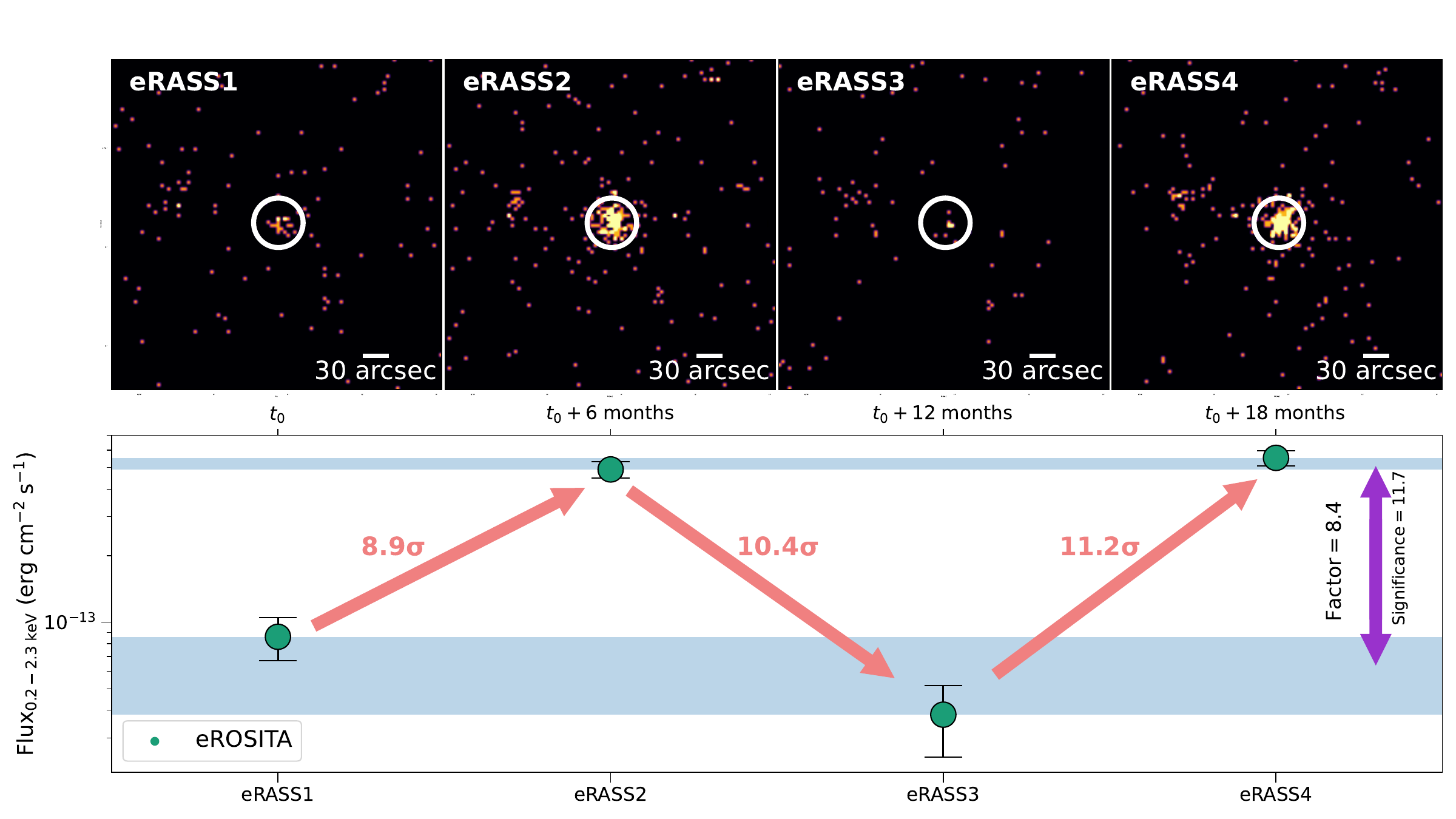}
\caption{Visualization of the selection for SMBHB candidates based on their \srgero light curves. {\it Top panel:} Observed counts at the position of the source eRASSt J0530-4125 (see Table \ref{table:bestcandidates}) at every \srgero scan in the 0.2--2.3 keV energy band. Each image is temporarily spaced every 6 months. The white circles correspond to a typical aperture with a radius of $30\arcsec$, similar to the scale bar displayed on the bottom right side of each panel. {\it Lower panel:} X-ray light curve of eRASSt J0530-4125. The colored arrows indicate the significance of the flux change between the consecutive scans. The blue-shaded areas indicate the bright and faint flux levels of the source, while the purple arrow shows the factor and significance between these two flux levels. }\label{fig:countsvariability}
\end{figure*}

\section{\srgero and the SMBHB quest}

\subsection{Overview of eROSITA all-sky scans}

\srgero combines a large field of view (FoV $\sim 1^{\circ}$), an effective scanning observation mode, and high sensitivity in the soft X-ray band (0.2--2.3 keV), making it the most efficient imaging survey telescope in X-rays \citep[][]{merloni+12,predehl+21}. By February 2022, \srgero completed four and one-third of the planned eight eROSITA all-sky scans \citep[eRASS;][]{predehl+21}. \srgero executes one all-sky scan by rotating around the observing axis (which is perpendicular to the axis that connects the spacecraft with the Sun) every 4 hours (referred to as one ``eROday'') and shifting its orbit by 1 degree per day, covering the whole sky in approximately 182 days (half a year). Each position close to the ecliptic equator is observed typically six times for up to 40 seconds per eROday, while sources close to the ecliptic poles are observed more often, creating regions of deep exposure \citep[see Fig. 3 in ][]{merloni+24}.

Following this strategy, eROSITA detected $\sim$1 million sources during the first all-sky survey on the Western Galactic hemisphere \citep[eROSITA-De: with Galactic longitudes between $180\lesssim l \lesssim 360$,][]{merloni+24}. For non-detected, variable, or transient sources that fall below the detection threshold of the eSASS pipeline, \cite{tubin+24} provide \srgero upper flux limits based on X-ray aperture photometry on the eROSITA standard calibration data products (counts image, background image, and exposure time). 
These detected fluxes and upper limits are used in the search for periodically variable SMBHB candidates.

\subsection{Sample selection}\label{sec:selection}

\subsubsection{eROSITA light curve selection}

When the four completed all-sky scans are stacked, the sensitivity increases, resulting in approximately 3 million detected sources. The eRASS:4 catalog (stacked version of the eRASS1 to eRASS4 catalogs — an internal eROSITA-De consortium product, version 230427) is the deepest, most sensitive, and most complete all-sky X-ray catalog available at 0.2--2.3 keV. It also provides relevant information such as positions, observation dates, and fluxes for each independent eRASS survey used in the stacking process. Therefore, the eRASS:4 catalog delivers immediate eROSITA X-ray light curves.

We explore this eROSITA stacked catalog with epoch information to search for quasi-periodic signals. Since eROSITA observed the sky every 6 months for almost two and a half years, we are sensitive to quasi-periods of $\sim 1$ year, as well as shorter period harmonics like $\sim4$ months. Therefore, we search for sources that show an `up-down-up-down' or `down-up-down-up' profile in their eROSITA light curve. The top panel of Fig.  \ref{fig:countsvariability} shows an example of an extragalactic source that follows such a flux pattern. 

Our eROSITA source selection is based on three criteria:

1) We request the consecutive `up' and `down' flux measurements to be different by $>$3$\sigma$ for an object to be preliminary included in our initial sample of candidates. Thus, 
\begin{align*}
    \frac{F_{i}-F_{i+1}}{\sqrt{F_{i,err}^{2}+F_{i+1,err}^{2}}}>3,
\end{align*}
\noindent where $i=1,2,3,4$ is the corresponding eRASS scan, $F_{i}$ and $F_{i,err}$ are the flux and flux error for the eRASS scan $i$ of a given source. In the case where the source falls into the sky region where we have five eROSITA scans, we require the above-mentioned criterion for eRASS1 to eRASS5 ($i=1,2,3,4,5$). Figure~\ref{fig:countsvariability} illustrates this selection criterion as the source varies by almost $10\sigma$ between the flux measurements spaced apart by 6 months. The `down' states are also allowed to be non-detections. In these cases, we use the corresponding 3$\sigma$ eROSITA upper flux limit values in the 0.2--2.3 keV band \citep{tubin+24}.

For the next two selection criteria, we first define `bright' and `faint' states based on the average flux of all the `up' data points and the average flux of all the `down' data points for each source. The uncertainty ranges of the `bright' and `faint' states ($F_{bright,err}$, $F_{faint,err}$, respectively) are given by the standard deviation of the individual fluxes contributing to the corresponding level and are represented by the blue-shaded areas in the bottom panel of Fig.~\ref{fig:countsvariability}. These uncertainties also give us an idea of how consistent the flux measurements are in either of the two flux states. 

2) We compute a factor $F$ that corresponds to the ratio between the average fluxes of the bright and the low flux levels for every source ($F=F_{bright}/F_{faint}$). Only sources with $F>3$ are selected.

3) We also calculate significance 
\begin{align*}
    S=\frac{F_{bright}-F_{faint}}{\sqrt{F_{bright,err}^{2}+F_{faint,err}^{2}}}
\end{align*}
\noindent that is the flux difference normalized by the error between the bright and faint flux levels for every source. Here, we require a value of at least $S=3$.

\subsubsection{Excluding galactic sources}
As a first attempt to remove a significant fraction of stellar sources from our light curve selection, we use the proper motion and parallax values from {\it Gaia} DR3 \citep{giadr3}.
We cross-match our eROSITA sources to {\it Gaia} sources using a 15-arcsecond matching radius. If multiple {\it Gaia} sources are found for one eROSITA source, we consider the closest {\it Gaia} source as the likely counterpart. Since we later visually inspect the remaining candidates, we emphasize that this matching procedure only helps to reduce the workload of the visual screening. 

We exclude stellar sources by requesting our sources to have a parallax and proper motion (R.A. and Dec. combined) significance consistent with zero within $3\sigma$ and $5\sigma$, respectively. Sources without {\it Gaia} counterparts are kept in our sample, as it is unclear whether they are faint, distant stellar objects or of extragalactic origin. 

As a next step, we visually inspected the Legacy Survey Data Release 10 \citep[LS DR10;][]{LSDR10} images of the remaining sources. For objects not covered in LS DR10, we inspected the Digitized Sky Survey \citep[DSS][]{dss} images. For a few sources, bright saturated stars were identified as their counterparts. We rejected these objects from our candidate sample. We also checked the SIMBAD database \citep[][]{simbad} for the remaining objects. If a stellar object was listed within 15 arcseconds of our X-ray position, we also excluded these objects from our candidate sample. The SIMBAD database was also used to identify known blazars and radio-loud AGN. The latter category was removed as well since jet precession could also produce periodic X-ray emission \citep[][]{liska+18}. As a result of these steps, we are confident that our sample contains little to no contamination from galactic objects.

\begin{figure}
    \centering
    \includegraphics[width=\linewidth]{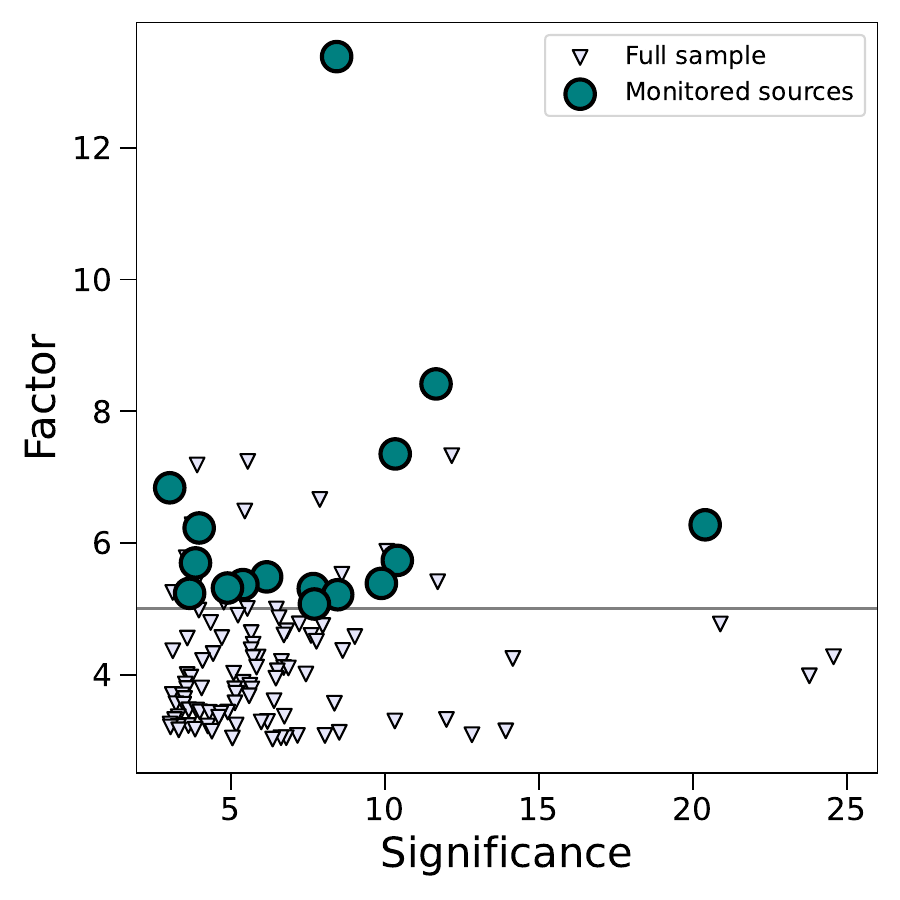}
    \caption{Distribution of the sample based on the factor ($F=F_{bright}/F_{faint}$) and significance ($S=\frac{F_{bright}-F_{faint}}{\sqrt{F_{bright,err}^{2}+F_{faint,err}^{2}}}$) values (see Sect. \ref{sec:selection}). For each source, the factor and significance are calculated between the `bright' and `faint' states represented by the blue-shaded area in Fig. \ref{fig:countsvariability}. We make a cut at $F>5$ and highlight the 16 sources that will be the focus of this paper. 
    The remaining extragalactic sources of the sample are displayed with triangles. }
    \label{fig:sample}
\end{figure}

\subsubsection{Primary SMBHB Candidate Sample}

The candidate sample consists of 119 extragalactic objects. Figure~\ref{fig:sample} shows the distribution of their factors $F$ and significances $S$. All objects shown fulfill our selection criteria of at least 3$\sigma$ difference in X-ray flux between each adjacent epoch, $F>3$, and a significance of at least $S>3$. 

Extended light-curve information for all sources is needed to confirm or rule out each quasi-periodic light curve signature. However, this would require a massive X-ray follow-up campaign. Therefore, we decided to focus our X-ray follow-up resources on the brightest and most variable objects. The objects must have a $F>5$, an X-ray flux in the bright state $F_{bright} \gtrsim 5\times \rm 10^{-13}\; erg\; s^{-1}\; cm^{-2}$ to allow a follow-up program with missions like \swi and \nic, or have already publicly available follow-up data. This then defines our primary SMBHB candidate sample of 16 sources highlighted in Fig.~\ref{fig:sample}. 
A summary of their properties is given in Table \ref{table:bestcandidates}.

\begin{table*}
\small\addtolength{\tabcolsep}{+1.0pt}
\caption{Primary SMBHB candidates catalog.} 
\label{table:bestcandidates} 
\centering
\scalebox{0.95}{
\begin{tabular}{lrrccrrrl}
\hline
\hline
Name & RA(2000) & Dec(2000) & Radec\_err& Separation & Sig. & Factor & Detection & X-ray follow-up\\
&($^{\circ}$)&($^{\circ}$)&($68\%$)&($\Delta$\arcsec)&&&Likelihood&\\
\hline \\
1 - eRASSt J0344-3327 & 56.107460 & -33.45521 & 1.09& 0.17 & 8.4 & 13.4 & 473.7 & \swix, \nic \\
2 - eRASSt J0530-4125 & 82.56909 & -41.42582 & 0.84 & 1.87 & 11.7 & 8.4 & 1106.2 & \swix, \nic \\
3 - eRASSt J1906-4850 & 286.53654 & -48.84064 & 0.49& 1.02 & 10.3 & 7.4 & 3788.3&  \swix, \nic \\
4 - eRASSt J0432-3023 & 68.07197 & -30.39961 & 0.63 &1.23& 3.0 & 6.8 & 1837.3 & \swix, \nic\\
5 - eRASSt J1130-0806 & 172.61730 & -8.10278 & 1.18&2.33 & 20.4 & 6.3 & 420.9 & -- \\
6 - eRASSt J1522-3722 & 230.56415 & -37.37989 & 0.85&1.24 & 4.0 & 6.2 & 732.1 & \nic\\
7 - eRASSt J1141+0635 & 175.47668 & 6.58614 & 0.94&1.07 & 10.4 & 5.7 & 774.8 & \swix, \nic, {\it XMM} \\
8 - eRASSt J0458-2159 & 74.66770 & -21.99204 & 0.34&0.98 & 3.9 & 5.7 & 8437.1 &  \nic\\
9 - eRASSt J2227-4333 & 336.98212 & -43.56079 & 1.42&1.35 & 6.2 & 5.5 & 271.5 & -- \\
10 - eRASSt J1124-0348 & 171.23409 & -3.81142 & 1.17&0.93 & 9.9 & 5.4 & 509.8 &  \nic\\
11 - eRASSt J0036-3125 & 9.05142 & -31.41707 & 1.26&1.30 & 5.4 & 5.4 & 413.2 & -- \\
12 - eRASSt J1003-2607 & 150.82466 & -26.12018 & 1.08&1.62 & 4.9 & 5.3 & 709.1 & \swix, \nic\\
13 - eRASSt J0818-2252 & 124.74048 & -22.87709 & 0.52&1.36 & 7.7 & 5.3 & 5886.2 & \swix, \nic\\
14 - eRASSt J0600-2939 & 90.23452 & -29.65689 & 1.08&2.88 & 3.7 & 5.2 & 580.3 &  --\\
15 - eRASSt J0044-3313 & 11.20658 & -33.23234 & 1.11&0.60 & 8.5 & 5.2 & 691.3 &  \swix\\
16 - eRASSt J0614-3835 & 93.71599 & -38.59458 & 0.73&1.33 & 7.7 & 5.0 & 1450.6 &  --\\
\hline
\end{tabular}}
\tablefoot{Table with the eROSITA names of the sources (column 1) and X-ray positions based on the stacked eRASS:4 coordinates (column 2 and 3). The ``Radec\_err'' values (column 4) correspond to the combined positional $1\sigma$ 68\% error derived from the PSF-fitting performed by the eROSITA source detection pipeline. The separation (column 5) indicates the distance, in arcseconds, between the eROSITA and {\it Gaia} DR3 positions.
We include the Significance (column 6) and Factor (column 7) parameters used in our SMBHB selection described in Sect. \ref{sec:selection}. 
We also show the detection likelihood (column 8) from the eRASS:4 catalog, which corresponds to the negative log-likelihood probability that the source counts are produced by background fluctuations. Larger detection likelihoods correspond to more observed counts relative to the background and, therefore, a smaller likelihood that the observed counts were produced by background fluctuations. 
Finally, we also report the X-ray missions used to monitor the flux evolution of the candidates (column 9). Five sources have not been observed. However, they will be targeted by \nic during cycle 7. The candidates are ranked according to their factor $F$ values.}
\end{table*}

\section{Data and data reduction}\label{followup}

Our X-ray monitoring programs aim to observe the candidates once every month for 5 months (six pointings in total) to fill in the eROSITA gaps and increase the observing cadence. We begin monitoring in the month that eROSITA would have observed the source again if the survey had continued after Feb. 2022. 
Therefore, assuming that the pattern of the eROSITA light curve continues over time, we can study the X-ray variability and emission properties of the candidates in more detail. We use these follow-up observations to discriminate between tentative quasi-periodic signals and red-noise-driven variability. Sources displaying tentative periodicities after an initial 1-month cadence follow-up are eligible for X-ray monitoring programs with a higher cadence.

In addition to X-ray follow-up, we obtained multi-wavelength imaging, photometry, and spectroscopic data for the final SMBHB candidate sample. Optical spectra were obtained for known and previously unclassified objects to confirm their extragalactic nature. Here, we describe the multi-wavelength data used in this paper. 

\subsection{Infrared data}

Mid-infrared (mid-IR) photometry data was collected from the NEOWISE-R mission \citep{Wright_2010,Mainzer+14} in the W1 and W2 bands, centered at 3.4 and 4.6 microns, respectively. The data were retrieved via the unTimely catalog \citep[][]{Meisner_2023} that are binned by $\sim180$ days to match the scanning pattern of the satellite. The infrared light curves are presented in Appendix \ref{appendix:particularcases}.

\subsection{Optical}\label{opt}

\subsubsection{Optical images}

\begin{figure*}
    \centering
    \includegraphics[width=1\linewidth]{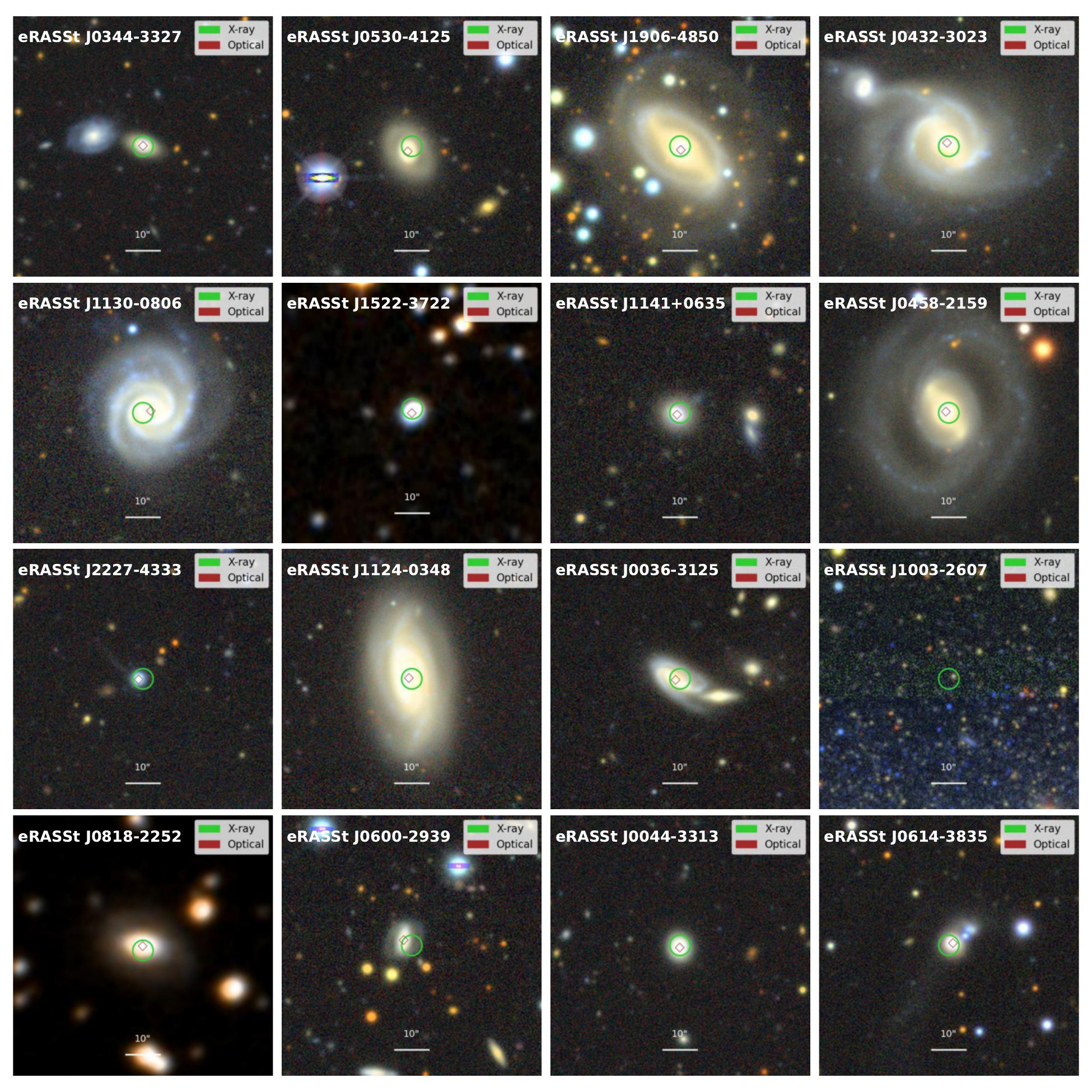}
    \caption{Optical images of the SMBHB candidates. The cutout images were retrieved from the Legacy Survey Viewer with a pixel scale of 0.15 arcseconds per pixel and are centered at the \srgero X-ray position of the candidates. We highlight the X-ray position with a {\it green} circle with a radius of 3\arcsec~that corresponds to a typical $3\sigma$ positional uncertainty and the optical {\it Gaia} DR3 positions with a {\it red} diamond when available. Two sources (eRASSt J1522-3722 and eRASSt J0818-2252) are outside the LS footprint, therefore, we show Digitized Sky Surveys (DSS) images. eRASSt J1003-2607 was not detected by {\it Gaia} DR3 since the tentative source position highlighted inside the green circle is a faint source with a LS DR10 magnitude of $m_r\sim 22$. A 10\arcsec~scale bar is also shown at the bottom of the images for reference. All images are oriented with North up and East to the left.}
    \label{fig:lsdr10images}
\end{figure*}

\begin{table*}
\small\addtolength{\tabcolsep}{+1.0pt}
\caption{Optical spectral properties of the main SMBHB candidates. } 
\label{table:opticaltable} 
\centering
\begin{tabular}{lcclllll}
\hline
\hline
Name & Telescope & Observation& z & X-ray Luminosity & \multicolumn{2}{c}{SMBH mass $\rm \log (M_{bh}/M_{\odot}$)} &\\
&&date&&($\times10^{42}\rm \; erg\; s^{-1}$)&$(\rm H\beta)$\tablefoottext{a}&($\rm H\alpha$)\tablefoottext{a}&\\
\hline \\
1 - eRASSt J0344-3327&NTT/EFOSC2&17/02/2024&0.090&$0.25 - 8.21$&--&($6.95\pm2.55$)\tablefoottext{b}&\\
2 - eRASSt J0530-4125&NTT/EFOSC2&17/02/2024&0.076&$0.54-7.76$&($7.26\pm0.48$)&($7.41\pm1.21$)\\
3 - eRASSt J1906-4850& Magellan/Baade&26/10/2021&0.049&$1.73-23.80$&$6.63\pm0.25$\tablefoottext{c}&($6.72\pm0.50$)\tablefoottext{c}&\\
4 - eRASSt J0432-3023&SAAO&28/02/2024&0.055&$0.47-9.03$&$7.31\pm0.19$&($7.08\pm0.78$)\\
5 - eRASSt J1130-0806&SAAO&28/02/2024&0.037&$0.23 - 1.87$&--&$7.24\pm0.30$\\
6 - eRASSt J1522-3722&SAAO&19/03/2023&0.064&$1.37-10.85$&$7.35\pm0.09$&$7.17\pm0.02$\\
7 - eRASSt J1141+0635&NTT/EFOSC2&16/02/2024&0.101&$3.28- 25.64$&($7.87\pm0.55$)&($7.41\pm0.48$)\\
8 - eRASSt J0458-2159& Magellan/Baade&11/12/2020&0.040&$2.13-17.64$&$6.89\pm0.05$&$6.85\pm0.08$\\
9 - eRASSt J2227-4333&6dF&29/06/2003&0.198&$8.82-59.32$&$7.53\pm0.22$&$7.46\pm0.21$\\
10 - eRASSt J1124-0348&NTT/EFOSC2&18/02/2024&0.021&$0.02-0.77$&($6.14\pm0.9$)&$6.65\pm0.18$\\
11 - eRASSt J0036-3125&SAAO&06/09/2023&0.108&$1.59-15.76$&$5.93\pm0.12$&$7.52\pm0.26$\\
12 - eRASSt J1003-2607&LBT/MODS&02/01/2025&0.001&0.0007--0.005&--&--\\
13 - eRASSt J0818-2252&SAAO&02/12/2023&0.034&$1.49-13.64$&$7.49\pm0.22$&($6.99\pm0.48$)\\
14 - eRASSt J0600-2939&SAAO&22/11/2024&0.104&$1.51-17.11$&-&$7.47\pm0.11$\\
15 - eRASSt J0044-3313&SAAO&19/10/2022&0.107&$2.09-20.96$&$7.45\pm0.09$&($6.64\pm0.47$)\\
16 - eRASSt J0614-3835&SAAO&30/11/2023&0.054&$1.14-6.83$&$6.03\pm0.16$&($6.49\pm0.58$)\\
\hline
\end{tabular}
\tablefoot{Table with sources (column 1) and the corresponding telescopes (column 2) and date (column 3) from which the optical spectra were obtained. The spectroscopic redshifts (column 4) were used to calculate the eROSITA 0.2--2.3 keV luminosities (column 5) in the faint and bright phases of the candidates. Single-epoch SMBH masses (column 6 and 7) were obtained based on the spectral fit of the broad emission lines. 
\tablefoottext{a}{The single-epoch SMBH masses are calculated based on the scaling relations presented in \cite{Vestergaard+06} and \cite{mejiarestrepo+22} for H$\beta$ and H$\alpha$, respectively. We report only statistical uncertainties for the SMBH mass measurements. An additional 0.3 dex should be included to consider the systematic uncertainties. }
\tablefoottext{b}{SMBH mass estimates in parenthesis are not well-constrained due to large uncertainties in the line measurements.}
\tablefoottext{c}{The source presents double-peaked broad Balmer lines. }
}
\end{table*}

Figure \ref{fig:lsdr10images} shows the optical images of the primary SMBHB candidates retrieved from the Legacy Survey Viewer\footnote{\url{https://www.legacysurvey.org/viewer}} \citep{LSDR10}. We mark the \srgero X-ray position of the candidates, listed in Table \ref{table:bestcandidates}, and the position of the optical counterpart detected by {\it Gaia}. All X-ray positions are consistent with their optical counterpart positions, considering $3\sigma$ eROSITA positional uncertainties. 
This gives strong support to the idea that the X-ray emission is likely coming from the nuclear regions of the galaxy.

\subsubsection{Spectroscopic data}

Optical spectroscopic observations were collected from the archive or newly obtained if missing. We aimed to confirm the extragalactic
nature of our candidates, obtain redshifts, and thus study the intrinsic AGN properties of our sources, such as luminosity and SMBH mass. 
Column 2 of Table \ref{table:opticaltable} lists the facilities used to obtain the optical spectroscopic observations of our SMBHB candidates. Four systems were observed with the ESO/La Silla EFOSC2 on the 3.6 m New Technology Telescope (NTT; program ID: 112.263K). The spectroscopic observations were performed with grating \#13 and a 1\arcsec~slit width, which corresponds to a wavelength resolution of 15.5 \AA~ in the range 3650--9250 \AA. The ESO observations were reduced with the dedicated software ESOReflex \citep[version 2.11.5,][]{freudling+13} to correct for the standard bias, flats, wavelength, and flux calibration. 

Eight candidates are bright and could, therefore, be observed with the SpUpNIC spectrograph \citep{crause+19} at the South African Astronomical Observatory (SAAO) 1.9 m telescope. 
The SAAO/SpUpNIC spectra were obtained with a 2.7\arcsec~slit width, a typical exposure time of 2 consecutive 1200 seconds, and a low-resolution grating to cover the entire optical range between $\sim3500$~\AA~and $\sim9000$~\AA. The data were reduced using standard bias corrections and flat-fielding. The wavelength and spectrophotometric calibrations were performed using an Argon arc-lamp and a standard star spectra taken on the same night of observations.

Two spectra were taken with the IMACS Short-Camera \citep{dressler+11} mounted on the 6.5 m Baade Magellan telescope located at Las Campanas Observatory. The Magellan/Baade observations were carried out using a 1\arcsec~and 0\farcs7 slit with total exposures of 600 and 1080 seconds for eRASSt J1906-4850 and eRASSt J0458-2159, respectively. The spectra were reduced with IRAF \citep{iraf+1986} following the usual procedure of overscan subtraction, flat-field correction, wavelength calibration using a He-Ne-Ar lamp, and a spectrophotometric standard star.

One optical spectrum (eRASSt J2227-4333) was retrieved from the archive of the 6dF Galaxy Survey \citep[][]{Jones+2004,Jones+2009}.

Given its faint nature, we observed eRASSt J1003-2607 with the 8-meter class Large Binocular Telescope (LBT) using long-slit spectroscopy with the MODS spectrograph \citep{Pogge+2010}. The observation was carried out on January 2nd, 2025, in one exposure of 3000 s with both the red and blue arms. The basic data reduction was done with the modsCCDred package\footnote{\url{https://github.com/rwpogge/modsCCDRed}}. 
We wavelength-calibrated the data, extracted the spectrum, and performed flux calibration using the {\it apall} package from IRAF. 

\begin{figure*}
    \centering
    \includegraphics[width=0.9\linewidth]{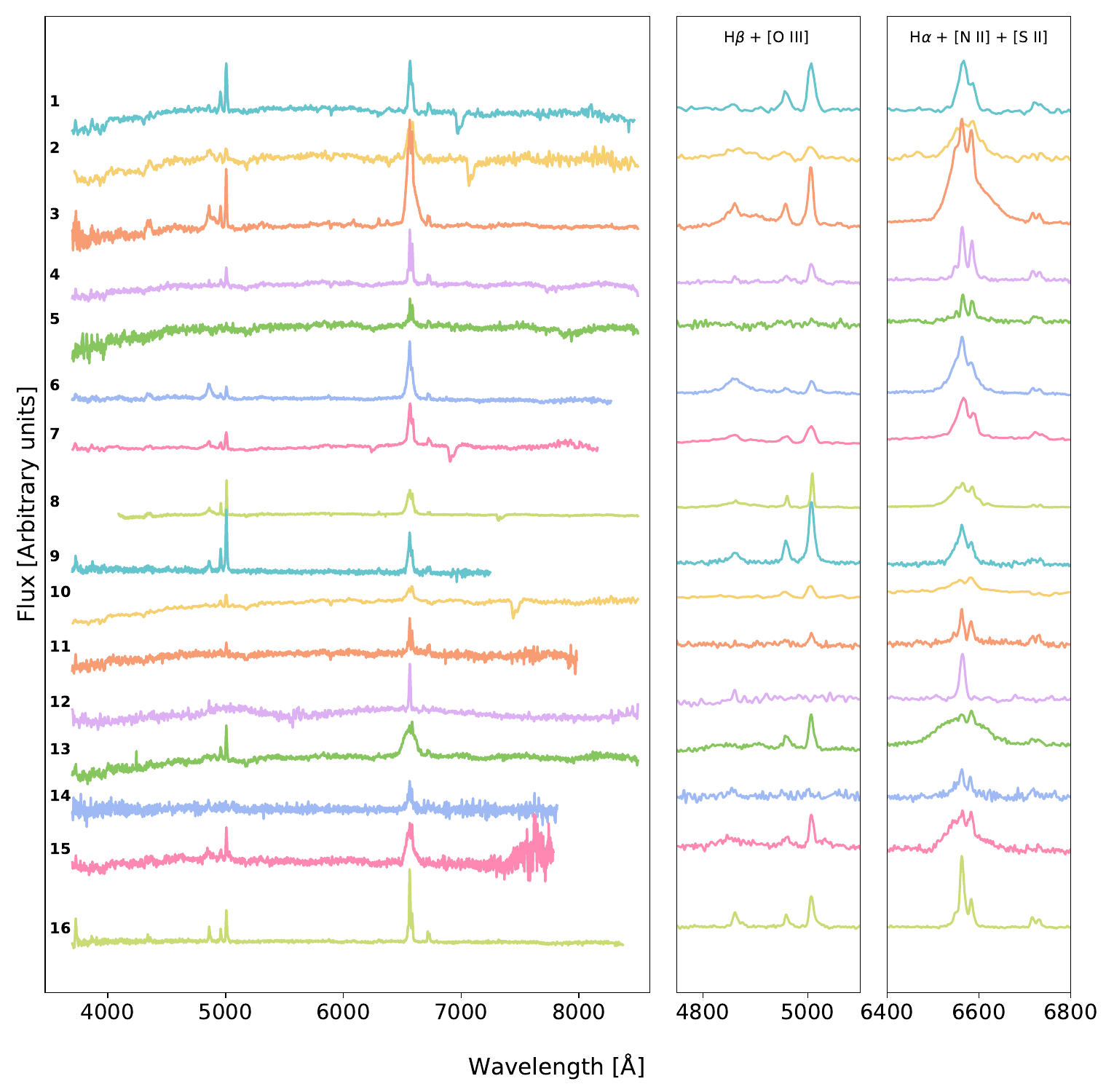}
    \caption{Optical spectra compilation of the best SMBHB candidates in their rest frame. The spectra are sorted from top to bottom based on the column factor $F$ from Table \ref{table:bestcandidates}, described in Sect. \ref{sec:selection}. The middle and right panels zoom into the H$\beta$ + [O \textsc{iii}] and the H$\alpha$ + [N \textsc{ii}] + [S \textsc{ii}] emission lines complex of each spectrum, respectively. The numbers in the plot correspond to every source listed in Table \ref{table:bestcandidates}. 
    The flux is given in arbitrary units where the spectra were scaled by different factors and added a constant offset of a few 10$^{-16}$ erg s$^{-1}$ cm$^{-2}$ $\AA^{-1}$ to visualize better the different spectral shapes of the SMBHB candidates. }
    \label{fig:opticalspectra}
\end{figure*}

Based on its LBT optical spectrum, we ruled out the AGN nature of eRASSt J1003-2607. We show in Fig. \ref{fig:opticalspectra} and in Sect. \ref{indiv:eRASSt_J1003-2607} that eRASSt J1003-2607 shows only two strong emission lines consistent with H$\beta$ and H$\alpha$ at the redshift of a nearby galaxy, NGC 3109 ($z=0.0014$). Therefore, the source is extragalactic but has a non-AGN off-nuclear origin, likely connected to an X-ray ultra-luminous object (see Sect. \ref{sec:discussion} and Sect. \ref{indiv:eRASSt_J1003-2607} for a detailed discussion about off-nuclear sources and this particular object, respectively). This source is, therefore, removed from the subsequent discussion of SMBHBs candidates and only described in Sect. \ref{indiv:eRASSt_J1003-2607}.

Figure \ref{fig:opticalspectra} shows the rest-frame optical spectra for the primary SMBHB candidates listed in Table \ref{table:bestcandidates}. We note that, visually, some sources show broad Balmer lines, indicating the presence of a type I AGN, while others show only narrow emission lines with a galaxy-like continuum.

\subsubsection{Optical photometric data}

Optical photometric data are available for most sources, providing helpful insight into the properties of the candidates. We obtained publicly available forced-photometric data on the target images from the Asteroid Terrestrial-impact Last Alert System \citep[ATLAS,][]{tonry+18,Smith+20} all-sky survey from their forced photometry pipeline\footnote{https://fallingstar-data.com/forcedphot/} \citep[][]{Shingles+21}. ATLAS allows us to have long optical photometric light curves of the candidates in the filters cyan (420--650 nm) and orange (560--820 nm). Atlas has been surveying the entire sky since $\sim2015$ with a cadence of $\sim2$ days. The data points were extracted by running forced photometry on the available reduced images and binned every 15 days. 

We also obtained Zwicky Transient Facility \citep[ZTF][]{Masci_2019} optical light curves in the $r$-band. Since ZTF is a Northern-equatorial sky survey, only five sources have data available. The light curves are shown for these individual cases in Figs.~\ref{fig:erasst_j0344-3327}, \ref{fig:multiwavelc}, and \ref{fig:eRASSt_J1141+0635} (see Appendix \ref{appendix:particularcases}).

\subsection{\swiu data}

SMBHB candidates that were observed with \swi (see Sect. \ref{subsect:xrays}) also have optical and UV photometric data in six filters: V, B, U, UVW1, UVM2, and UVW2 (with central wavelengths 5468, 4392, 3465, 2600, 2246, and 1928 \AA, respectively) thanks to the Ultra-violet Optical Telescope \citep[UVOT,][]{uvot+2005} mounted in the spacecraft. We obtained magnitudes and fluxes using the task \textsc{uvotsource}, which performs aperture photometry at the position of the source. The routine \textsc{uvotsource} extracts counts from an aperture of 5\arcsec~and corrects them by the background counts extracted from a 20\arcsec~circular aperture in a source-free region close to the source of interest. The UVOT light curves are shown in Figs.~\ref{fig:multiwavelc} and \ref{fig:eRASSt_J1141+0635}.

\subsection{X-rays}\label{subsect:xrays}

We use X-ray data from several different missions, including eROSITA, {\it Neil Gehrels Swift Observatory} \citep{Gehrels+2004}, {\it Neutron star Interior Composition Explorer}, and \textit{XMM-Newton} \citep[][]{xmm}. Here, we describe their data reduction. 

The eROSITA fluxes in the 0.2--2.3 keV band are extracted from the stacked eRASS:4 catalog with epoch information. For the two most extensively monitored sources, we also extract 2.3--5.0 keV and 0.2--5 keV fluxes by reducing and analyzing their eROSITA spectra (eSASS task \texttt{srctool}). This data is processed with the eROSITA standard processing pipeline version c020, which is based on the eROSITA Standard Analysis Software System \citep[eSASS,][]{brunner+22}. 
The source and background spectra are extracted from a 60\arcsec~circular region and a source-free annulus region with an inner radius of 140\arcsec~and an outer radius of 240\arcsec, respectively. We only retrieve X-ray spectra of the individual eRASS observations in the bright flux levels since there are not enough counts in the faint flux states of the sources. In these low-count scans, we report the corresponding eROSITA 3$\sigma$ upper limits in the 2.3--5.0 keV and 0.2--5 keV bands.

\swix observations were performed during cycles 19 and 20 (Target ID: \#16217, \#97405) and using the director's discretionary time (DDT) in the standard photon counting (PC) mode. For each observation, we retrieve publication-ready images, light curves, and spectra, perform source detection, and obtain X-ray positions of the sources using the online \swix data products generator\footnote{\url{https://www.swift.ac.uk/user_objects/index.php}} \citep{evans+07,goad+07,evans+09,evans+14,evans+20}. The \swix data are processed using HEASOFT v6.32.

\nic observations were obtained during cycle 4 (\#5141), cycle 6 (\#7124), and with several DDT campaigns. \nic data were reduced with the software NICERDAS v12 distributed with HEAsoft version 6.33.2. We follow the \nic analysis guidelines and excluded data taken during orbit day, an undershooting rate below 300 counts s$^{-1}$, and a default overshooting rate of less than 30 counts s$^{-1}$ per detector. We model the background with the parameterized X-ray background model SCORPEON that consists of a non-X-ray background ({\sc nxb}: related to trapped electrons, the South Atlantic Anomaly, or cosmic rays) and an astrophysical X-ray background ({\sc sky}: corresponding to Cosmic X-ray Background, diffuse local or galactic X-ray emission, solar winds, and neutral oxygen). Unlike other X-ray background library models, SCORPEON is parameterized and can be fitted together with the source model as a function of time and sky position.

We observed eRASSt J1141+0635 with \xmm (ObsID 0881880101) on 28 Dec. 2023 with a net exposure time of 112 ks. Pile-up did not affect the observations. The data was analyzed with the \xmm science analysis system (SAS) version 21.0.0 and HEASOFT version 6.33.2. The EPIC pn and MOS camera were operated in full-frame mode with the medium optical blocking filter. 
We extracted the source spectra from a $40\arcsec$ circular aperture and the background spectra from source-free regions on the same CCD chip. For the pn camera, we selected data from pattern 0 and patterns 1--4 separately.

\section{Data analysis}

\subsection{Full optical spectrum fitting}

\begin{figure}
    \centering
    \includegraphics[width=1\linewidth]{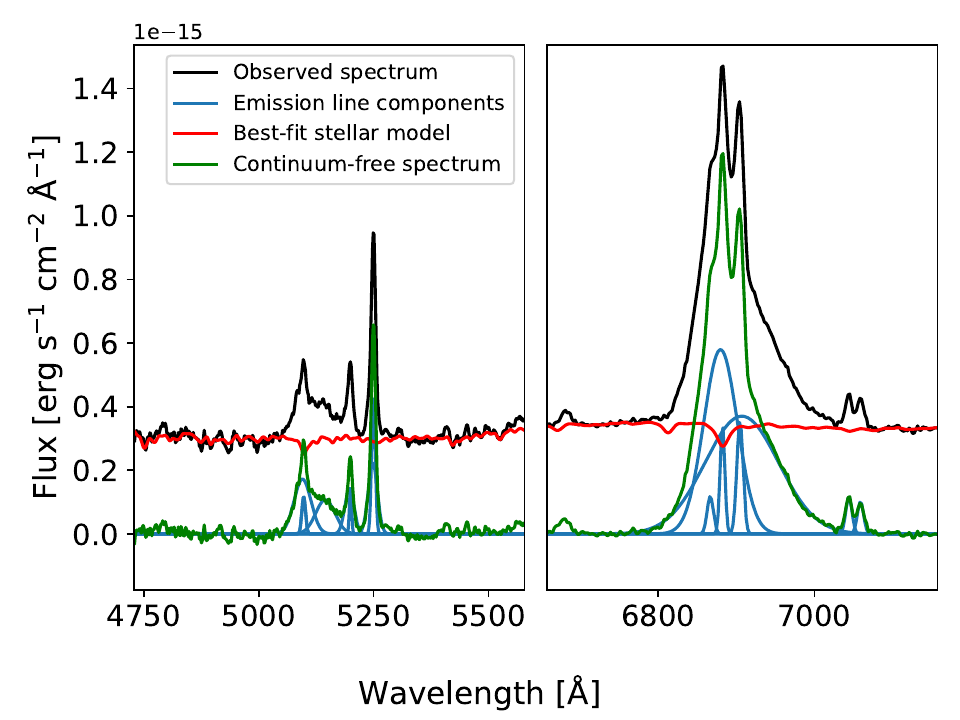}
    \caption{Visualization of the full spectral fit of eRASSt J1906-4850 Magellan/Baade spectrum. The observed spectrum of the galaxy is shown in black, while the pPXF best-fit stellar model is shown in red. A continuum-free emission-line spectrum is shown in green, and the individual Gaussian components are shown in blue. We show the H$\beta+$[\textsc{O iii}] complex and the  H$\alpha+$[\textsc{N ii}]$+$[\textsc{S ii}] in the left and right panel, respectively.  }
    \label{fig:fullfit}
\end{figure}

To characterize the nature of the sources and obtain reliable flux measurements from the broad and narrow emission lines, we correct for stellar absorption features in the spectra. We perform full spectral fitting using the pPXF package originally described in \citep{Cappellari2004C} and upgraded in \cite{Cappellari2017} and \cite{Cappellari2023} to model and subtract the stellar contribution to obtain continuum-free emission-line spectra. The tool pPXF fits the stellar contribution to the spectrum by minimizing the residuals between the observed galaxy and the model. The model of the galaxy is defined by a convolution between stellar population templates and a line-of-sight velocity distribution (LOSVD) parametrized by a Gauss-Hermite series \citep[see][for a detailed description of pPXF]{Cappellari2017}. We use single stellar populations (SSP) templates from the extended MILES (E-MILES) library, which covers the full spectral range between 1680--50,000 \AA~at moderately high-resolution \citep{Vazdekis+16}.

We include additive and multiplicative Legendre polynomial functions to account for discrepancies in shape between the galaxy and the templates. The degrees of the polynomial functions are chosen on a spectrum-to-spectrum basis. Since the main goal of the fit is to subtract the stellar contribution from the spectrum and not to derive stellar parameters, we choose slightly high values for the polynomial degrees (ranging between 5 and 15). We perform a regularized (\texttt{regul}=100) fit where the most prominent emission lines of ionized gas and sky were previously masked. 
Figure \ref{fig:fullfit} shows an example of the full spectral fitting routine on the Magellan/Baade spectrum of eRASSt J1906-4850. The best-fit stellar model, shown in red, is a combination of single stellar templates broadened to match the resolution of the observed spectrum. Once it is subtracted from the observed spectrum (black), we obtain the continuum-free spectrum (green) that traces the ionized gas component of the galaxy. The model is interpolated at the location of the bright emission lines that were initially masked based on the overall best stellar population fit. Thus, the continuum-free spectrum provides an absorption-corrected and model-dependent estimate of the total flux of the ionized gas lines, particularly for H$\beta$ and H$\alpha$.

Considering the redshift distribution of our sources and the wavelength coverage of the spectra, we always find the following emission lines: H$\beta$ $\lambda4861$, [\textsc{O iii}] $\lambda4959,\lambda5007$, H$\alpha\;\lambda6563$ [\textsc{N ii}] $\lambda6549,\lambda6583$ and [\textsc{S ii}] $\lambda6717,\lambda6730$. These emission lines provide valuable information about the dominant source ionizing the gas in the host galaxy. Additionally, their broad components can be used to calculate single-epoch SMBH masses. 

We model the emission lines with Gaussian profiles using Pyspeckit, the Python Spectroscopic Toolkit package \citep{2011ascl.soft09001G}. One narrow ($\sigma < 500$ km s$^{-1}$) Gaussian component was used to fit each of the mentioned emission lines in each spectrum. Additionally, to reproduce the Broad Line Region (BLR) emission of the Balmer lines, we incorporated one broad ($300<\sigma < 3300$ km s$^{-1}$) component for H$\beta$ and one for H$\alpha$. We fixed the ratios of the [\textsc{O iii}] and [\textsc{N ii}] doublet to their theoretically-determined values of 3 \citep{2006agna.book.....O} to reduce the degrees of freedom of the fit. Since the spectral resolutions of our spectra are variable depending on the used instrument, we usually cannot separate the H$\alpha$ component from the [\textsc{N ii}] lines accurately. Therefore, we use the strong and well-isolated [\textsc{O iii}]$\lambda5007$ line as a template to define the widths for all the narrow lines, particularly the [\textsc{N ii}] lines. This choice does not leave significant residuals in the other lines. The emission line fitting process is also shown in Fig. \ref{fig:fullfit}. We note that eRASSt J1906-4850 can only be well-fitted with two broad Gaussian components per Balmer line and an extra, broader, and blue-shifted component for the [\textsc{O iii}] lines. Similarly, eRASSt J1522-3722 needs two Gaussian components to fit the broad and asymmetric [\textsc{O iii}]$\lambda5007$ line.

We obtain the flux, velocity, and velocity dispersion in the line-of-sight of the ionized gas based on the 0th, 1st, and 2nd moment of the Gaussian profiles. Once we perform an initial fit, we retrieve reliable error measurements to these quantities using a bootstrapping technique. Here, we resample the residuals, assuming they are Gaussian distributed and centered in the flux of the model. We add these new residuals to the initial best-fit model of the emission lines, and we perform a new fit. 
We repeat this process 1000 times and store the fluxes, velocities, and velocity dispersions of each realization. Thus, the final emission line measurement will be given by the median and standard deviation of the 1000 realizations. We obtain reliable redshifts based on the wavelength shift of the narrow [\textsc{O iii}]$\lambda5007$\AA~ emission line and the 1000 realizations of the bootstrapping method. The typical errors on the redshifts are of the order of $\sim 10^{-5}$. These redshifts are reported in Table \ref{table:opticaltable} (column 3), and they were also used to compute the X-ray luminosity of the SRG/eROSITA data (column 4).

\subsection{X-ray data analysis}

The \srgero, \swix, \nic, and \xmm spectra were fitted using the Python interface of the \texttt{XSPEC} \citep{arnaud+96} spectral-fitting program, \texttt{PyXspec}. We use Poisson
statistics \citep{cash+79}, cosmic abundances from \cite{wilms+00}, photoelectric absorption cross sections provided by \cite{verner+96}, and we quote all the X-ray parameter errors at the 1$\sigma$ (68\%) confidence level. Typical requested exposure times for \swix and \nic spectra are 3 and 8 ks, respectively. We fit the spectra and obtain X-ray fluxes using a single power-law model with a fixed Galactic absorption (\texttt{tbabs$_{Gal}$*powerlaw}) that depends on the position on the sky of the SMBHB candidate and is obtained from measurement of the Galactic neutral atomic hydrogen column density\footnote{\url{https://heasarc.gsfc.nasa.gov/cgi-bin/Tools/w3nh/w3nh.pl}} \citep{HI4PI+16}. We also fitted an absorbed power-law (\texttt{tbabs$_{Gal}$*tbabs$_{Intr}$*powerlaw}) to the different spectra to test for the intrinsic absorption of the X-ray sources. Spectral parameters such as intrinsic column density N$_{\rm H}$, photon index $\Gamma$, and hardness ratio are calculated for every observation. The hardness ratio ($\rm HR=(hard-soft)/(hard+soft)$) is calculated using the soft (0.2--2.3 keV) and hard (2.3--5.0 keV) energy bands defined by the eROSITA bands. Figures \ref{fig:allxraycandidates} and  \ref{fig:allxraycandidates1} show the light curves of the monitored objects with fluxes and luminosities retrieved from the different X-ray missions. Observed luminosities are computed by considering the luminosity distance of each source that depends on the optical redshift and the assumed $\Lambda$CDM cosmology \citep{2009ApJS..180..225H}.

\subsection{Stacked X-ray analysis}

We stacked the \nic and \swix observations of each source to constrain their time-averaged spectral properties. \nic observations were stacked using the NICERDAS v12 tool \texttt{niobsmerge} that combines cleaned and calibrated products of several observation segments. \swix stacked spectra were directly downloaded from the Living Swift XRT Point Source Catalogue \citep[LSXPS][]{evans+23}. LSXPS is a living catalog that gets updated in nearly real-time and contains position, fluxes, spectral details, and variability information for hundreds of thousands of X-ray point sources detected by the Swift X-ray Telescope. 

We consider two simple X-ray models to study properties like the line-of-sight absorption and the overall spectral shape of the source. We fit a redshifted power-law absorbed by Galactic and intrinsic hydrogen atoms (\texttt{tbabs$_{Gal}$ $\ast$ tbabs$_{Intr}$ $\ast$ zpowerlaw}) and an absorbed thermal accretion disk consisting of multiple blackbody components (\texttt{tbabs$_{Gal}$ 
$\ast$ tbabs$_{Intr}$ $\ast$ diskbb}).

\section{Results}\label{sec:results}
\subsection{Optical}

We confirm that 15 out of the 16 observed SMBHB candidates are extragalactic sources with redshifts ranging between $z\sim 0.02 - 0.2$. Most of the sources display bright emission lines such as [O \textsc{iii}], H$\alpha$, and [N \textsc{ii}] and a unique variety of line profiles, especially in the broad components.

\begin{figure}
    \centering
    \includegraphics[width=1.0\linewidth]{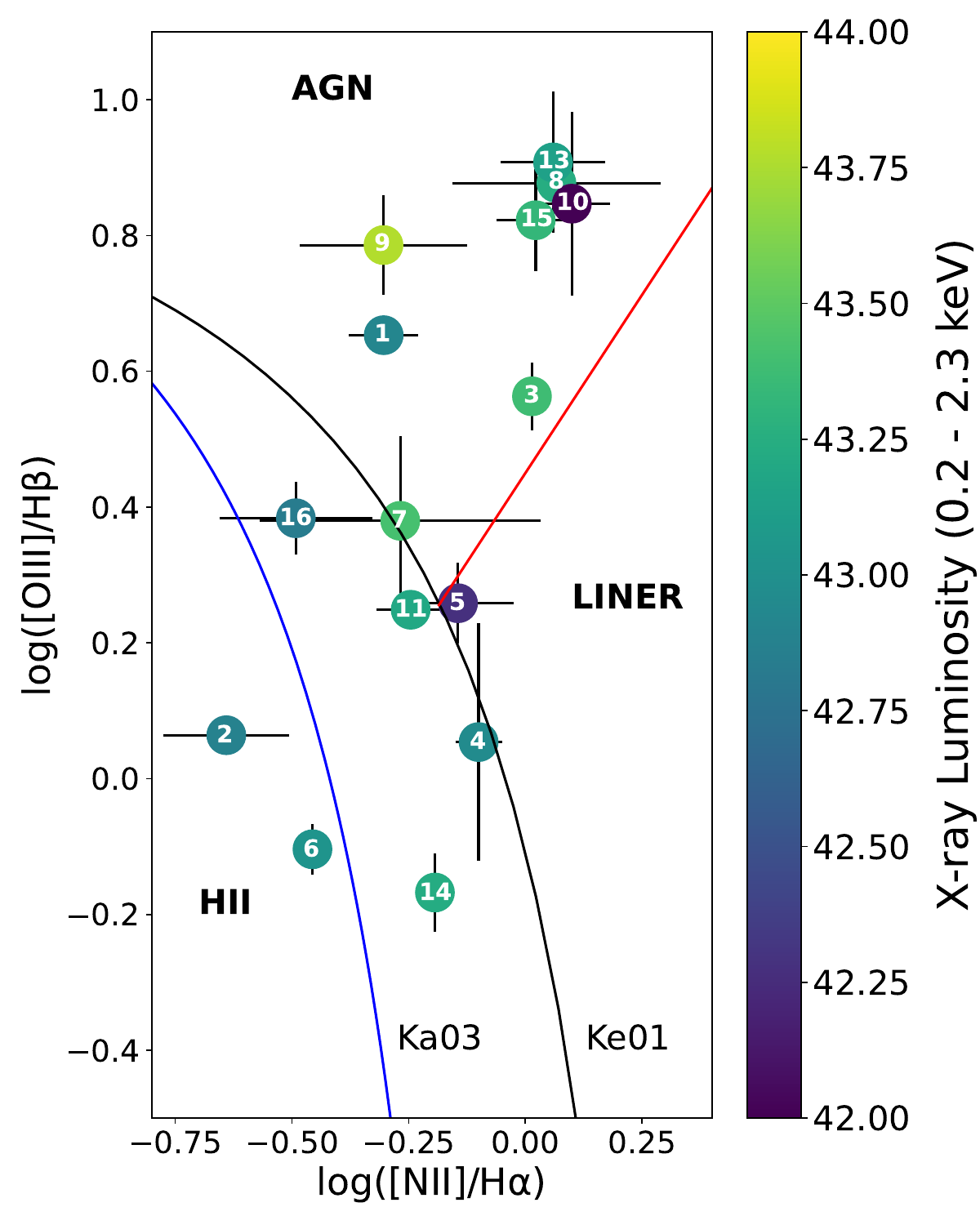}
    \caption{[\textsc{N ii}]-BPT ([\textsc{O iii}]$\lambda5007$/H$\beta$  versus [\textsc{N ii}]$\lambda6583$/H$\alpha$) diagnostic diagram for the SMBHB candidates. The black curve marks the separation proposed by \citet[][denoted as Ke01 in the diagrams]{kewley+01}, between the theoretical maximum ionization driven by pure star-formation in HII regions and those regions ionized by AGN. The red solid line on the BPT diagram marks the separation between AGN and Low Ionization Narrow Emission-line Regions (LINERs), as proposed by \cite{Schawinski+07}, while the blue solid curve represents the separation between AGN and star formation reported by \cite{Kauffmann+03} denoted as Ka03 on the [NII]-BPT diagram. The markers are color-coded according to the observed eROSITA luminosity of the bright level in the 0.2--2.3 keV band. The numbers of the markers correspond to the sources listed in Table \ref{table:bestcandidates} }   
    \label{fig:bpt}
\end{figure}

The Baldwin, Phillips \& Telervich (BPT; \citealp{bpt})  diagram considers optical emission-line flux ratios to provide a unified classification of the nature of the dominating ionization mechanism \citep{bpt,Veilleux+87,kewley+00,kewley+01,kewley+06}. Figure \ref{fig:bpt} shows the [\textsc{N ii}]-BPT diagnostic diagram for the spectroscopically observed sources. The region delimited by the blue \citep[denoted as Ka03,][]{Kauffmann+03} and the black curve \citep[Ke01 in the diagram,][]{kewley+01} is defined as a composite region where AGN and SF can be responsible for the ionization of the gas in the host galaxy.  
Despite all of the sources being X-ray selected, we note that not all of them show strong AGN features based on their optical spectra and their location on the BPT diagram. eRASSt J0530-4125 ($\#2$ in the diagram) displays a galaxy-like spectrum with noticeable stellar absorption features and a weak H$\beta$ broad emission line. On the other hand, eRASSt J1522-3722 ($\#6$ in the diagram) shows strong broad components proper of AGN. However, the source falls into the star-forming portion of the diagram because the [\textsc{O iii}] emission line was fitted with two Gaussian components due to the line being broad and asymmetric. Therefore, at the moment of taking the flux ratio, only the bright and narrower component was considered. Broad and asymmetric [\textsc{O iii}]$\lambda5007$ lines could indicate the presence of outflows in the nuclear region of the galaxy.

Following the $M_{bh}-H\beta$ and $M_{bh}-H\alpha$ scaling relations from \cite{Vestergaard+06} and \cite{mejiarestrepo+22}, respectively, we calculate single-epoch SMBH masses assuming that the broad line region size is larger than any binary separation and that the gas is virialized around the central black hole(s). The SMBH masses of the candidates based on the FWHM and luminosity of the broad H$\beta$ and H$\alpha$ lines are presented in Table \ref{table:opticaltable}. Sources without SMBH mass measurements do not show significant broad H$\beta$ in their spectrum. Instead, we derive SMBH masses using the H$\alpha$ line, following \cite{mejiarestrepo+22}. eRASSt J1906-4850 is the only source with a particularly interesting double-peaked broad Balmer profile, as shown in Fig. \ref{fig:fullfit}. We derive a velocity shift of $\Delta v\rm = 2915\pm736\; km\; s^{-1}$ based on the wavelength separation of the corresponding line centroids. 

Based on their spectroscopic observations, we confirm that all objects (except eRASSt J1003-2607, which displays an off-nuclear nature) are extragalactic sources. The galaxies show evidence of AGN activity with relatively low single-epoch SMBH masses ($\sim10^{7}\; \rm M_{\odot}$). Sources falling into the composite and the star-forming regions of the BPT diagram have typical X-ray luminosities of the order of $10^{43}\; \rm erg\; s^{-1}$ in the 0.2--2.3 keV, suggesting that the mechanism ionizing the gas in the host galaxy is unrelated to the X-ray emitting source. See Sect. \ref{subsect:nature} for a detailed discussion on the origin of the X-rays in our sources.

\subsection{X-rays}

\begin{figure*}
    \centering
    \includegraphics[width=1.0\linewidth]{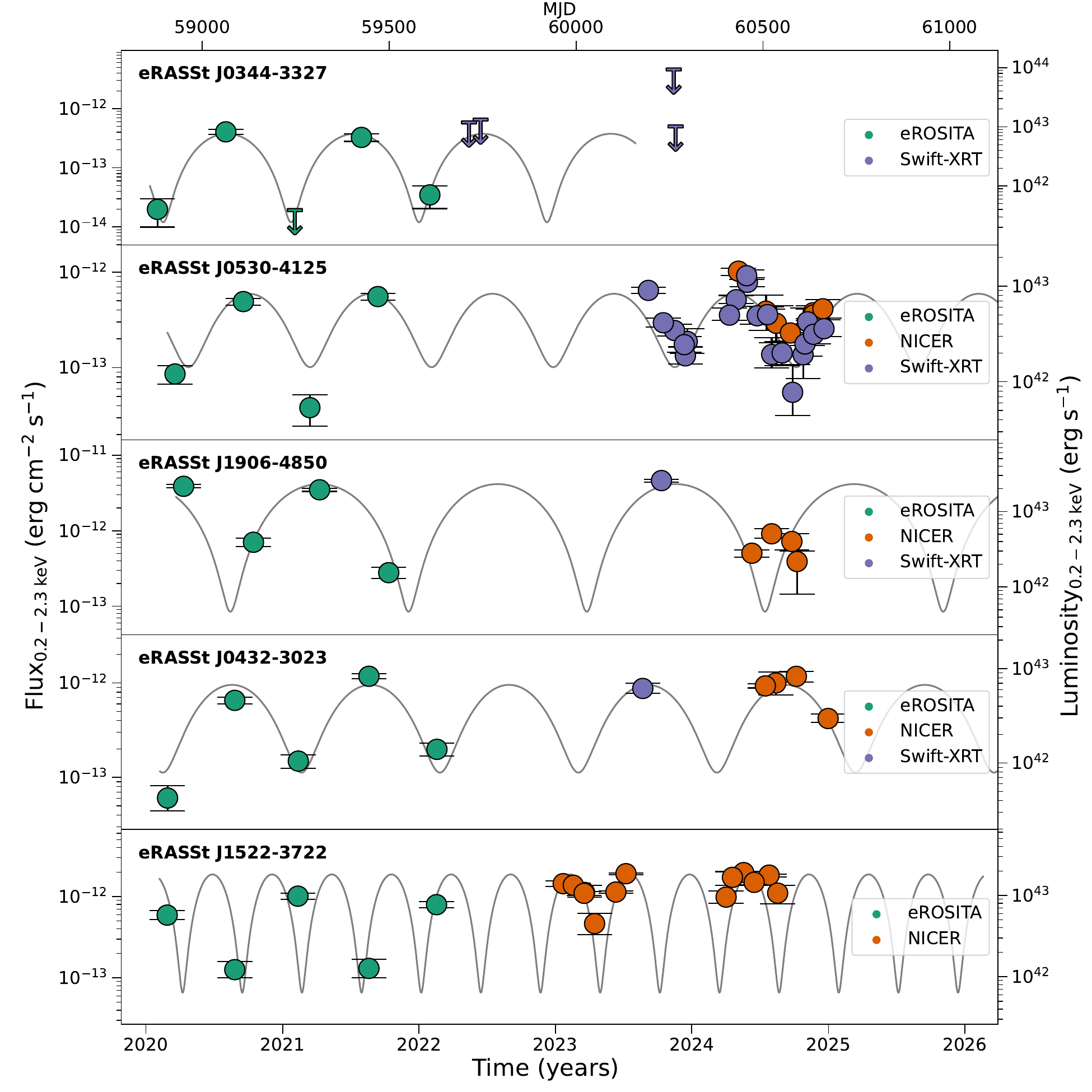}
    \caption{X-ray light curves of the X-ray monitored SMBHB candidates. SRG/eROSITA, \swix, and \nic data are shown in green, purple, and orange, respectively. We display a sinusoidal curve (gray) to visualize the contrast between the observed X-ray variability and periodically variable signal. The secondary y-axis shows the observed frame 0.2--2.3 keV X-ray luminosities of the sources based on their observed X-ray fluxes and redshifts derived from their optical spectra. }
    \label{fig:allxraycandidates}
\end{figure*}

\begin{figure*}
    \centering
    \includegraphics[width=1\linewidth]{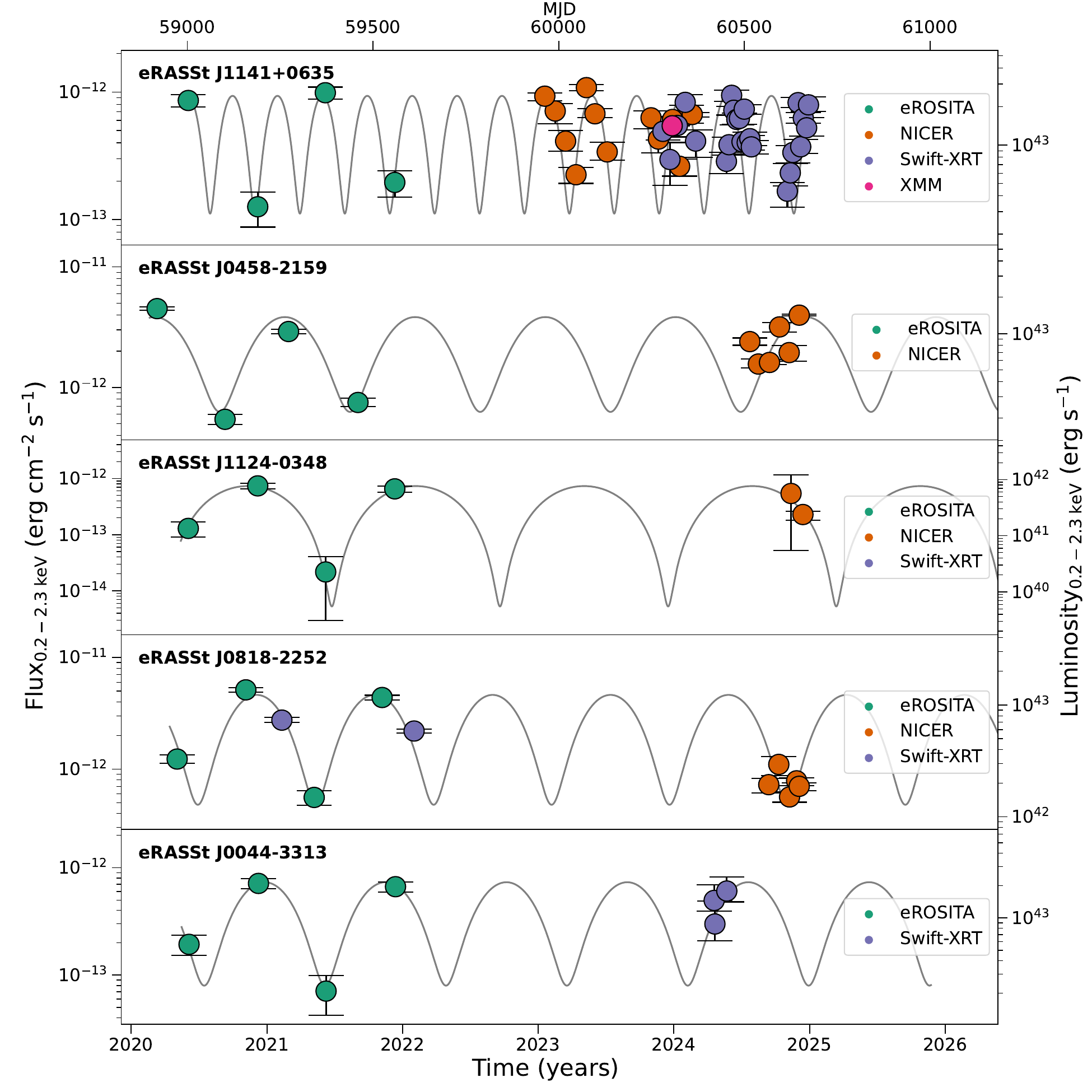}
    \caption{Continuation of Fig. \ref{fig:allxraycandidates}. X-ray light curves of the second half of the X-ray monitored SMBHB candidates. 
    }
    \label{fig:allxraycandidates1}
\end{figure*}

Figures \ref{fig:allxraycandidates} and  \ref{fig:allxraycandidates1} compile every X-ray observation performed during our search for periodically varying light curves. We only show the sources that were observed before Dec. 31st, 2024. 

The observed-frame 0.2--2.3 keV X-ray luminosities of the monitored galaxies range between $10^{42}-10^{43}\; \rm erg\; s^{-1}$. These values strongly suggest that the X-ray emission is produced by accretion onto the SMBH in the center of the galaxy since no other objects or phenomena in the host galaxy can produce such energetic and persistent X-ray emission. We note that some sources support the quasi-periodic behavior in Fig. \ref{fig:allxraycandidates} and \ref{fig:allxraycandidates1}. However, longer X-ray monitoring programs are needed to confirm or rule out the periodic nature of the sources. We emphasize that since there are sources with few X-ray data points, the tentative periodicities are mainly driven by the eROSITA data points. A stringent periodic analysis should be performed with a well and more populated light curve.

We display a sinusoidal curve (gray) based on the period found by the Lomb-Scargle periodogram \citep{lomb+76,scargle+82}. This periodogram technique is designed to detect periodic signals in unevenly spaced observations. The period of the sinusoidal curve is obtained from the frequency at which each source has a maximum in their periodogram. We emphasize that this curve is only to guide the visualization of the tentative periods and highlights the contrast between the X-ray variability and the periodic curve. We also emphasize that the periods derived might not be statistically significant since the number of data points are still insufficient to perform a proper modeling and false-positive probability analysis of the periodogram. The periodogram is designed to find periods in light curves by comparing the observed signal against pure white noise (null hypothesis of not having any signal at all). However, AGNs are variable, and their variability is stochastic, dominated by red noise. Therefore, using the periodogram to confirm periodicities in AGN should be accompanied by a detailed analysis that considers the red noise as a null hypothesis, as described in \cite{Vaughan+2005}. See Sect. \ref{subsect:rednoise} for a detailed discussion of red noise and stochastic AGN variability.

\subsubsection{The two most well-monitored sources in X-rays}

eRASSt J0530-4125 (\#2) and eRASSt J1141+0635 (\#7) are the two most well-monitored sources in our SMBHB candidate sample. Here, we present extended X-ray properties for these objects, and we discuss their X-ray variability.   

\begin{figure*}[h!]
    \centering
    \includegraphics[width=1.0\linewidth]{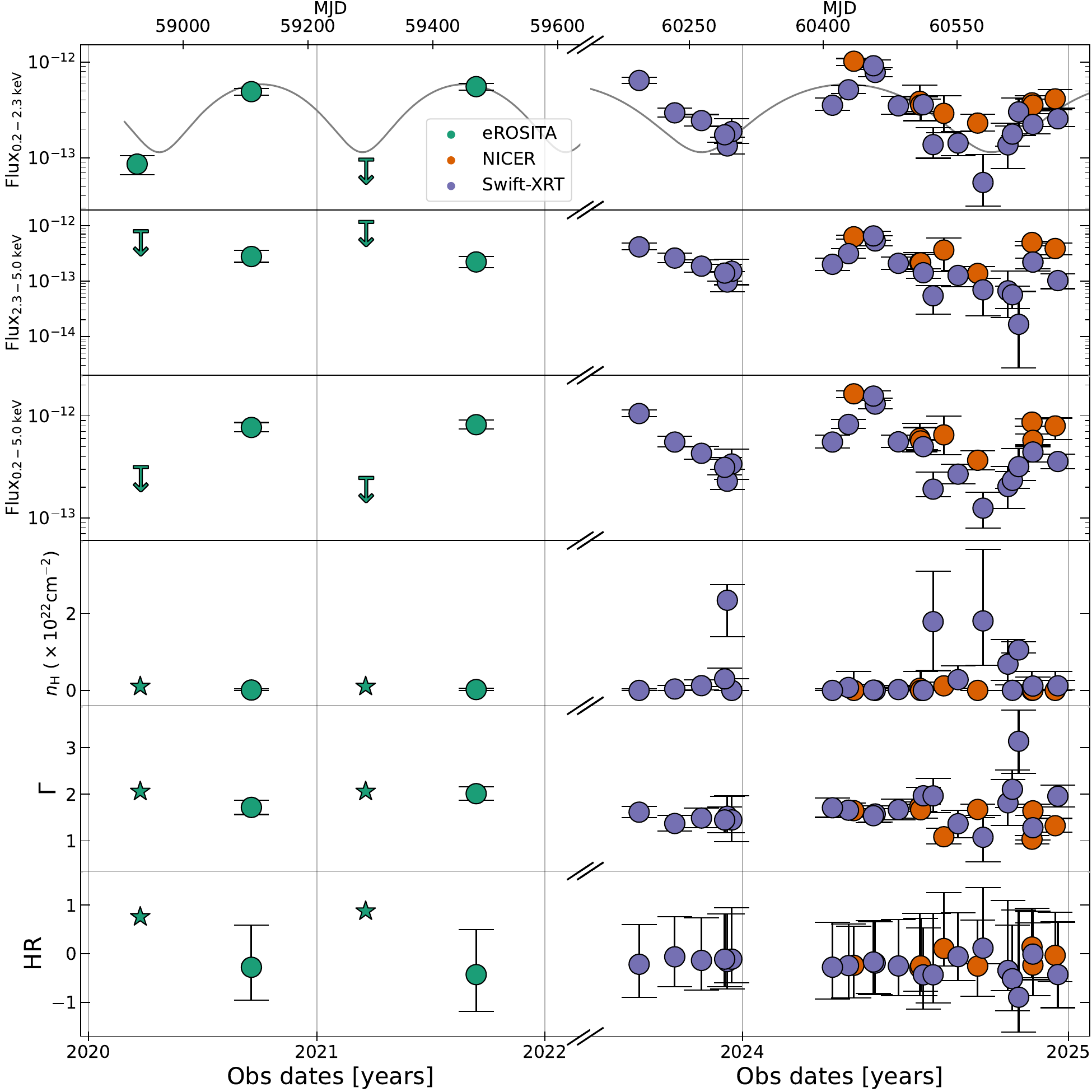}
    \caption{X-ray properties of eRASSt J0530-4125 observed with eROSITA, \swix, and \nic. Light curves at 0.2--2.3 keV, 2.3--5.0 keV, and 0.2--5.0 keV bands are shown in the first, second, and third panels, respectively. Spectral X-ray properties such as intrinsic N$_{H}$, photon index $\Gamma$, and hardness ratio are also shown in the fourth, fifth, and sixth panels. 
    }
    \label{fig:bestsmbhbcandidate}
\end{figure*}

{\it eRASSt J0530-4125: } The source displays an interesting variability pattern that seems to be persistent, even after 2 years of the last eRASS4 observation. Figure \ref{fig:bestsmbhbcandidate} shows the soft (0.2--2.3 keV), hard (2.3--5.0 keV), and total (0.2--5.0 keV) X-ray light curve for eRASSt J0530-4125. 
The light curves in different energy bands follow the same trend, indicating that the variability is not produced by changes in the spectral shape. This is also visualized in the subsequent panels where the column density $n_{\rm H}$, photon index $\Gamma$, and hardness ratio are rather constant in time. Apparent outliers in the column density and photon index measurements are associated with low exposure times while in the faint flux state, therefore having large uncertainties.

The X-ray light curve reveals a tentative quasi-periodic variability with a period of $\sim12$ months, as highlighted with the gray sine curve in the top panel of Fig. \ref{fig:bestsmbhbcandidate}. We emphasize that the periodic signal needs to be carefully tested and interpreted. We do not discard the possibility that the variability pattern is the result of stochastic processes (see also Sect. \ref{subsect:rednoise}).

In the case that the periodic signal is caused by a SMBHB, we can derive the distance between the SMBHs. Assuming a 12-month period and a combined SMBH mass of 10$^{7.26}$\,M$_{\odot}$ (Table\ref{table:opticaltable}), the distance would be $1.2$ milli-pc (considering a Keplerian orbit:
$d=[GMP_{0}^{2}/4\pi^2]^{1/3}$; where $d$ is the distance, $M$ is the total mass of both SMBHs, $P_0$ is the measured period, and $G$ is the gravitational constant). 
We obtain an orbital velocity  of $v \sim 8000$ km s$^{-1}$, using $v =2\pi d P^{-1}$ \citep{Serafinelli+20}.
The velocity can also be used to estimate the energy shift of the 6.4 keV Fe K$\alpha$ lines in the X-ray spectrum. We find that $\Delta \rm E_{Fe\; K\alpha} = 0.17$ keV. This energy shift would barely be resolved by \xmm.

\begin{figure*}[h!]
    \centering
    \includegraphics[width=1.0\linewidth]{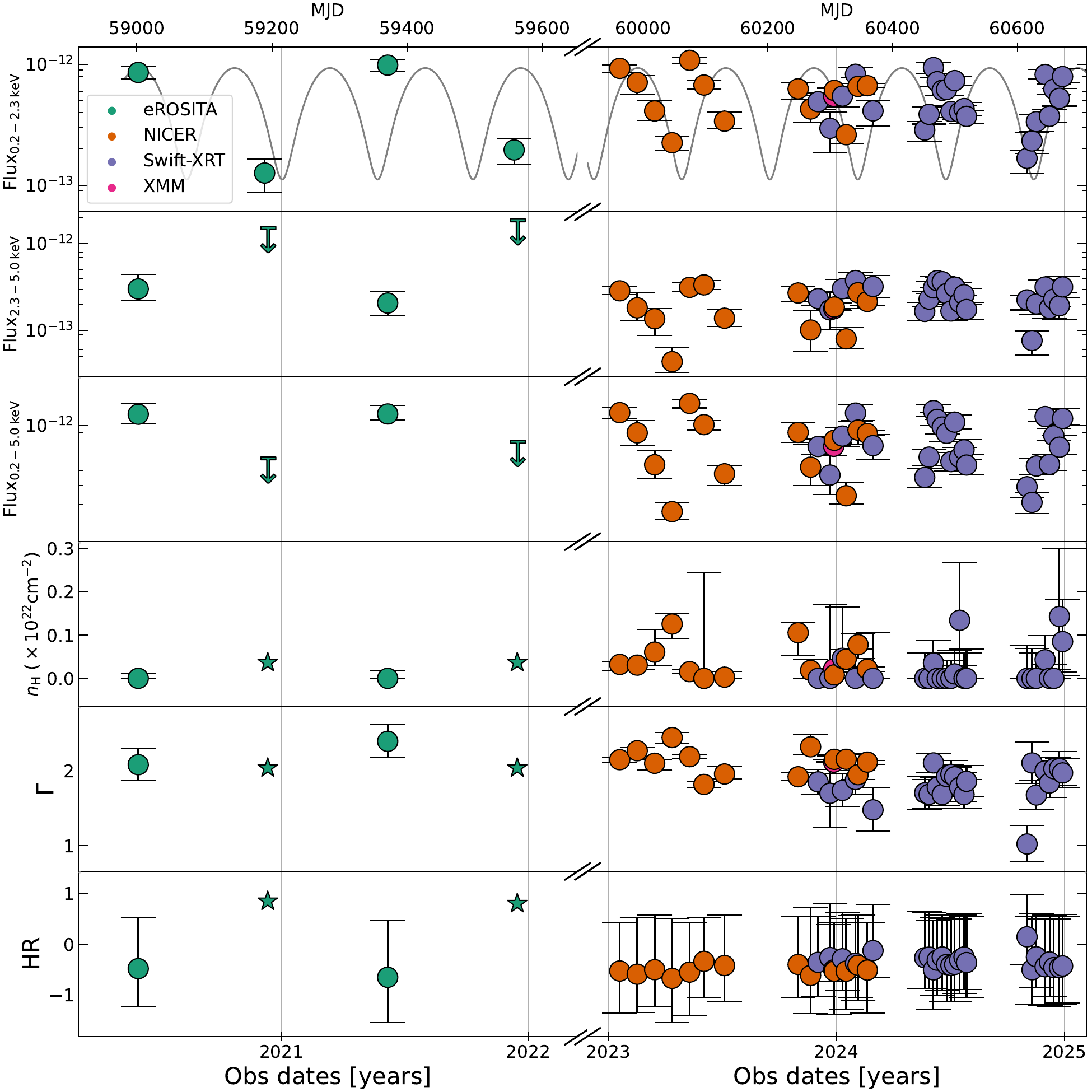}
    \caption{X-ray properties of eRASSt J1141+0635 observed with eROSITA, \nic, \swix, and \xmm. The panels are similar to those shown in Fig. \ref{fig:bestsmbhbcandidate}. 
    }
    \label{fig:bestsmbhbcandidate2}
\end{figure*}

{\it eRASSt J1141+0635: } The \srgero data of this source shows a clear `up-down-up-down' pattern. However, the {\it NICER} monitoring (orange markers in Fig. \ref{fig:allxraycandidates1}) suggests a more complex X-ray variability where the flux decreases in a timescale of three months and rises again after one month. Until the first half of 2023, the data matched the periodic behavior of the sinusoidal curve with a period of $T\sim 120$ days.

At later times, any hint of periodic variability is lost. On 28 Dec. 2023, we obtained a pre-approved AO20 ToO \xmm observation for a net total of 112 ks, aiming to detect a double-peaked profile in the Fe k$\alpha$ emission line. We detect a faint and barely resolved emission line at an observed line energy of $\rm E_{line}=6.22_{-0.24}^{+0.29}$ keV, a width of $\sigma_{\rm line}=0.8_{-0.2}^{+0.3}$ keV, and normalization of $n_{\rm line} = 3.54_{-0.9}^{+1.0}\times10^{-6}$ photons cm$^{-2}$ s$^{-1}$. A detailed analysis of the \xmm observation will be presented in future works.

eRASSt J1141+0635 was further monitored during \swix cycle 20 with a cadence of one week. The high-cadence \swix data show a more structured variability than in the previous observation block; however, it is unclear whether or not the initial quasi-periodic signal detected with \nic is again present. Further X-ray observations are needed to disentangle the nature of this source. If the source is confirmed to be a SMBHB by future observations, the disappearance of the periodic signal could imply that the accretion of matter in binaries is not always stable and that there could be episodes of less or more accretion that would disrupt the periodic behavior of the sources.

\subsubsection{Stacked X-ray properties of the sample}

\begin{table*}
\fontsize{20}{10}\selectfont
\caption{Stacked X-ray spectral properties of the monitored SMBHB candidates. } 
\label{table:stackedxray} 
\centering
\resizebox{\textwidth}{!}{%
\rotatebox{0}{
\begin{tabular}{ccc|cccc|cccc}
\hline
\\
\hline
&&&&&&&&&\\
&&&\multicolumn{4}{c|}{\texttt{tbabs$_{Gal}$ $\ast$ tbabs$_{Intr}$ $\ast$ zpowerlaw}} & \multicolumn{4}{c}{\texttt{tbabs$_{Gal}$ 
$\ast$ tbabs$_{Intr}$ $\ast$ diskbb}}\\
&&&&&&&&&\\
Source\tablefoottext{a}&Obs.&Exposure&$N_{\rm H}$\tablefoottext{b} & $\Gamma$ & Norm. & $\rm stat/dof=\chi^{2}$ & $N_{\rm H}$  & T$_{in}$& Norm. & $\rm stat/dof=\chi^{2}$ \\

&&(s)&($\times 10^{20}\;\rm cm^{-2}$)&&($\times10^{-4}$)&&($\times 10^{20}\;\rm cm^{-2}$)&(eV)&&\\
&&&&&&&&&\\
\hline
&&&&&&&&&\\
1 & \nic & 15981 &
$1.57_{-1.57}^{+141}$\tablefoottext{c} &
$5.35_{-0.94}^{+1.64}$ &
$2.13_{-1.49}^{+2.13}\times10^{-2}$ &
$180.33/144=1.25$ &
$0.0_{-0.0}^{+0.0}$ &
$69_{-9.93}^{+11.6}$ &
$5.24_{-3.02}^{+8.18}\times10^{2}$ &
$180.16/147=1.23$\\ 
&&&&&&&&&\\
\hline
&&&&&&&&&\\
\multirow{2}{*}{2} & \nic & 35802 & 
$4.48_{-4.48}^{+0.6}$ &
$1.97_{-0.08}^{+0.06}$ &
$1.99_{-0.05}^{+0.05}$ &
$251.61/169=1.49$ &
$0.0_{--0.0}^{+-0.0}$ &
$478_{-12}^{+13}$ &
$0.48_{-0.05}^{+0.05}$ &
$305.93/173=1.77$ 
\\&&&&&&&&&\\
 &\swix& 70119 & 
$0.0_{-0.0}^{+5.0}$ &
$1.67_{-0.07}^{+0.08}$ &
$1.22_{-0.075}^{+0.088}$ &
$697.14/1020=0.68$ &
$0.0_{--0.0}^{+0.0}$ &
$1260_{-66}^{+72}$ &
$113_{-20}^{+23}$ &
$750.66/1020=0.74$ 
\\&&&&&&&&&\\
\hline
&&&&&&&&&\\
3& \nic & 24130 & 
$19.3_{-11.2}^{+2.42}$ &
$2.84_{-0.18}^{+0.19}$ &
$3.19_{-0.19}^{+0.19}$ &
$246.64/176=1.40$ &
$5.92_{-2.73}^{+994}\times10^{5}$ &
$998_{-991}^{+514}$ &
$4.72_{-2.39}^{+200000}\times10^{11}$ &
$349.12/180=1.94$ 
\\&&&&&&&&&\\
4 &\nic & 40670 & 
$0.75_{-0.75}^{+1.29}$ &
$1.94_{-5.99}^{+6.38}$ &
$1.85_{-0.045}^{+0.045}$ &
$241.88/181=1.34$ &
$0.0_{-0.0}^{+0.0}$ &
$353_{-8.36}^{+9.12}$ &
$1.44_{-0.14}^{+0.141}$ &
$238.09/185=1.29$ 
\\&&&&&&&&&\\
6 &\nic & 63663 & 
$0.0_{-0.0}^{+3.87}$ &
$2.06_{-0.035}^{+0.05}$ &
$6.52_{-0.09}^{+0.09}$ &
$233.7/171=1.37$ &
$0.0_{-0.0}^{+0.0}$ &
$539_{-4.32}^{+4.63}$ &
$0.98_{-0.03}^{+0.03}$ &
$700.15/175=4.00$ 
\\&&&&&&&&&\\
\hline
&&&&&&&&&\\
\multirow{3}{*}{7} &\nic & 108169 & 
$3.82_{-0.39}^{+0.14}$ &
$2.26_{-0.02}^{+0.02}$ &
$4.27_{-0.03}^{+0.03}$ &
$528.7/186=2.84$ &
$0.0_{-0.0}^{+0.0}$ &
$354_{-1.74}^{+3.21}$ &
$3.46_{-0.12}^{+0.07}$ &
$1922.37/190=10.12$ 
\\&&&&&&&&&\\
 &\swix &89832 & 
$0.0_{-0.0}^{+2.22}$ &
$1.8_{-0.05}^{+0.05}$ &
$1.91_{-0.08}^{+0.09}$ &
$766.03/1020=0.75$ &
$0.0_{-0.0}^{+0.0}$ &
$1060_{-36}^{+38}$ &
$3.15_{-0.4}^{+0.4}\times10^{-2}$ &
$989.81/1020=0.97$ 
\\&&&&&&&&&\\
&\xmm & 112600 & 
$2.75_{-0.18}^{+0.2}$ &
$2.3_{-0.014}^{+0.015}$ &
$2.23_{-0.017}^{+0.018}$ &
$8601.12/9166=0.94$ &
$0.0_{-0.0}^{+0.0}$ &
$426_{-0.53}^{+0.52}$ &
$0.96_{-0.005}^{+0.005}$ &
$14694.14/9166=1.60$ 
\\&&&&&&&&&\\
\hline
&&&&&&&&&\\
8&\nic & 39761 & 
$0.0_{-0.0}^{+0.0}$ &
$1.95_{-0.03}^{+0.02}$ &
$9.11_{-0.07}^{+0.1}$ &
$219.20/185=1.18$ &
$0.04_{-0.04}^{+0.49}$ &
$463_{-4.88}^{+4.8}$ &
$2.61_{-0.11}^{+0.11}$ &
$546.38/189=2.89$ 
\\&&&&&&&&&\\
10&\nic & 17335 & 
$6.36_{-2.36}^{+2.76}$ &
$1.34_{-0.12}^{+0.13}$ &
$1.23_{-0.071}^{+0.061}$ &
$228.17/155=1.47$ &
$0.04_{-0.04}^{+2.45}$ &
$1030_{-93}^{+72}$ &
$2.35_{-0.48}^{+0.88}\times10^{-2}$ &
$231.39/158=1.46$ 
\\&&&&&&&&&\\
\hline
&&&&&&&&&\\
13&\nic & 31238 & 
$5.56_{-3.43}^{+8.64}$ &
$2.03_{-0.13}^{+0.14}$ &
$3.70_{-0.14}^{+0.16}$ &
$250.41/176=1.42$ &
$2.06_{-0.43}^{+97.90}\times10^{4}$ &
$1000_{-3.95}^{+-17.60}\times10^{3}$ &
$1.0_{-0.9}^{+10000}\times10^{20}$ &
$501.73/180=2.79$ 
\\&&&&&&&&&\\
\hline
\end{tabular}}}

\tablefoot{
\tablefoottext{a}{The number of the sources refer to Table \ref{table:bestcandidates}.} 
\tablefoottext{b}{Intrinsic column density.}
\tablefoottext{c}{The errors in the measurements correspond to the 68\% confidence interval.}
}
\end{table*}

The time-averaged X-ray spectra were also analyzed to study the spectral shape properties of the sources. Stacking the available data increases the signal-to-noise of the observations and can provide valuable information about their nature. Table \ref{table:stackedxray} summarizes the best-fit values for the two models tested. The models have a fixed Galactic absorption component (\texttt{tbabs$_{Gal}$}) derived from the \cite{HI4PI+16}. We note that all sources have low intrinsic absorption and are mostly well characterized by power-law models with typical AGN-like photon index ($\Gamma\sim2$). Only one object (eRASSt J0344-3327) shows a soft power law, likely associated with the TDE nature of the source (see Appendix \ref{appendix:particularcases}).

\section{Discussion}\label{sec:discussion}

\subsection{Stochastic AGN variability as a source of quasi-periodic light curves}\label{subsect:rednoise}

Throughout the paper, we have been cautious with interpreting the quasi-periodic signals displayed by the galaxies as SMBHBs since single-SMBH accretion is stochastic and can display artificial periodic features. Here, we give a more detailed explanation of the red-noise phenomena and emphasize that in the second paper of this program, we will properly quantify the effect of stochastic accretion in our sample (Tubín-Arenas et al. in prep.). 

Accretion onto single SMBHs produces highly variable and aperiodic X-ray emission with a stochastic behavior: the so-called red noise \cite[see, e.g.,][]{Press+1978,Vaughan+03}. The power spectral density (PSD) indicates how strong the variability is as a function of temporal frequency (timescale$^{-1}$), and it has been used to study variability across different timescales. In red-noise-dominated sources, the X-ray PSD of these light curves is commonly represented by a power law with slopes ranging between 1--2 \citep[][]{Lawrence+1987,McHardy+1987,Markowitz_2003,McHardy+2006,Paolillo+23}, opposed to random white-noise PSDs with slope values near zero. 

The search for periodic AGN features has been limited by the data quality and poor time resolution of the observations. Additionally, single-AGN stochastic variability can mimic quasi-periodic variability \citep[][]{Krishnan+21,witt+22}. \cite{Vaughan+16} claim that the periodic variability of the quasar PG 1302-102 found by \cite{graham+15} can be explained by stochastic variability. They also found a few-cycle `phantom' periodicities in simulated red-noise light curves based on steep PSD shapes. `Phantom' periodicities are found in large optical quasar surveys \citep[see, e.g.,][]{MacLeod+2010} where the tentative periods of the light curve tend to be similar to the total length of the observations \citep[][]{Vaughan+16}. 
False positives can be reduced by investigating the range of parameter space in which binary systems can be detected. 
\cite{Krishnan+21} found that discarding timescales greater than (roughly) $1/3$ of the total observation length is an effective way to reduce false positives. 
\cite{witt+22} found that periodic signals are more easily detectable if the period is short or the amplitude of the signal is large compared to the contribution of the noise.

We cannot rule out the presence of red-noise variability in our sample until further follow-up campaigns provide strong evidence of either periodic or stochastic variability. Real periodic processes should continue to show periodic behavior in future observations, while AGNs dominated by red-noise processes will gradually deviate from the periodic scenario. Therefore, continuous and extended X-ray monitoring programs will be needed when searching for periodic light curves that are expected of SMBHBs. 

We will investigate in detail the effect of red-noise-induced variability in the selection of our sources by obtaining false positive rates (paper II). These rates will be computed by simulating millions of pure red-noise light curves and over-imposing the \srgero scanning pattern (i.e., one datapoint every six months). Thus, we will identify the number of spurious sources that show the same variability pattern selected in this paper. We will also include quasi-periodic signals mixed with different PSD slopes and red-noise levels to test our selection method of the \srgero candidates. These results will be presented in Paper II (Tubín-Arenas et al. in prep.).

\subsection{Possible origin of the X-ray quasi-periodicity in our sample}\label{subsect:nature}

In the following section, we assume that the quasi-periodic signatures found in the X-ray light curves are not caused by red noise but are real. Given the obtained multi-wavelength data, we discuss which scenarios are still plausible for the individual sources. These scenarios include off-nuclear sources, objects interacting with the central SMBH (e.g., TDEs, QPEs), and SMBH binaries. Under each scenario, we explain why we consider it for the given source. To summarize our current understanding of our sources, we give an overview of the already excluded and still possible scenarios in Table \ref{table:nature}. If several scenarios are still listed for a source, the current data do not allow us to favor one possible scenario over another.

\textit{Off-nuclear sources:} Ultra-luminous X-ray sources are extragalactic objects located outside the nucleus of the host galaxy with bolometric luminosities ranging between $10^{38}\; \rm and\; 10^{42}\; erg\; s^{-1}$. The most extreme cases are also known as hyper-luminous objects (HLXs). The source population of ULXs and HLXs are expected to be diverse, covering both neutron-star binaries and black-hole binaries \citep[][]{Kaaret+2017,Gong_2016}. However, HLXs are more likely to require an intermediate-mass black hole (IMBH) to power their emission \citep[][]{Webb+2012}. The brightest known HLX is HLX-1, located in ESO 243-49. It has a maximum 0.2--10 keV luminosity of $\sim10^{42}\; \rm erg\; s^{-1}$ \citep[][]{Farrell+2009}. This luminosity range is of particular interest as it covers the overlap of ULXs and low-luminosity AGN. 
Additionally, the X-ray light curve of HLX-1 shows possible recurrence in its variability \citep[see, e.g.,][]{Lasota+2011,servillat+2011}, which could be interpreted as `up-down' patterns when sampled with low cadence, such as with \srgero. Therefore, HLXs could contaminate our sample, given their luminosity and variability. 

Two of the most used methods to discard the presence of HLXs in our sample are X-ray positions and luminosities. HLXs are expected to be off-nuclear sources, so precise X-ray positions compared against the optical position of the galactic nuclei can provide an initial assessment of the nature of these objects. Figure \ref{fig:lsdr10images} and Table \ref{table:bestcandidates} show the separation, in arcseconds, between the optical {\it Gaia} positions, centered in the nuclei of the galaxies and the position of the X-ray emission. All of the candidates have X-ray emissions consistent with being emitted by the central parts of the galaxy (except for eRASSt J1003-2607. See Appendix \ref{appendix:particularcases}). The sources with the largest separation (eRASSt J1130-0806 and eRASSt J0600-2939, separated by $2\farcs33$ and $2\farcs88$, respectively) are still consistent with being nuclear within the eROSITA positional uncertainty. However, we note that this method does not discard off-nuclear sources with projected separations smaller than the X-ray spatial resolution. 
Most of the luminosities in the faint flux state of our sources listed in Table \ref{table:opticaltable} are consistent with the maximum luminosity of HLX-1. However, when considering the observed luminosities (only between 0.2--2.3 keV) of our candidates in the bright phase, we note that all of the galaxies (except for eRASSt J1130-0806 and eRASSt J1124-0348) are almost one order of magnitude brighter than HLX-1. Therefore, the luminosity argument makes it unlikely that the X-ray origin in our sources comes from an off-nuclear, close-to-center HLX object. 
We show in Table \ref{table:nature} the two sources from our sample that could be HLXs based on their 0.2--2.3 keV X-ray luminosities in their bright states (eRASSt J1130-0806 and eRASSt 1124-0348).
The HLX scenario for object eRASSt J1003-2607 is supported by its X-ray luminosity and its off-center position. Since the X-ray luminosities of eRASSt J1130-0806 and eRASSt 1124-0348 are also consistent with the ranges expected for low luminosity AGN, we still consider them under the other possible scenarios. If they are of HLX origin, it is interesting to note that they would be among the most luminous HLXs ever detected. 

\textit{Objects interacting with the central SMBH:} i) TDEs produce energetic flares originating in the nuclei of galaxies due to the accretion of disrupted stellar debris onto the SMBH. TDEs usually display soft, thermally-dominated X-ray spectra, and their light curves typically display a sharp rise in brightness followed by a slow decline with timescales of months or years ($\propto t^{-5/3}$) \citep[see][and reference therein]{saxtontde+2020,Gezari+2021G}. Lately, more exotic phenomena, such as repeating TDEs, have been discovered \citep[see, e.g.,][]{Payne+2021,Liu+2023L,Malyali+2023,liupTDE+24}. In this scenario, the star survives the first encounter with the SMBH and only loses a fraction of its mass. The system will produce new accretion events in the subsequent encounters with the SMBH. The recurrence time of the X-ray flares in the known pTDEs ASASSN-14ko \citep[][]{Payne+2021} and eRASSt J045650.3-203750 \citep[][]{Liu+2023L,liupTDE+24} are consistent with the timescales probed by our SMBHB search with timescales of 100 to 300 days, respectively.  

ii) Quasi-periodic eruptions (QPEs) are newly-discovered X-ray transients that originate in the nuclei of nearby galaxies where only the X-ray light curve shows persistent, soft, repeating, and narrow X-ray flares with timescales of hours to weeks \citep[see, e.g.,][]{Miniutti+2019,Giustini+2020,Arcodia+2021,Miniutti+23b,Quintin+23,arcodia+2024,Nicholl+24,guolo+2024}. Possible scenarios to explain QPEs are accretion disk instabilities \citep[e.g.,][]{pan+2022} or stellar objects impacting the accretion disk of the SMBH in close and elliptical orbits \citep[][]{xian+2021}. QPEs and TDEs have been found to prefer similar host galaxies \citep[][and references therein]{Wevers+2022,Wevers+2024}. Additionally, QPEs have been detected in galaxies years after TDEs \citep[][]{Miniutti+23b,Quintin+23,Nicholl+24,Pasham+2024}, suggesting a connection in the formation mechanism of both transients. In this scenario, a main-sequence star (brought into the nucleus as an extreme mass-ratio inspiral; EMRI) emits flares every time it crosses the accretion disk formed by the debris of a TDE \citep{linial+2023,sukova+2024}. 

Repeating X-ray transients (such as TDEs and QPEs) could potentially explain the initial pattern of the \srgero-selected objects in this work. Repeating TDEs present longer timescales, consistent with the timescales of our SMBHB search. In fact, the source with the largest ratio between the faint and bright X-ray states in our sample (eRASSt J0344-3327) is likely a TDE, based on the sharp rise and slow decline in its optical and IR light curves, as well as its steep and soft X-ray spectrum (see Fig. \ref{fig:erasst_j0344-3327}). Other sources such as eRASSt J1906-4850 (Fig. \ref{fig:eRASSt_J1906-4850}), eRASSt J1130-0806 (Fig. \ref{fig:eRASSt_J1130-0806}), eRASSt J1124-0348 (Fig. \ref{fig:eRASSt_J1124-0348}), eRASSt J0036-3125 (Fig. \ref{fig:eRASSt_J0036-3125}), and eRASSt J0044-3313 (Fig. \ref{fig:eRASSt_J0044-3313}) show weak but noticeable multi-wavelength variability features that might resemble the behavior presented by eRASSt J0344-3327. In these sources, we can neither claim nor discard their TDE nature (see Table \ref{table:nature}). For the remaining sources, the lack of TDE-like features in the multi-wavelength data and the presence of hard X-rays disfavor the TDE scenario.

On the other hand, the longest repeating time detected for a QPE is 22 days \citep[][]{guolo+2024}, which is not long enough to explain the long-term variability of our candidates, such as eRASSt J0530-4125 or eRASSt J1141+0635. Under the repeating transients framework, the sources with eROSITA-only X-ray data or those with short monitoring programs could belong to a very particular branch of the QPEs population. In this scenario, there would be an EMRI object orbiting the SMBH with long periods ($\sim1$ year), crossing the accretion disk in quasi-regular intervals producing more energetic, harder, and long-lasting eruptions. Most of the sources in Table \ref{table:nature} are still marked as possible QPEs since this scenario is more complicated to prove or rule out with the available data. In case these sources are found to be QPEs, they would belong to an extreme class with one of the longest periods detected. This is assuming that there were no eruptions during the six months that eROSITA did not observe the source. Further X-ray observations with a higher cadence than eROSITA would provide better constraints of any possible eruption timescales. 

\textit{SMBH binaries:} The variability landscape of the X-ray light curves presented in Fig. \ref{fig:allxraycandidates} and \ref{fig:allxraycandidates1} is diverse and intricate. eRASSt J0530-4125 shows a smoother and recurrent quasi-periodic behavior with timescales of the order of 1 year. This object is interesting since AGNs with (quasi) periodic behavior are rare, especially in X-rays. \cite{DOrazio+23} summarize the observational signatures of SMBHBs and provide a compilation of the known SMBHB candidates together with the techniques employed to select them.  

Testing and confirming the binary nature of the best candidates is challenging and requires additional multi-wavelength evidence. The last column in Table \ref{table:nature} lists the sources in which we have not discarded the SMBHB scenario based on their multi-wavelength properties. We note that the X-ray periodic scenario needs further observational evidence with additional extended X-ray monitoring programs, complemented by optical and IR photometric data. Once the periodicity of the objects can be confirmed for several cycles, deep X-ray observations aiming to detect double-peaked Fe K$\alpha$ emission line profiles will be needed to further test the SMBHB scenario.

\begin{table*}
\small\addtolength{\tabcolsep}{+1.0pt}
\caption{Light curve status and possible nature of the primary sample of SMBHB candidates. 
} 
\label{table:nature} 
\centering
\begin{tabular}{c|cccc|crc}
\hline
\hline
 & \multicolumn{4}{c}{Light curve status} & \multicolumn{3}{c}{Possible nature of the X-ray emission}\\
\hline
\multirow{2}{*}{Source}&eROSITA & eROSITA  & eROSITA  & \multirow{2}{*}{Periodic signal} & \multirow{2}{*}{HLXs} & \multirow{2}{*}{TDE/QPE} & \multirow{2}{*}{SMBHB} \\ 
&only&$+$ short program&$+$ long-term program&&&&\\
\hline
%$&&&&&&&\\
1 &&\ding{51}&&&&\ding{51}/\ding{51}&\ding{51}\\
2 &&&\ding{51}&\ding{51}&&&\ding{51}\\
3 &&\ding{51}&&&&\ding{51}/\ding{51}&\ding{51}\\
4 &&\ding{51}&&\ding{51}&&\;\;/\ding{51}&\ding{51}\\
5 &\ding{51}&&&\ding{51}&\ding{51}&\ding{51}/\ding{51}&\ding{51}\\
6 &&&\ding{51}&\ding{51}&&&\ding{51}\\
7 &&&\ding{51}&\ding{51}&&&\ding{51}\\
8 &&\ding{51}&&\ding{51}&&\;\;/\ding{51}&\ding{51}\\
9 &\ding{51}&&&\ding{51}&&\;\;/\ding{51}&\ding{51}\\
10 &&\ding{51}&&&\ding{51}&\ding{51}/\ding{51}&\ding{51}\\
11 &\ding{51}&&&\ding{51}&&\ding{51}/\ding{51}&\ding{51}\\
12 &&&\ding{51}&\ding{51}&\ding{51}&&\\
13 &&\ding{51}&&\ding{51}&&\;\;/\ding{51}&\ding{51}\\
14 &\ding{51}&&&\ding{51}&&\;\;/\ding{51}&\ding{51}\\
15 &\ding{51}&&&\ding{51}&&\ding{51}/\ding{51}&\ding{51}\\
16 &\ding{51}&&&&&\;\;/\ding{51}&\ding{51}\\
\hline
\end{tabular}
\tablefoot{Summary of the X-ray monitoring campaigns for each source and their tentative nature. We categorized the sources according to their amount of X-ray follow-up. eROSITA-only (column 2) are sources that only have eROSITA data and will be observed in future programs. eROSITA+short program (column 3) and eROSITA+long program (column 4) are the sources where we have less or more than 6 X-ray pointings after the eROSITA data, respectively. We mark as periodic signals (column 5) those sources where the data have not ruled out the possibility of periodicity. This column also includes sources where we only have eROSITA data since more data is required to confirm or rule out quasi-periodic signals. The possible nature of the sources is divided into three categories. HLXs (column 6), nuclear transients (such as TDEs/QPEs; column 7), and SMBHBs (column 8). We mark the sources where the data have not entirely ruled out the possibility that the source possesses a given nature.} 
\end{table*}

\subsection{How rare are these SMBHB candidates?}
We aim to obtain an X-ray observational estimate of the fraction of extragalactic sources that could potentially host SMBHBs, and we compare it to predictions. For this purpose, we need to compare the number of SMBHB candidates (Fig. \ref{fig:sample}) to the number of sources that could have been selected if they had met the specific variability criteria. This implies that not every source in the eROSITA-De sky is bright enough to detect the specific variability pattern and thus search for SMBHB candidates.

We consider sources that are bright enough to have room for flux changes consistent with our selected sample. However, the exposure time across the sky is not uniform, leading to a non-uniform flux distribution. 
We note that 15 (out of 16) selected monitored sources in Fig. \ref{fig:sample} have more than 50 counts in the bright state. Using this criterion, we find $\sim40.000$ sources among the 2.3 million sources that according to their {\it Gaia} properties and consistent with our selection criteria as set out in Sec. \ref{sec:selection} are likely extragalactic and have at least one eROSITA scan with more than 50 detected counts. Thus, we estimate that one out of $\sim2600$ sources (a fraction of $\sim3.7\times10^{-4}$) has a $F>5$, a $S>3$ and a flux in the bright state larger than $\sim5\times10^{-13}\rm \; erg\; s^{-1}\; cm^{-2}$ (monitored sources in Fig. \ref{fig:sample}). 
This fraction corresponds to the most optimistic upper limit on the number of X-ray-selected sources. However, we also note that this calculation is limited by the cadence and scanning strategy of eROSITA.

Thus, we also have to obtain an estimate, under several assumptions, of the number of SMBHBs with all possible periods. We note that we are sensitive to windows of approximately 1 and 2 months around the tentative periods of $\sim4$ and $\sim12$ months, respectively. Sources whose tentative periods do not perfectly match the eROSITA cadence will show a larger discrepancy between the flux values of the corresponding bright or faint state, translating into smaller significances and factors in our selection criteria. Therefore, we consider this time window of $\sim3$ months (2 from the sources with periods $\sim12$ months and 1 from sources with periods of $\sim4$ months) as our eROSITA-sensitive time window.

According to theoretical models, when the separation of the SMBHs in a binary system is of the order of the BLR size, the BLR starts suffering truncation \citep[][]{POPOVIC201274,derosa+18}. We assume that at smaller separations, the SMBHs are already embedded in the circumbinary disk, giving us a typical separation value for SMBHBs that could be emitting periodic X-ray emission. We find a BLR size of $\sim0.01$ pc (12 lt-days) following the BLR luminosity-size relation from \cite{Greene+2010}. Here, we have assumed an average total SMBH mass of our sample of $\sim10 ^{7}\;\rm M_{\odot}$, an average X-ray flux of $\sim7\times10^{-13}\rm\; erg\; s^{-1}\; cm^{-2}$ in the eROSITA 0.2--2.3 keV band, an average X-ray luminosity in the 2--10 keV band of $\sim9.4\times10^{42}\rm\; erg\; s^{-1}$, and an average redshift of $z=0.075$. These are typical values obtained for our primary SMBHB sample (see, Sect. \ref{sec:results}).

Binaries with these separations of $0.01$ pc have periods of $\sim 30$ years, assuming Keplerian motion, an equal mass ratio, and circular orbits. Assuming also that SMBHBs have a uniform distribution of orbital periods over these 30 years, we notice that the scanning pattern of eROSITA is only sensitive to 3 out of these 360 months, corresponding to a fraction of $8\times10^{-3}$ (based on an average total SMBHB mass of $\sim10 ^{7}\;\rm M_{\odot}$).  Therefore, combining these fractions, we estimate an upper limit on the probability of detecting X-ray variable SMBHBs of $\sim0.05$ when we consider the 15 sources of Fig. \ref{fig:sample}. This fraction will likely decrease when we account for the real origin of the sources and the false positive rate of the red-noise-induced variability in AGNs.

Theoretical and observational studies have predicted an abundance of SMBHBs based on several assumptions, selections, and techniques. \cite{Volonteri_2009} predicted $\sim0.01\;\rm deg^{-2}$ sub-pc binary quasars at $z < 0.7$ based on the hierarchical assembly, growth, and dynamics of SMBHs in a $\Lambda$CDM cosmology. This result corresponds to a fraction of $\sim6\times10^{-4}$ SMBHBs in their volume and flux-limited theoretical sample. Another study predicts an apparent occurrence of $\sim10\%$ sub-pc SMBHBs in SDSS based on bulk radial velocity shifts of the broad emission lines \citep[][]{Guo+2019}. \cite{Hayasaki+2010} found that $\sim1\%$ of AGNs have close binary massive black holes with an orbital period of less than 10 years based on the SMBH mass function and the theory of evolution of SMBHBs interacting with a circumbinary disk. Using a combination of cosmological, hydrodynamic simulations, binary merger models, and AGN spectra and variability prescriptions, \cite{kelley+19} found that the expected fraction of SMBHB is $\sim1\%$ at $z<0.6$. However, only a few (fractions about $10^{-7}-10^{-5}$) can be identifiable in the Catalina survey (CRTS) or LSST. From an observational point of view, \cite{chen+2020} found $\sim1$ quasar per deg$^{2}$ with likely periodicity in their optical light curves in the overlapping region of the Dark Energy Survey Supernova (DES-SN) fields and the Sloan Digital Sky Survey Stripe 82 (SDSS-S82). 

Although several assumptions were made, our optimistic observational probability of $\sim$5\% is fully consistent with theoretical predictions and previous observational estimates. Longer X-ray monitoring campaigns might reduce the number of possible SMBHB candidates in this sample, thus reducing the currently derived observational probability.

\subsection{Gravitational waves expectations}

In their latest stages, SMBHBs will lose most of their angular momentum due to GW production. Therefore, if our sources are the progenitors of such close GW-emitting SMBHBs, they will also be ideal targets of GW missions and their multi-wavelength follow-up. The predictions from both methods would then be expected to be consistent with each other. However, we note that our X-ray-selected SMBHB candidates are only a subsample of the GW-SMBHB sample, as our objects need to be accreting to see their X-ray emission. Gravitational-wave missions, on the other hand, will also be sensitive to non-accreting SMBHBs. 

In principle, SMBHBs with masses calculated in our sample ($\sim10^{7}\; \rm M_{\odot}$) are detectable by space-borne interferometer GW missions such as the Laser Interferometer Space Antenna \citep[LISA;][]{Amaro-Seoane+2017} or TianQin \citep[][]{Luo_2016}. However, the tentative orbital periods of our candidates are longer than those that will be probed by these missions.

Following \cite{petersandmathews+1963} and \cite{peters+1964a}, we derive a GW-driven merging timescale for eRASSt J0530-4125 of $4\times10^{6}$ years by assuming that the X-ray emission is effectively coming from a SMBHB in a GW-driven regime with circular orbits, a distance between the SMBHs of 1.2 milli-pc, and a total mass of $M_{H\beta}=10^{7.23}\; M_{\odot}$ with equal mass ratios. In general, this GW-driven decay is faster for eccentric binaries but slower for binaries of unequal mass.

\section{Conclusions}\label{sec:conclusions}

We have explored the largest time-domain X-ray catalog, searching for quasi-periodic light curves in extragalactic X-ray sources, aiming to discover a hidden population of SMBH binaries. Using the \srgero X-ray catalog of the Western Galactic hemisphere, we selected X-ray sources with `up-down-up-down' or `down-up-down-up' profiles (from eRASS scan to scan) in their 0.2--2.3 keV flux light curves.

Initially, we selected a total of 16 SMBHB candidates. Optical spectroscopic analysis confirmed the extragalactic and nuclear nature of 15 out of the 16 candidates. We found that most of the sources host low single-epoch SMBH masses ($\sim10 ^{7}\;\rm M_{\odot}$) based on the width and luminosity of the broad Balmer lines. Most of the sources are characterized as AGN, given their location in the BPT diagnostic diagram. By using their optical redshifts, we derived observed X-ray luminosities for the sources ranging between $\sim10^{40-43}\rm\; erg\; s^{-1}$ in the 0.2--2.3 keV band. We claim that their luminous X-ray emission is very likely to originate in the vicinity of the SMBH at their centers, thus ruling out an off-nuclear origin. 

Extensive \swix and \nic X-ray monitoring campaigns followed up the primary SMBHB candidates with a typical cadence of 1 month. Given the monitoring strategy and the number of data points in the light curves, we found support for a 12-month periodicity in the galaxy eRASSt J0530-4125 at redshift $z=0.076$. Unlike the optical and IR emission, the UV bands of the \swiu light curves tend to follow the variability trend defined by the X-ray light curve with a tentative period of $\sim1$ year. The lack of optical and IR variability is consistent with the scenario where the SMBHB is surrounded by a colder circumbinary disk that does not vary with the orbital period of the binary.
Assuming that the periodic signal is caused by a SMBHB, we calculate a distance of $1.2$ milli-pc between the orbiting SMBHs.  
The optical spectral analysis of the source shows that eRASSt J0530-4125 has a galaxy-like spectrum with weak emission lines and a lack of strong broad emission lines, leading to an uncertain single-epoch SMBH mass measurement of $\log\; M_{bh}(H\beta)/M_{\odot}=7.26\pm0.48$. Based on the optical narrow emission lines, the gas in the host galaxy can be explained by an ionizing continuum from star-forming regions rather than AGN emission. This, however, does not explain the X-ray emission.

We emphasize that the effect of stochastic red-noise X-ray variability produced by accretion onto a single SMBH can mimic a few cycles of periodic behavior, affecting and biasing our sample and interpretation of the X-ray light curves (possibly including eRASSt J0530-4125). The effect of red-noise variability in our sample will be investigated in detail in the second paper of this program (Tubín-Arenas et al. in prep.). Simulations of stochastic AGN light curves coupled with an eROSITA-like observation pattern will provide a quantifiable estimation of the false positives in our sample and will help us improve the source selection. However, longer and extended monitoring programs will also be needed to discriminate between the SMBHB scenario and red-noise-induced variability.

Under the assumption that stochastic red-noise effects do not drive the variability, we observed that our sources share features with some of the rarest transient events. We noted that repeating TDEs might potentially meet our initial eROSITA selection criteria. However, most of the sources in our sample exhibit X-rays consistent with AGN emission ($\Gamma\sim2$), and their multi-wavelength properties do not indicate evidence of optical or IR flares. The absence of a multi-wavelength response to the X-ray emission resembles the scenario recently unveiled by QPEs, where they are only detectable in the soft X-ray regime. Our sources, however, display harder X-ray emission with longer timescale variability. Therefore, 
our current multiwavelength data  
make it unlikely that for the majority of our sources a TDE and QPE scenario is the origin of the quasi-periodic light curve signature. The SMBHB scenario remains a likely origin for the X-ray emission. 

The number of selected SMBHB candidates with quasi-periodic X-ray signals was used to derive an optimistic estimate of $\sim0.05$ SMBHB per X-ray emitting galaxy.
This value is consistent with existing theoretical predictions that suggest that $<1-10\%$ of the galaxies host a SMBHB.

\srgero is the first repeating all-sky X-ray survey that has opened a window to detect new and rare extragalactic events, such as TDEs, QPEs, and SMBHBs. Combined with spectroscopic and multi-wavelength follow-up, it is paving a new road toward finding and characterizing the demographics of SMBHBs. We have seen that multi-wavelength monitoring programs covering multiple cycles are needed to fully test the scenarios proposed in this paper. Unfortunately, given the timescales displayed by the objects, X-ray programs are expensive and not always available. This highlights the need for missions such as \srgero that were designed to perform all-sky monitoring of X-ray sources at regular intervals. Current and future X-ray missions will revolutionize the search for SMBHBs by observing with unprecedented spectral resolution the Fe K$\alpha$ emission lines. XRISM \citep[][]{xrism+2018} and NewAthena \citep[][]{newathena+2025} can and will provide high-throughput imaging with high X-ray spectral resolution down to a $\sim5$ eV in the 0.3--12 keV band. High-resolution observations at these energies will easily resolve any double-peaked lines or other effects of the binary in the Fe K$\alpha$ line profile \citep[][]{McKernan+15}.

\begin{acknowledgements}
We thank the referee for their useful comments that helped
to improve the manuscript.
D.T. acknowledges support by DLR grant FKZ 50 OR 2203. 
M.K. is supported by DLR grant FKZ 50 OR 2307.
G.L. acknowledges support from the German DLR under contract 50 QR 2104.
D.H. acknowledges support from DLR grant FKZ 50 OR 2003.
AGM acknowledges support from Narodowe Centrum Nauki (NCN) grant 2018/31/G/ST9/03224, and partial support from NCN grant 2019/35/B/ST9/03944.

This work is based on data from eROSITA, the soft X-ray instrument aboard SRG, a joint Russian-German science mission supported by the Russian Space Agency (Roskosmos), in the interests of the Russian Academy of Sciences represented by its Space Research Institute (IKI), and the Deutsches Zentrum f\"{u}r Luft- und Raumfahrt (DLR). The SRG spacecraft was built by Lavochkin Association (NPOL) and its subcontractors, and is operated by NPOL with support from the Max Planck Institute for Extraterrestrial Physics (MPE). The development and construction of the eROSITA X-ray instrument was led by MPE, with contributions from the Dr. Karl Remeis Observatory Bamberg \& ECAP (FAU Erlangen-Nuernberg), the University of Hamburg Observatory, the Leibniz Institute for Astrophysics Potsdam (AIP), and the Institute for Astronomy and Astrophysics of the University of T\"{u}bingen, with the support of DLR and the Max Planck Society. The Argelander Institute for Astronomy of the University of Bonn and the Ludwig Maximilians Universit\"{a}t Munich also participated in the science preparation for eROSITA. The eROSITA data shown here were processed using the eSASS/NRTA software system developed by the German eROSITA consortium.

This work made use of data supplied by the UK Swift Science Data Centre at the University of Leicester. We acknowledge the use of public data from the Swift data archive.

Based on observations collected at the European Organisation for Astronomical Research in the Southern Hemisphere under ESO programme 112.263K.

This paper uses observations made at the South African Astronomical Observatory (SAAO).

This work has made use of data from the Asteroid Terrestrial-impact Last Alert System (ATLAS) project. The Asteroid Terrestrial-impact Last Alert System (ATLAS) project is primarily funded to search for near earth asteroids through NASA grants NN12AR55G, 80NSSC18K0284, and 80NSSC18K1575; byproducts of the NEO search include images and catalogs from the survey area. This work was partially funded by Kepler/K2 grant J1944/80NSSC19K0112 and HST GO-15889, and STFC grants ST/T000198/1 and ST/S006109/1. The ATLAS science products have been made possible through the contributions of the University of Hawaii Institute for Astronomy, the Queen’s University Belfast, the Space Telescope Science Institute, the South African Astronomical Observatory, and The Millennium Institute of Astrophysics (MAS), Chile.

The Legacy Surveys consist of three individual and complementary projects: the Dark Energy Camera Legacy Survey (DECaLS; Proposal ID \#2014B-0404; PIs: David Schlegel and Arjun Dey), the Beijing-Arizona Sky Survey (BASS; NOAO Prop. ID \#2015A-0801; PIs: Zhou Xu and Xiaohui Fan), and the Mayall z-band Legacy Survey (MzLS; Prop. ID \#2016A-0453; PI: Arjun Dey). DECaLS, BASS and MzLS together include data obtained, respectively, at the Blanco telescope, Cerro Tololo Inter-American Observatory, NSF’s NOIRLab; the Bok telescope, Steward Observatory, University of Arizona; and the Mayall telescope, Kitt Peak National Observatory, NOIRLab. Pipeline processing and analyses of the data were supported by NOIRLab and the Lawrence Berkeley National Laboratory (LBNL). The Legacy Surveys project is honored to be permitted to conduct astronomical research on Iolkam Du’ag (Kitt Peak), a mountain with particular significance to the Tohono O’odham Nation. NOIRLab is operated by the Association of Universities for Research in Astronomy (AURA) under a cooperative agreement with the National Science Foundation. LBNL is managed by the Regents of the University of California under contract to the U.S. Department of Energy. This project used data obtained with the Dark Energy Camera (DECam), which was constructed by the Dark Energy Survey (DES) collaboration. Funding for the DES Projects has been provided by the U.S. Department of Energy, the U.S. National Science Foundation, the Ministry of Science and Education of Spain, the Science and Technology Facilities Council of the United Kingdom, the Higher Education Funding Council for England, the National Center for Supercomputing Applications at the University of Illinois at Urbana-Champaign, the Kavli Institute of Cosmological Physics at the University of Chicago, Center for Cosmology and Astro-Particle Physics at the Ohio State University, the Mitchell Institute for Fundamental Physics and Astronomy at Texas A\&M University, Financiadora de Estudos e Projetos, Fundacao Carlos Chagas Filho de Amparo, Financiadora de Estudos e Projetos, Fundacao Carlos Chagas Filho de Amparo a Pesquisa do Estado do Rio de Janeiro, Conselho Nacional de Desenvolvimento Cientifico e Tecnologico and the Ministerio da Ciencia, Tecnologia e Inovacao, the Deutsche Forschungsgemeinschaft and the Collaborating Institutions in the Dark Energy Survey. The Collaborating Institutions are Argonne National Laboratory, the University of California at Santa Cruz, the University of Cambridge, Centro de Investigaciones Energeticas, Medioambientales y Tecnologicas-Madrid, the University of Chicago, University College London, the DES-Brazil Consortium, the University of Edinburgh, the Eidgenossische Technische Hochschule (ETH) Zurich, Fermi National Accelerator Laboratory, the University of Illinois at Urbana-Champaign, the Institut de Ciencies de l’Espai (IEEC/CSIC), the Institut de Fisica d’Altes Energies, Lawrence Berkeley National Laboratory, the Ludwig Maximilians Universitat Munchen and the associated Excellence Cluster Universe, the University of Michigan, NSF’s NOIRLab, the University of Nottingham, the Ohio State University, the University of Pennsylvania, the University of Portsmouth, SLAC National Accelerator Laboratory, Stanford University, the University of Sussex, and Texas A\&M University. BASS is a key project of the Telescope Access Program (TAP), which has been funded by the National Astronomical Observatories of China, the Chinese Academy of Sciences (the Strategic Priority Research Program “The Emergence of Cosmological Structures” Grant \# XDB09000000), and the Special Fund for Astronomy from the Ministry of Finance. The BASS is also supported by the External Cooperation Program of Chinese Academy of Sciences (Grant \# 114A11KYSB20160057), and Chinese National Natural Science Foundation (Grant \# 12120101003, \# 11433005). The Legacy Survey team makes use of data products from the Near-Earth Object Wide-field Infrared Survey Explorer (NEOWISE), which is a project of the Jet Propulsion Laboratory/California Institute of Technology. NEOWISE is funded by the National Aeronautics and Space Administration. The Legacy Surveys imaging of the DESI footprint is supported by the Director, Office of Science, Office of High Energy Physics of the U.S. Department of Energy under Contract No. DE-AC02-05CH1123, by the National Energy Research Scientific Computing Center, a DOE Office of Science User Facility under the same contract; and by the U.S. National Science Foundation, Division of Astronomical Sciences under Contract No. AST-0950945 to NOAO.

\end{acknowledgements}

\bibliographystyle{aa}
\bibliography{bib.bib}

\begin{appendix}

%\onecolumn

\section{Individual properties of the SMBHB candidates}\label{appendix:particularcases}

\subsection{eRASSt J0344-3327}

\begin{figure}[h!]
    \centering
    \includegraphics[width=1\linewidth]{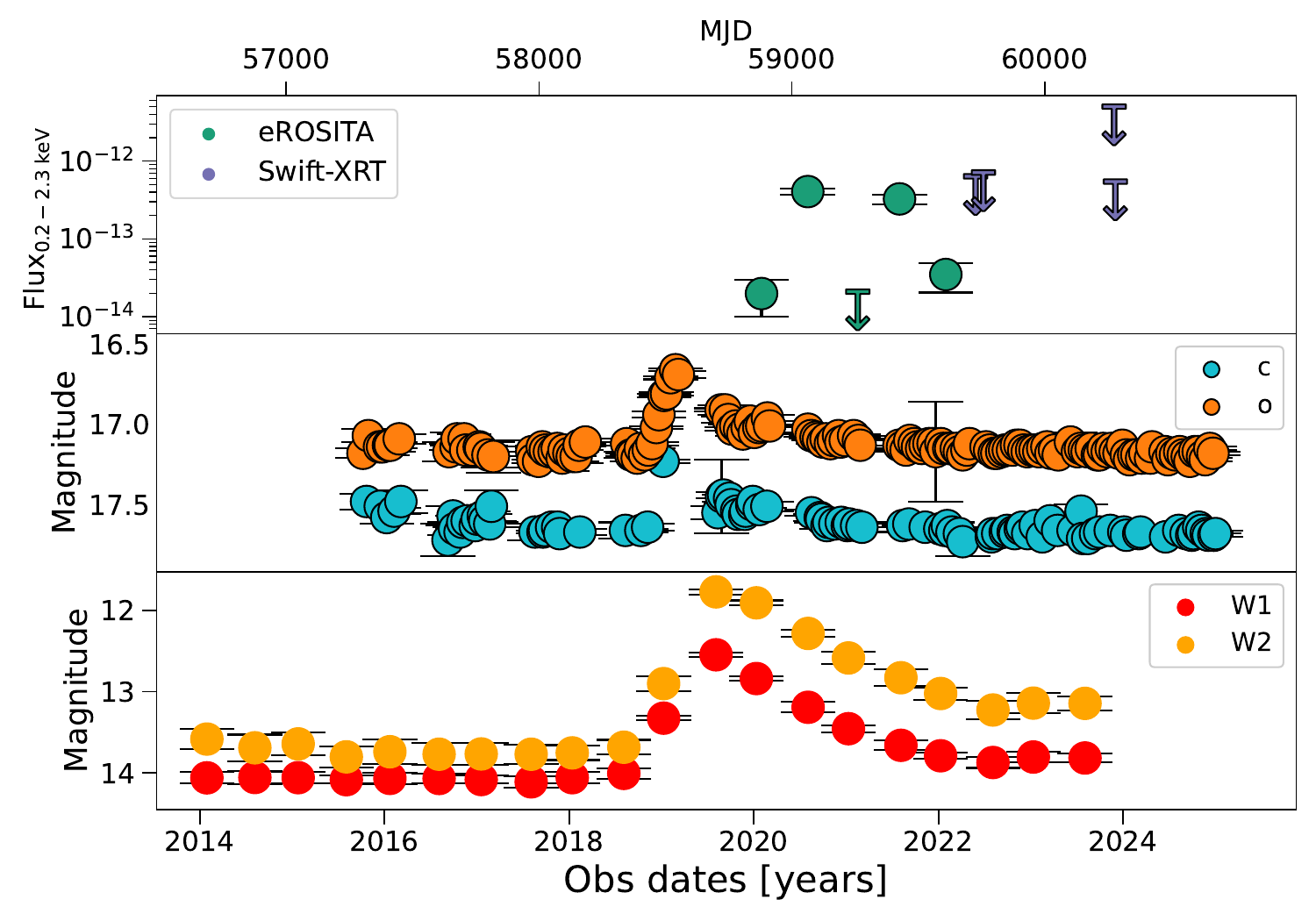}
\caption{Multi-wavelength light curve of eRASSt J0344-3327. X-ray data is given by \srgero (green markers) and \swix (purple upper limits) in the top panel. Optical data is retrieved from ATLAS in their ``c'' and ``o'' bands (light-blue and orange markers) and presented in bins of 15 days in the middle panel. The WISE light curve (red and yellow markers) is shown in the bottom panel. }
    \label{fig:erasst_j0344-3327}
\end{figure}

eRASSt J0344-3327 ($z=0.09$) is the source showing the largest flux ratio between the bright and faint states of the eROSITA light curve. The source is a galaxy with AGN features, given the relatively strong [\textsc{O iii}]$\lambda5007$\AA~line and the BPT diagram. 

eRASSt J0344-3327 was observed 4 times by \swix between 2022-05-05 and 2023-11-09, and no X-ray source was detected at the position of eRASSt J0344-3327. 
The \swix LSXPS Upper limit server reports a $3\sigma$ upper limit of $2.0\times10^{-3}$ cts s$^{-1}$ in the 0.3--10.0 keV stacked image that corresponds with a $3\sigma$ upper flux limit of $1.8\times10^{-13}\rm \; erg\; s^{-1}\; cm^{-2}$ assuming a Galactic-absorbed power-law spectral model with a column density and photon index given by the values in Table \ref{table:stackedxray} (n$_{\rm H}=1.57\times10^{20}\; \rm cm^{-2}$; $\Gamma=5.35$). Assuming this Galactic-absorbed power-law spectral model, we scaled the 0.3--10.0 keV upper limit flux to the \srgero energy bands, as shown in Fig. \ref{fig:allxraycandidates}. We note that the \swix upper limits of 2022 are still consistent with the periodic prediction. However, deeper observations are needed to confirm or rule out any possible rebrightening expected from our SMBHB predictions. 

\nic observed the galaxy between July and October 2022. The stacked observations have a net exposure time of 16 ks. The spectral fitting shows that it is a soft source ($\Gamma\sim5$) that can also be characterized by an accretion disk consisting of multiple blackbody components ($kT\sim70$ eV).

ATLAS and WISE light curves in Fig. \ref{fig:erasst_j0344-3327} show a sharp brightening between August 2018 and January 2019 and reached its peak in August 2019 according to the WISE data. The optical and IR fluxes have decayed exponentially since then. This behavior suggests a tidal disruption event \citep[TDE,][]{Hills+1975,rees+1988} where the soft X-ray emission seen by \srgero is likely produced by the repeated accretion of the debris material onto the SMBH after the disruption event. 

There is a 4-year window between the tentative start of the TDE and the lack of detection on the \swix monitoring observations where the source might have been detected in X-rays. This window matches the decay time seen in the ATLAS and WISE light curves. The X-ray variability on eRASSt J0344-3327 could be explained by a repeating partial tidal disruption event (pTDE), similar to the variability found in eRASSt J045650.3-203750 \citep[][]{Liu+2023L,liupTDE+24}. The TDE and the later `down-up-down-up-down' variability on eRASSt J0344-3327 matches the scenario proposed by \cite{shutdeinsmbhb+20} where several flux dips occurred during the decay phase. The variability in \cite{shutdeinsmbhb+20} is associated with X-ray flares from the stellar tidal disruption by a SMBHB candidate. A more detailed discussion about TDEs and X-ray quasi-periodic events is presented in Sect. \ref{subsect:nature}.

\subsection{eRASSt J0530-4125}

eRASSt J0530-4125 ($z=0.076$) is one of the most variable and best-studied eROSITA-selected SMBHB candidates. Only a \ros all-sky survey upper flux limit at a level of $<3\times10^{-13}\; \rm erg\; s^{-1}\; cm^{-2}$ in the 0.2--2.0 keV band is found in archival X-ray data, consistent with the faint flux level of the source. 

The optical analysis of the source shows that eRASSt J0530-4125 has a galaxy-like spectrum with weak broad and narrow emission lines. The broad emission lines found in the spectrum, in particular H$\beta$, are not significantly detected, leading to an uncertain single-epoch SMBH mass measurement of $\log\; M_{bh}/M_{\odot}=7.26\pm0.48$. Based on the optical narrow emission lines, the gas in the host galaxy appears to be ionized by star-formation processes instead of AGN. 

\begin{figure}[h!]
    \centering
    \includegraphics[width=1\linewidth]{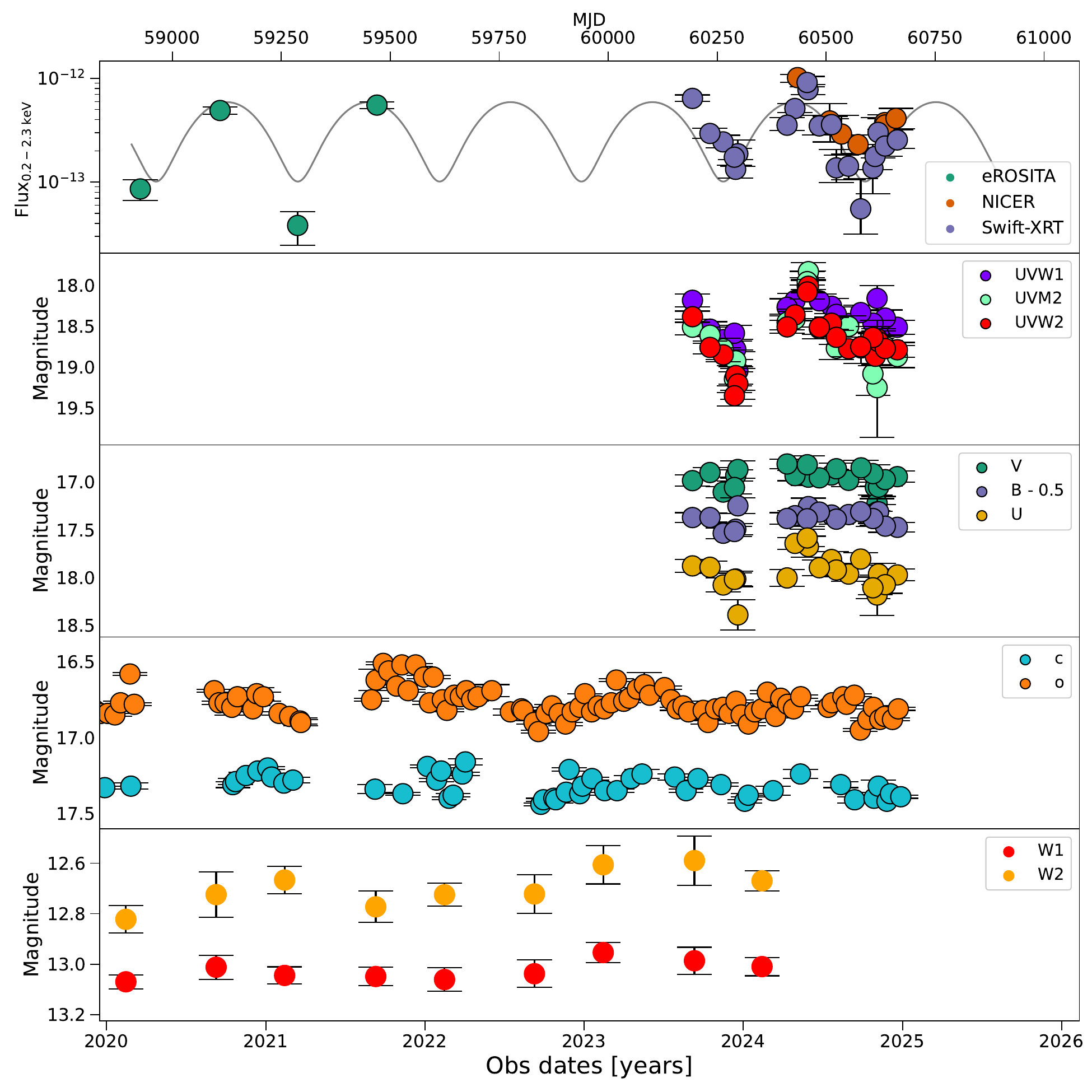}
    \caption{Multi-wavelength light curve of eRASSt J0530-4125. The first panel shows the X-ray data from \srgero+ \swix+ \nic at the 0.2--2.3 keV band. UV and optical light curves from \swiu are shown in the second and third panels, while optical and IR light curves, obtained from ATLAS and WISE, are displayed in the fourth and fifth panels, respectively. 
    }
    \label{fig:multiwavelc}
\end{figure}

Figure \ref{fig:multiwavelc} shows the multi-wavelength data collected for eRASSt J0530-4125. We note that the UV bands of the \swiu light curves follow the variability trend defined by the X-ray light curve. The hot and more energetic photons usually come from the vicinity of the SMBHs. Therefore, an X-ray-UV correlation is expected under the SMBHB scenario. ATLAS data provides a preliminary assessment of the host-galaxy contribution by subtracting a reference image from every observation. We do not notice any significant optical variability in this new (AGN-only) ATLAS light curve.

\subsection{eRASSt J1906-4850}

eRASSt J1906-4850 (also known as 	
LEDA 3904185) is a nearby ($z=0.049$) and luminous ($L_{0.2-2.3\; \rm keV} = (1.73-23.8)\times 10^{42}$ erg s$^{-1}$) Seyfert I galaxy with a double-peaked broad Balmer line in the optical spectrum, as shown in Fig. \ref{fig:fullfit}. One of the broad components is consistent in kinematics with the narrow Balmer line while the other component is redshifted by $\Delta v\rm = 2900\pm740\; km\; s^{-1}$. According to the narrow line diagnostic diagram and the spectral shape of the optical spectrum, the source is consistent with being an AGN.

\begin{figure}[h!]
    \centering
    \includegraphics[width=1.0\linewidth]{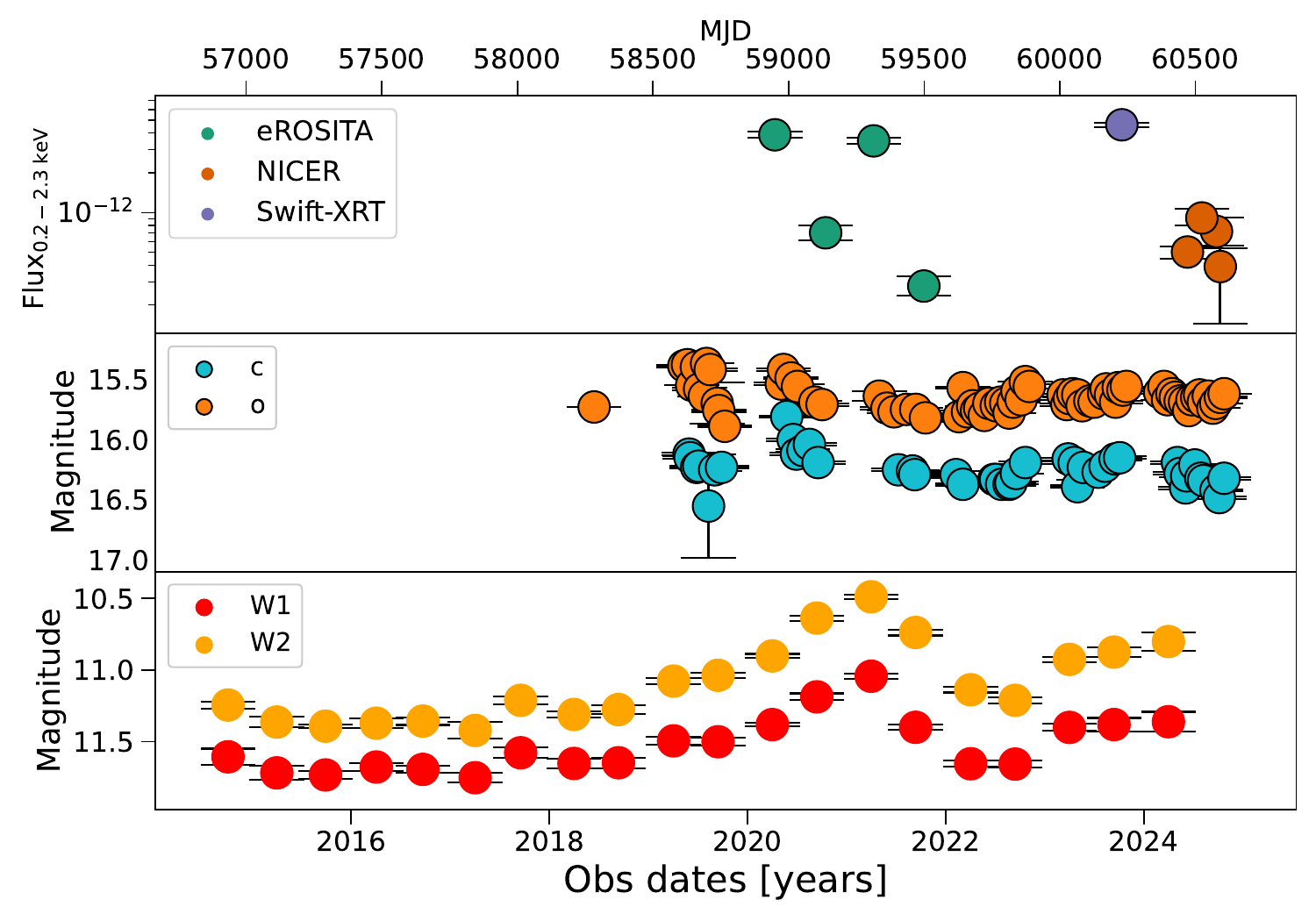}
    \caption{Multi-wavelength observations of eRASSt J1906-4850. }
    \label{fig:eRASSt_J1906-4850}
\end{figure}

Two single-epoch SMBH masses were derived by assuming that each hypothetical SMBH is surrounded by an independent BLR, leading to rather low-mass measurements. The mass estimations based on the second and redshifted component of H$\beta$ and H$\alpha$ lines are $\rm \log (M_{bh}/M_{\odot})=6.64\pm0.38$ and $\rm \log (M_{bh}/M_{\odot})=7.50\pm0.23$, respectively.
Double-peaked broad Balmer lines have been proposed to indicate the presence of SMBHBs \citep[see, e.g.,][]{doan+20,terwel+22}. However, a disk-like BLR on a single SMBH can also reproduce double-peaked profiles tracing the gas located in the outer edge of the accretion disk and the inner edge of the BLR \citep[][]{Eracleous+94,Eracleous+09,Storchi-Bergmann_2017}.

Figure \ref{fig:eRASSt_J1906-4850} 
shows the X-ray, optical, and IR light curves of the galaxy eRASSt J1906-4850. Similar to eRASSt J0344-3327, the optical and IR light curves seem to have a pronounced peak before and during the \srgero observations. However, the rise in brightness in the IR light curve takes longer than in eRASSt J0344-3327, making a TDE scenario less likely. The IR seems to lag the optical emission by 2 years, according to the peaks of the corresponding light curves. The multi-wavelength landscape seems to match the behavior presented in eRASSt J0344-3327, where the X-ray `up-down' emission corresponds with the flux dimming after an increase in the brightness in optical and IR. However, the TDE scenario is disfavored based on the slow rise of the IR emission and the lack of features in the optical data. eRASSt J1906-4850 was detected by the \ros all-sky survey in September 1990 with a flux of $\sim3\times10^{-12}\; \rm erg\; s^{-1}\; cm^{-2}$ in the 0.2--2.0 keV band, consistent with the faint flux levels of eRASSt J1906-4850.

\begin{figure}[h!]
    \centering
    \includegraphics[width=1.0\linewidth]{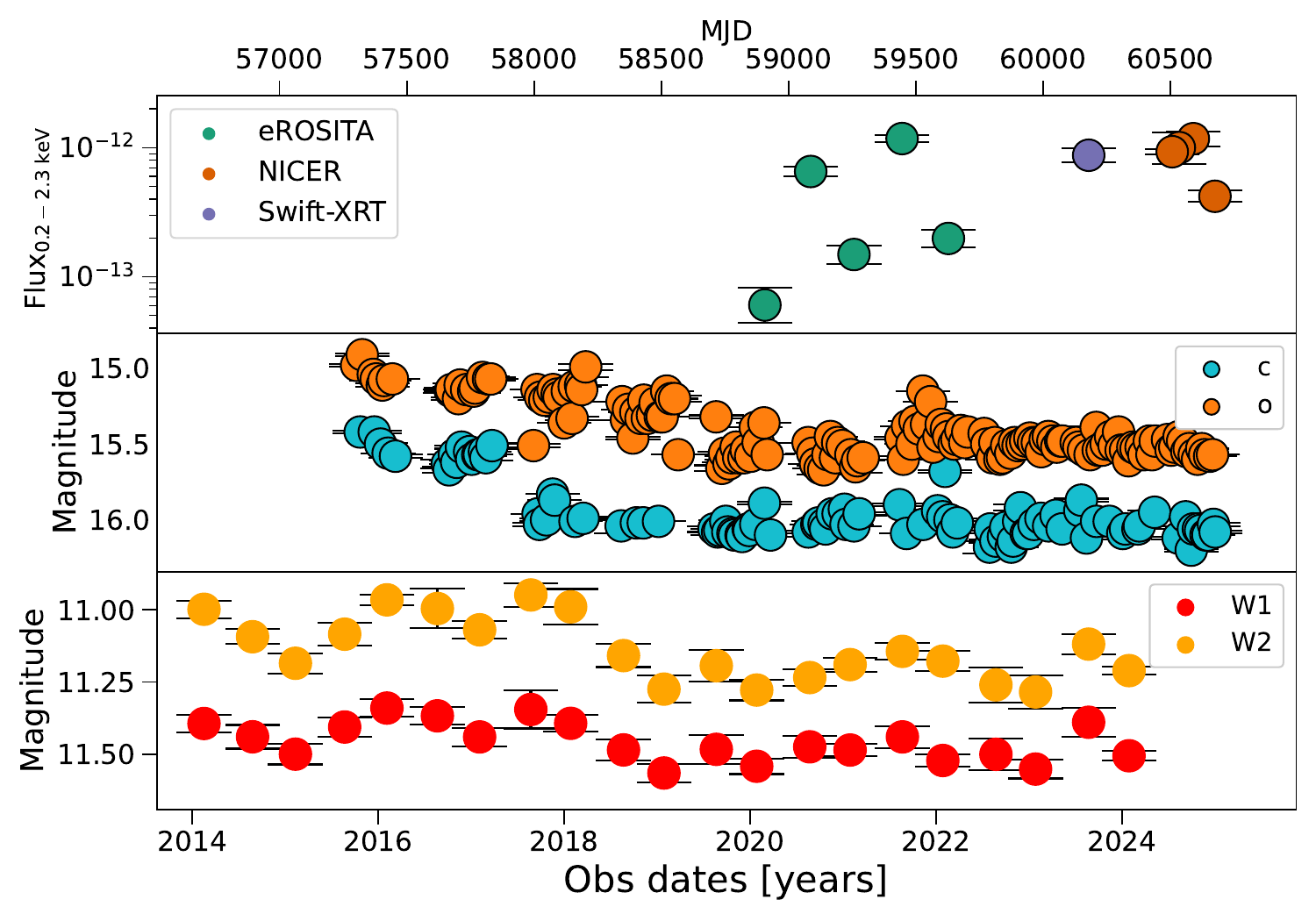}
    \caption{Multi-wavelength observations of eRASSt J0432-3023. The data is similar to the data of Fig. \ref{fig:erasst_j0344-3327}. The X-ray panel also displays \nic data shown with orange markers. }
    \label{fig:eRASSt_J0432-3023}
\end{figure}

\subsection{eRASSt J0432-3023}

eRASSt J0432-3023 (ESO 421-9) is a Seyfert 2 Galaxy at a redshift of $z=0.055$. It has been lately monitored by \nic, but further X-ray observations are needed to confirm or rule out the hint of periodic variability that this source presents. The multi-wavelength light curve in Fig. \ref{fig:eRASSt_J0432-3023} shows no noticeable correlation between the X-ray, optical, and IR. 

\begin{figure}[h!]
    \centering
    \includegraphics[width=1.0\linewidth]{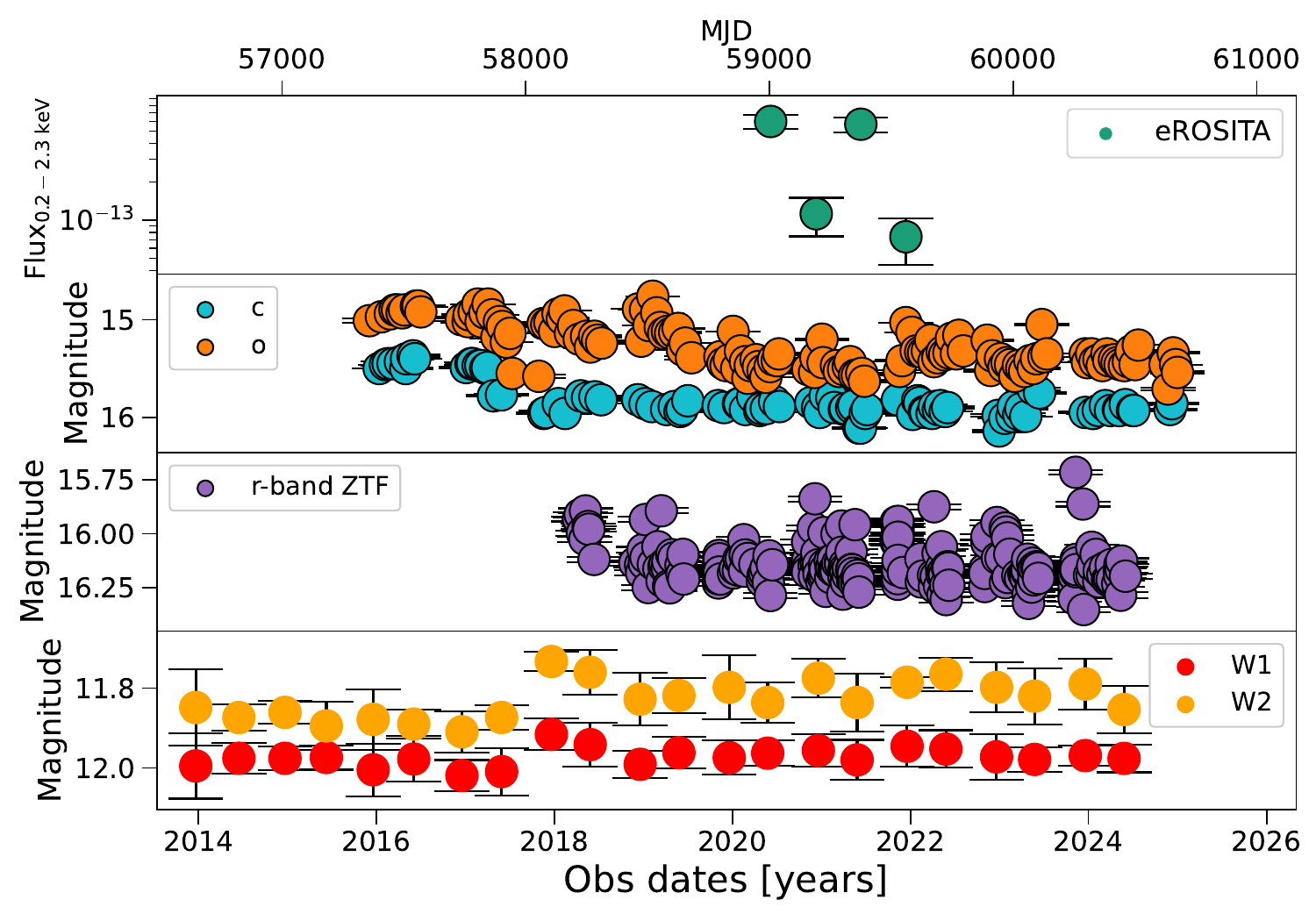}
    \caption{Multi-wavelength observations of eRASSt J1130-0806. The data is similar to the data of Fig. \ref{fig:erasst_j0344-3327} with additional ZTF r-band data in the third panel.  }
    \label{fig:eRASSt_J1130-0806}
\end{figure}

\subsection{eRASSt J1130-0806}

eRASSt J1130-0806 (MCG-01-29-027: $z=0.037$) shows a galaxy-like spectrum with faint emission lines and the absence of broad Balmer lines. The optical counterpart of the \srgero coordinates corresponds to a face-on spiral galaxy. The multi-wavelength data (Fig. \ref{fig:eRASSt_J1130-0806}) show optical variability. However, there is no correlation between the eROSITA and the optical data.
The low luminosity X-ray data and the slightly large separation between the X-ray and optical coordinates could indicate the presence of off-nuclear sources, such as hyper-luminous objects (HLXs). For a discussion about non-nuclear objects that produce variable X-ray emission, see Sect. \ref{subsect:nature}.

\begin{figure}[h!]
    \centering
    \includegraphics[width=1.0\linewidth]{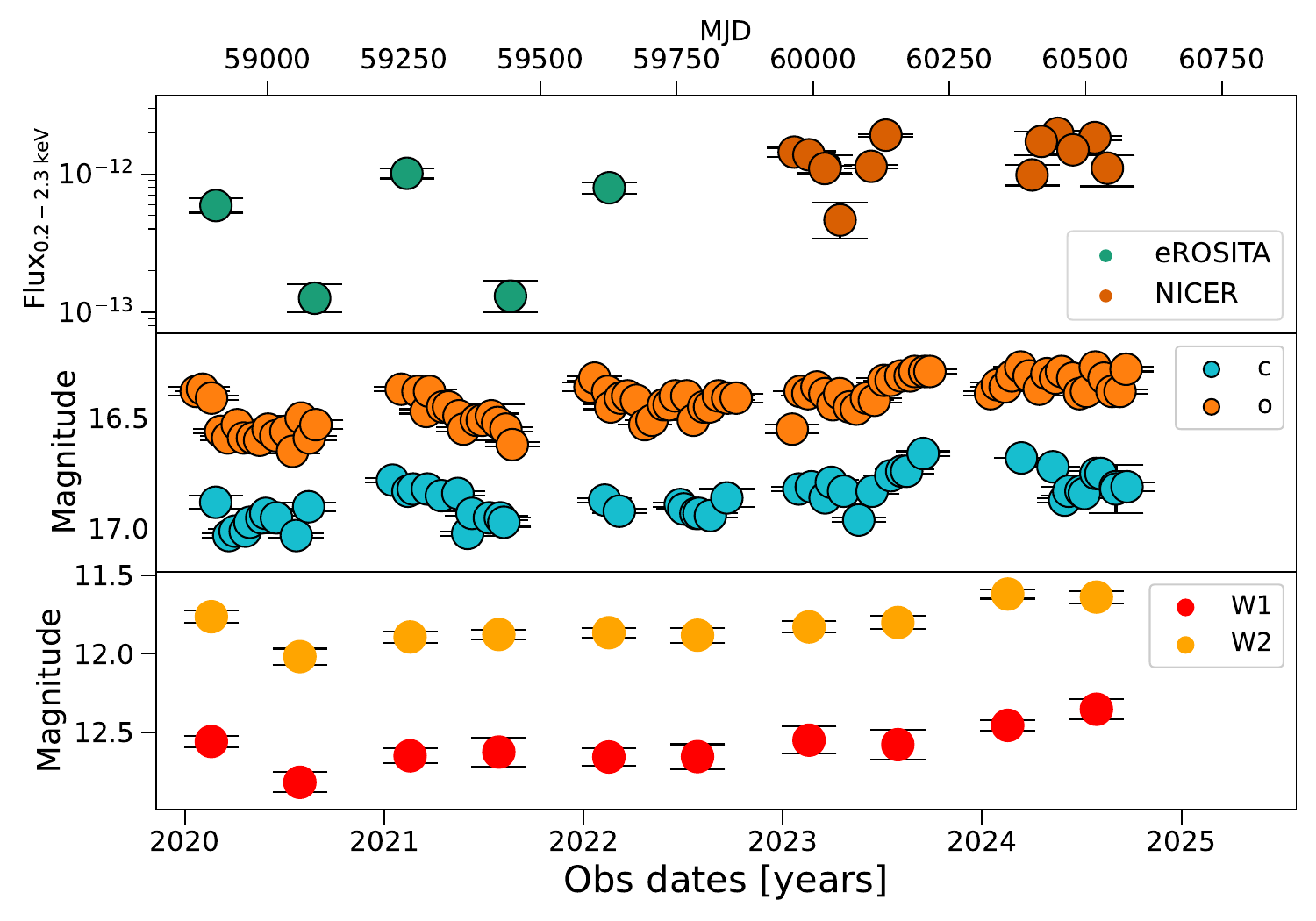}
    \caption{Multi-wavelength observations of eRASSt J1522-3722. The data is similar to the data of Fig. \ref{fig:eRASSt_J0432-3023}. }
    \label{fig:eRASSt_J1522-3722}
\end{figure}

\subsection{eRASSt J1522-3722}

eRASSt J1522-3722 was observed under our NICER program (\#5141) to test its X-ray variability under the SMBHB scenario. The \nic monitoring program shows an X-ray dimming in the light curve, followed by an increase in the flux. 
However, our monitoring programs have been unable to catch the source in its faint flux state. The optical ATLAS data in Fig. \ref{fig:eRASSt_J1522-3722} seems to follow the pattern established by the X-ray data, especially between eRASS1 and eRASS2 and during the \nic monitoring program of early 2023. 

eRASSt J1522-3722 (or 2MASS J15221541-3722488) is classified as an AGN candidate by the Strasbourg Astronomical Data Center (CDS-SIMBAD). Our newly obtained SAAO spectrum of eRASSt J1522-3722 shows broad Balmer emission lines, confirming the AGN nature of the source and a redshift of $z=0.064$. The [\textsc{O iii}]$\lambda5007$\AA\ line was fitted with one narrow and one broad Gaussian line, indicating the presence of outflows. The galaxy falls into the HII region of the BPT because the [\textsc{O iii}] line possesses multiple components, and only the narrow line was used for the diagnostic diagram. 

\subsection{eRASSt J1141+0635}

\begin{figure}[h!]
    \centering
    \includegraphics[width=1.0\linewidth]{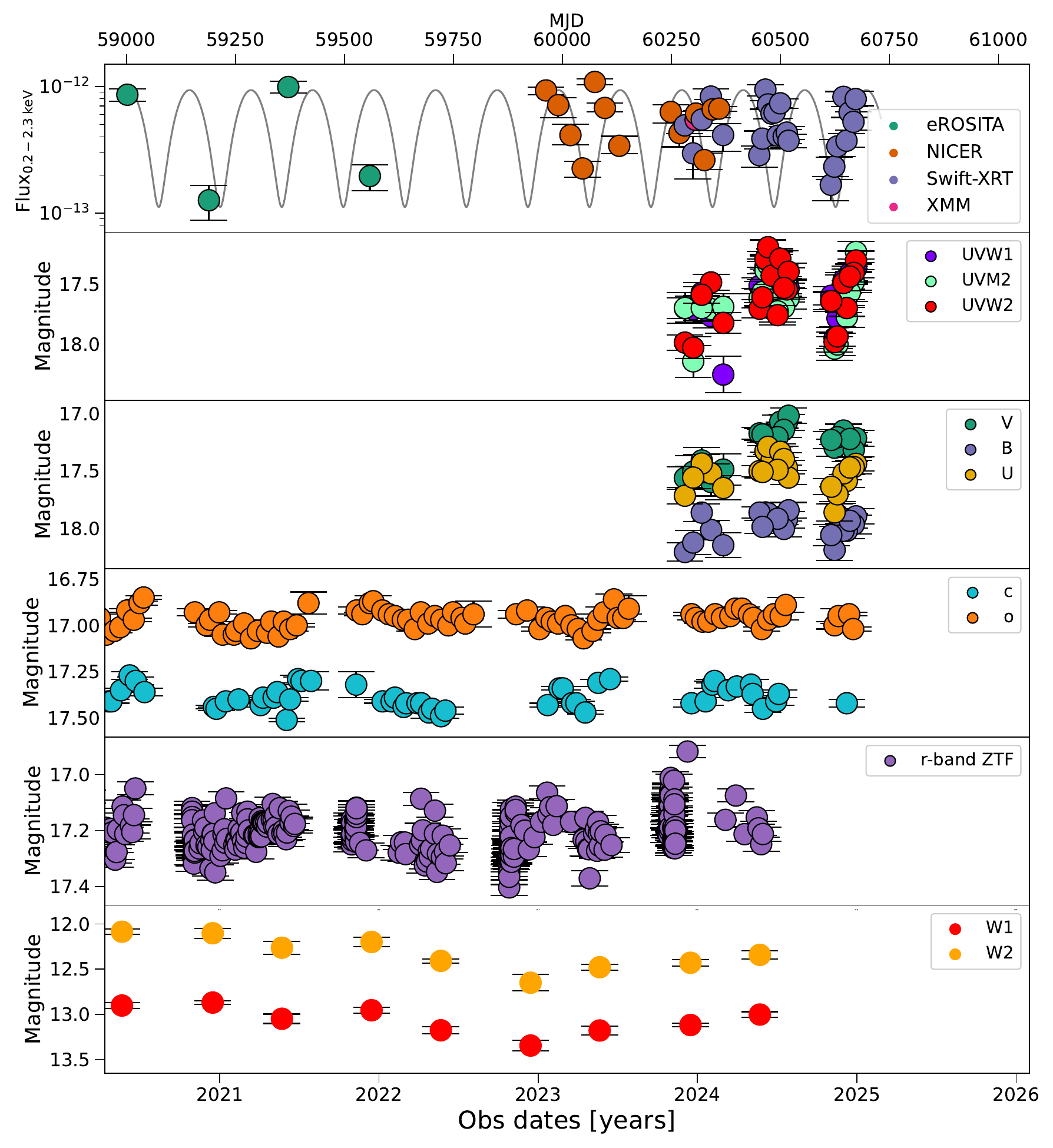}
    \caption{Multi-wavelength observations of eRASSt J1141+0635. Similar to Fig. \ref{fig:multiwavelc} with additional ZTF data in the fifth panel.
    }
    \label{fig:eRASSt_J1141+0635}
\end{figure}

eRASSt J1141+0635 (or 2MASS J11415445+0635096; $z=0.101$) is one of the most monitored sources in our sample. Initially, the X-ray data visually suggested a periodic behavior with a tentative period of $T\sim 120$ days. However, the source does not show any hint of periodic variability anymore. Lately, the \swix monitoring program showed a more ordered variability than in the previous observations. We emphasize that more X-ray observations are needed to disentangle the nature of this source. We note that only the UV data follow the X-ray variability (Fig. \ref{fig:eRASSt_J1141+0635}).

\subsection{eRASSt J0458-2159}

\begin{figure}[h!]
    \centering
    \includegraphics[width=1.0\linewidth]{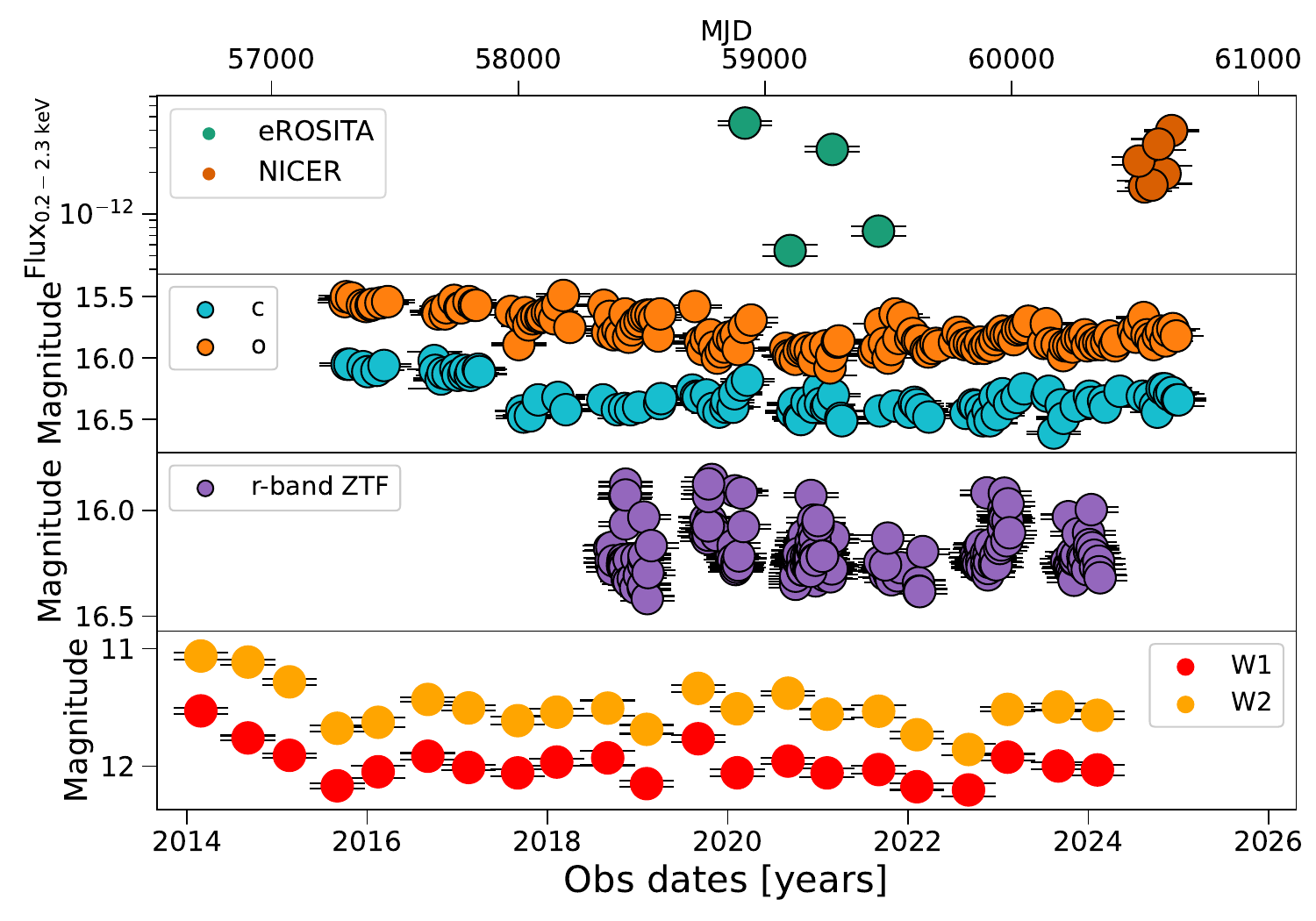}
    \caption{Multi-wavelength observations of eRASSt J0458-2159. The data and markers are similar to those in Fig. \ref{fig:multiwavelc} with additional r-band ZTF optical data.}
    \label{fig:eRASSt_J0458-2159}
\end{figure}

eRASSt J0458-2159 (ESO 552-39; $z=0.040$) is a Seyfert 1 Galaxy monitored by \nic. The source was observed with the \ros all-sky survey, \ros-PSPC, and \ros-HRI pointed observations with fluxes of 9.1, 2.0, and 19.80$\times10^{-13}\rm \; erg\; s^{-1}\; cm^{-2}$, respectively. The ROSAT data are consistent with the faint flux level of the source shown in Fig. \ref{fig:eRASSt_J0458-2159}.

The galaxy shows strong broad Balmer lines and [\textsc{O iii}]$\lambda5007$\AA~emission line consistent with AGN emission, as shown by the BPT diagnostic diagram. 

\subsection{eRASSt J2227-4333}

\begin{figure}[h!]
    \centering
    \includegraphics[width=1.0\linewidth]{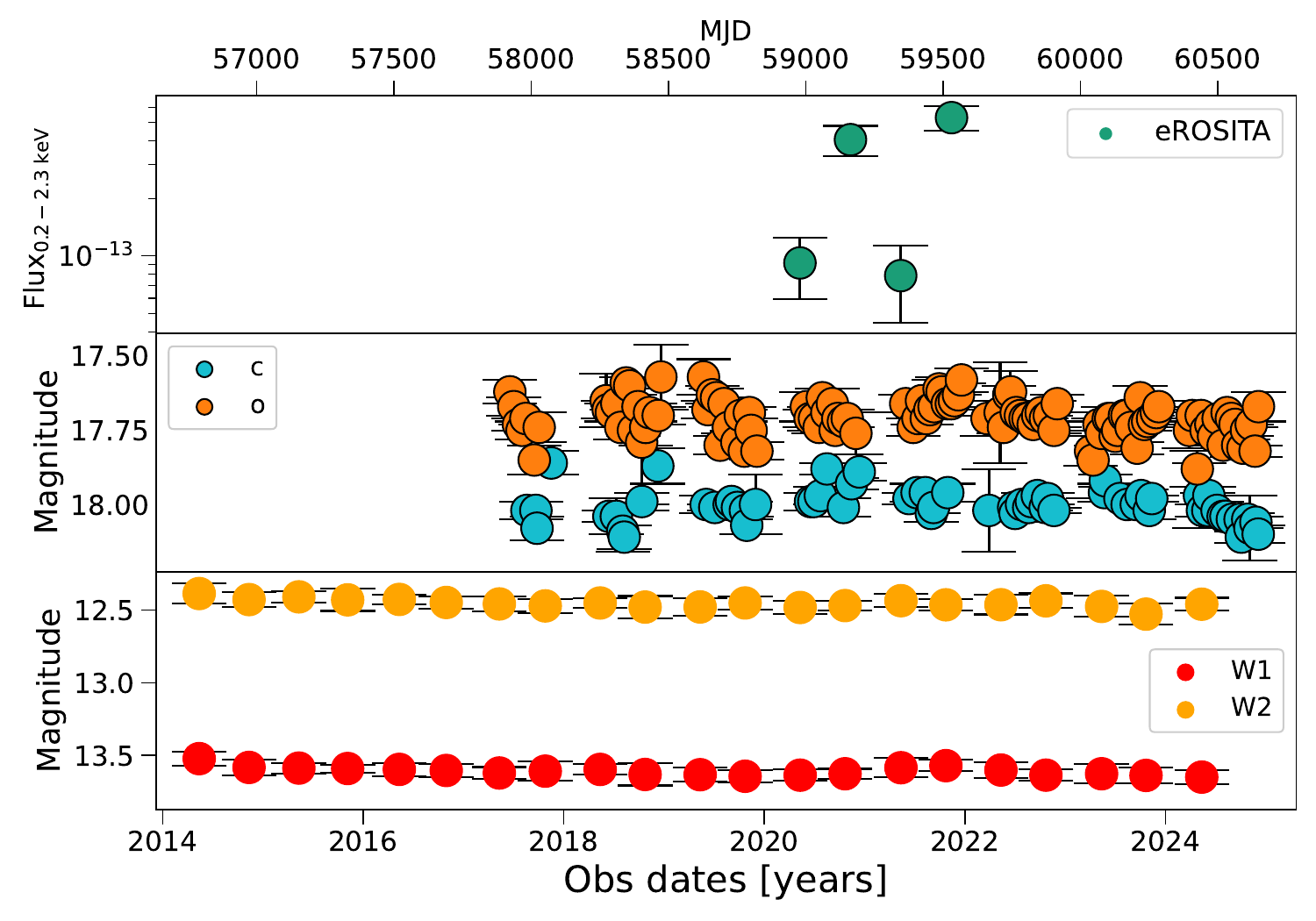}
    \caption{Multi-wavelength observations of eRASSt J2227-4333. Data similar than in Fig. \ref{fig:erasst_j0344-3327}.}
    \label{fig:eRASSt_J2227-4333}
\end{figure}

eRASSt J2227-4333 (6dFGS gJ222755.8-433339) is a galaxy at redshift $z=0.198$ with a strong [\textsc{O iii}]$\lambda5007$\AA\ emission line but without very prominent broad Balmer lines, indicating a type II AGN. 

The optical and IR light curves of the source (Fig. \ref{fig:eRASSt_J2227-4333}) are rather constant, and there is no clear relation with the eROSITA data.

\subsection{eRASSt J1124-0348}

\begin{figure}[h!]
    \centering
    \includegraphics[width=1.0\linewidth]{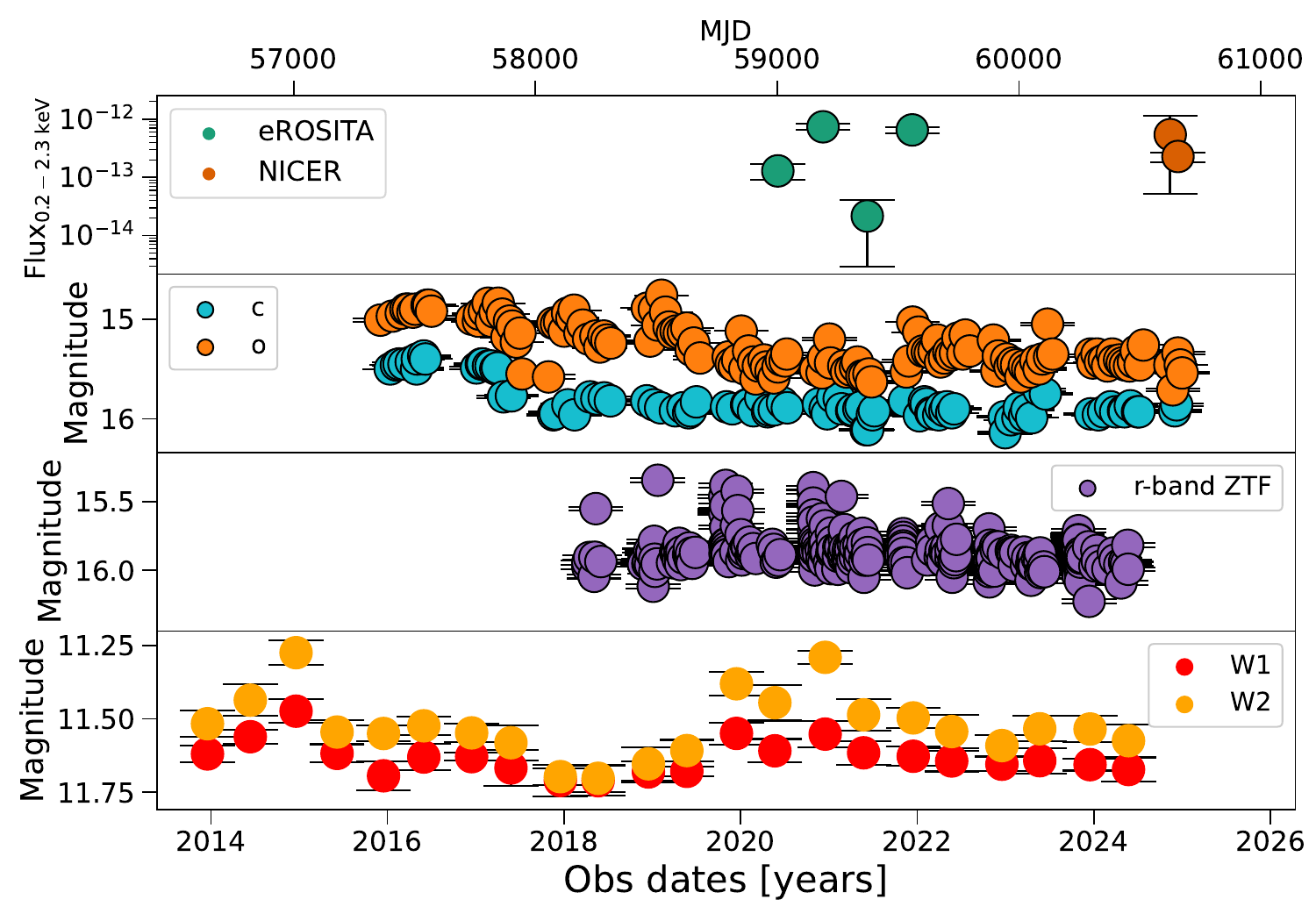}
    \caption{Multi-wavelength observations of eRASSt J1124-0348. Data and markers similar as in Fig. \ref{fig:eRASSt_J1130-0806}.}
    \label{fig:eRASSt_J1124-0348}
\end{figure}

eRASSt J1124-0348 (or 6dFGS gJ112456.3-034840) is a low luminosity Seyfert II Galaxy with a redshift of $z=0.021$. It only has a \ros all-sky survey upper limit of $\sim4\times10^{-13}\; \rm erg\; s^{-1}\;cm^{-2}$ on the 0.2--2.0 keV band, consistent with the faint flux state of the eROSITA data. 
The optical spectrum reveals a galaxy-like spectral shape with a strong [\textsc{O iii}] emission line relative to the H$\beta$ line. The single-epoch SMBH can only be constrained using the slightly broad component of H$\alpha$. 

The multi-wavelength data in Fig. \ref{fig:eRASSt_J1124-0348} show interesting optical variability that apparently matches the IR peaks of 2021. It seems that these peaks in optical and IR are related to the X-ray variability. However, more data is needed to understand the nature of these features. 

\subsection{eRASSt J0036-3125}

eRASSt J0036-3125 (LEDA 172477) is a galaxy at $z=0.108$ with weak emission lines and weak AGN features in the optical spectrum. Despite this, eRASSt J0036-3125 shows a luminous bright state in its X-ray light curve, reaching a luminosity of $1.5\times10^{43}\rm \; erg\; s^{-1}$ in the 0.2--2.3 keV band. 

\begin{figure}[h!]
    \centering
    \includegraphics[width=1.0\linewidth]{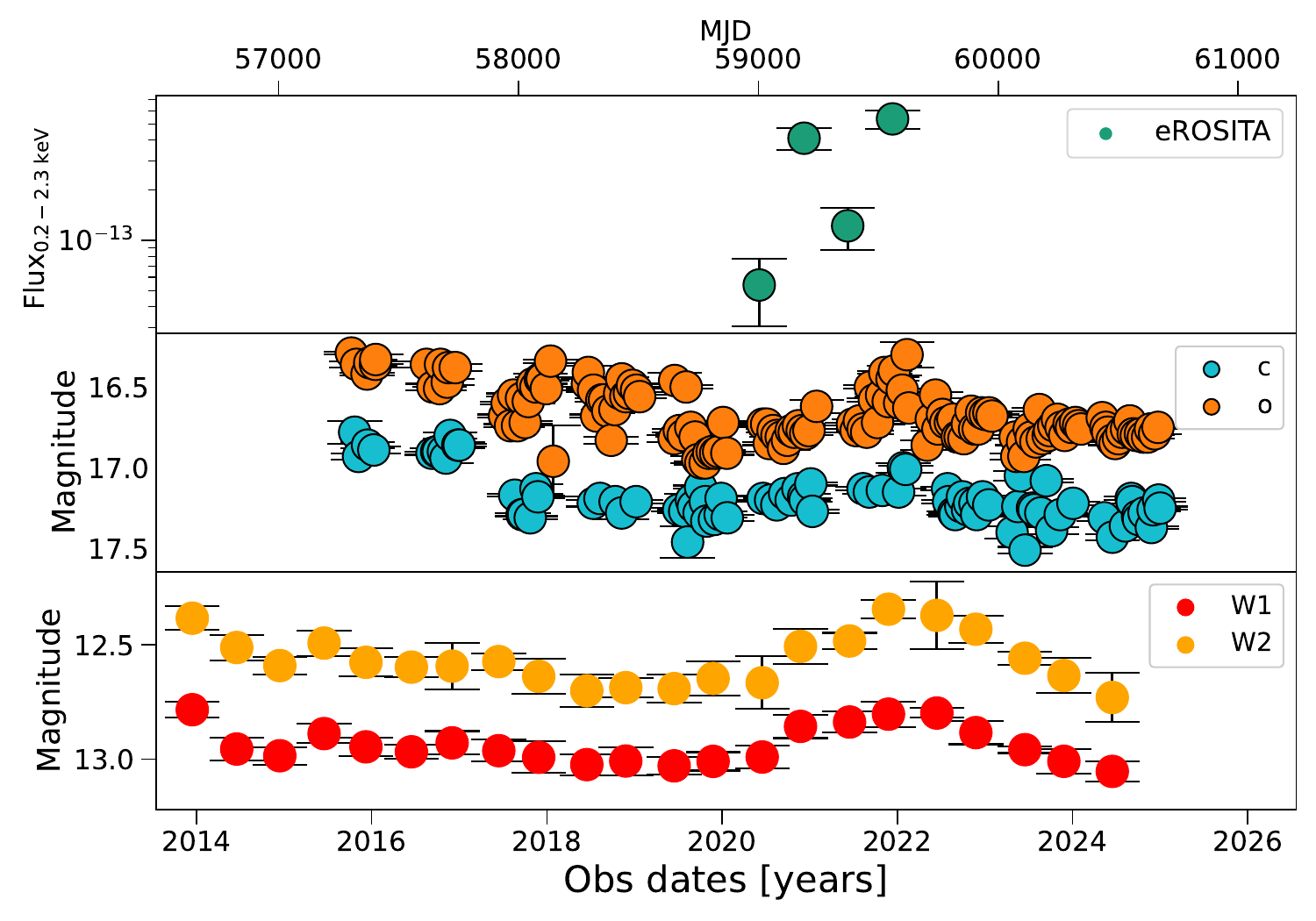}
    \caption{Multi-wavelength observations of eRASSt J0036-3125. Data and markers similar as in Fig. \ref{fig:erasst_j0344-3327}. }
    \label{fig:eRASSt_J0036-3125}
\end{figure}

The optical and IR emission appears to be lagging an increase in the X-ray emission traced by the eROSITA data (Fig. \ref{fig:eRASSt_J0036-3125}). Additional X-ray data are needed to understand if the X-ray emission reaches a stable level again, similar to the optical and IR light curves.  

\begin{figure}[h!]
    \centering
    \includegraphics[width=1.0\linewidth]{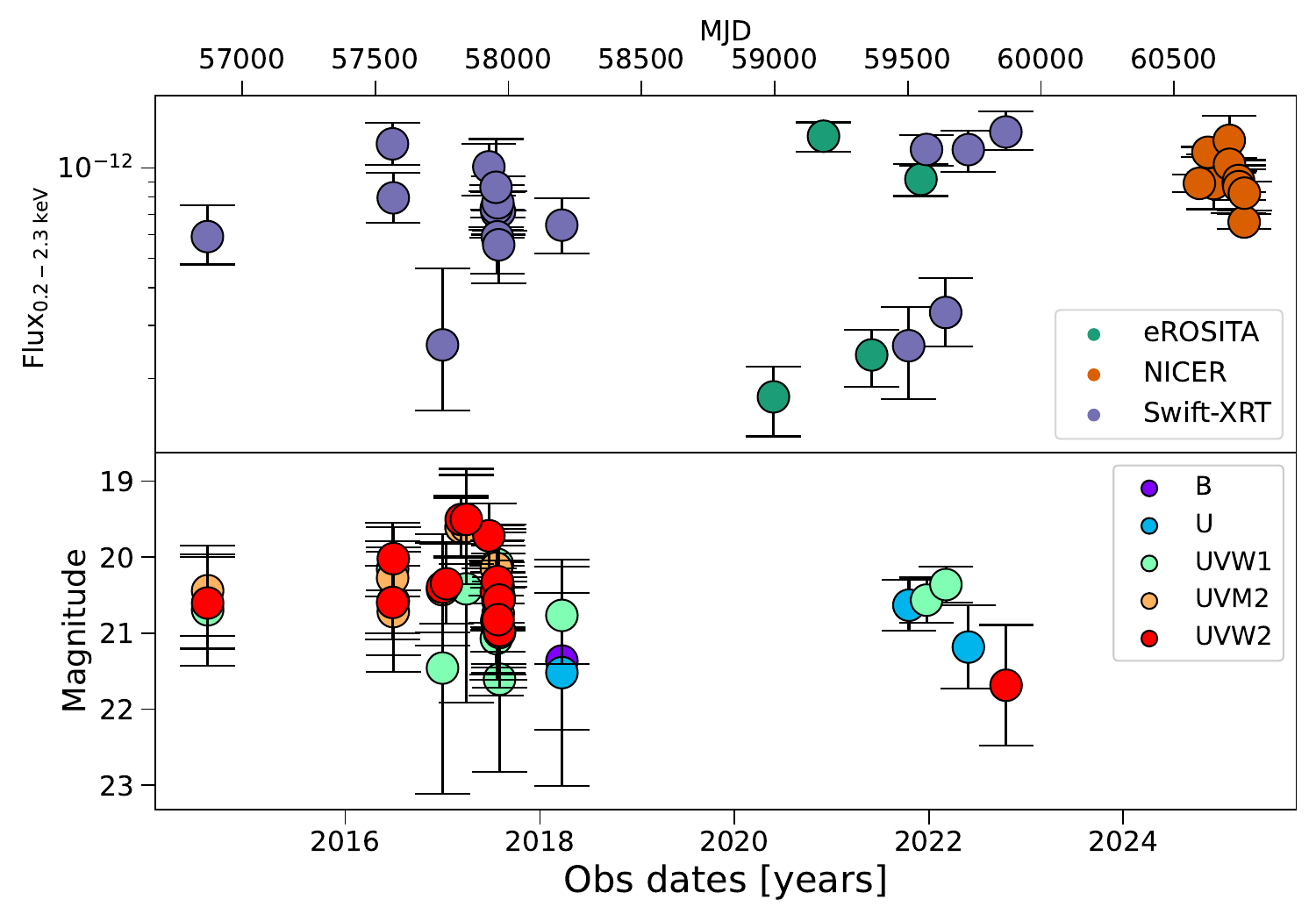}
    \caption{Multi-wavelength observations of eRASSt J1003-2607. Only \swix (purple markers in top panel) and \swiu (bottom panel) observations in different bands were previously available for this object, covering almost 10 years of data. Additional \nic observations (orange markers) were obtained to monitor the X-ray evolution of the source. 
    }
    \label{fig:eRASSt_J1003-2607}
\end{figure}

\subsection{eRASSt J1003-2607}\label{indiv:eRASSt_J1003-2607}

eRASSt J1003-2607 (RX J1003.2-2607) was classified as AGN by \cite{Kahabka+00} while studying the gas content of the dwarf galaxy NGC 3109 using \ros PSPC observations of background X-ray sources. No obvious extragalactic counterparts are detected at the X-ray position of the AGN, suggesting a very faint or distant object. Only a point source with a magnitude of $m_r=22.28$ is found within a radius of 1.62\arcsec~from the X-ray position in the LS DR10 images. The X-ray and UV light curve is shown in Fig. \ref{fig:eRASSt_J1003-2607}.

\begin{figure}[h!]
    \centering
    \includegraphics[width=1.0\linewidth]{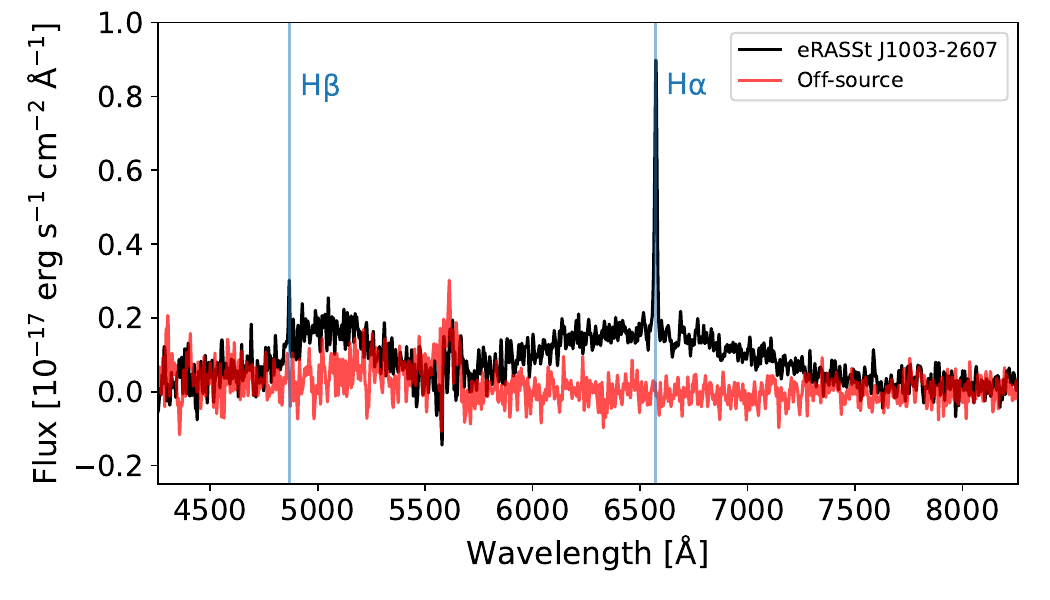}
    \caption{LBT optical spectrum of the tentative eRASSt J1003-2607 counterpart (black) and an off-source position (red). The source shows two emission lines consistent with being H$\beta$ and H$\alpha$ lines at a redshift of $z=0.001345$. The lack of emission lines in the off-source position indicates that the lines belong to the tentative eRASSt J1003-2607 counterpart. The source is likely to be an HLX.
    }
    \label{fig:spec_eRASSt_J1003-2607}
\end{figure}

Figure \ref{fig:spec_eRASSt_J1003-2607} shows the LBT spectrum of the likely optical counterpart of eRASSt J1003-2607 and the spectrum extracted at an offset position. We extracted a spectrum of an off-source position to assess whether the observed emission lines are from the interstellar medium of the foreground galaxy or not. The source spectrum exhibits only two strong emission lines consistent with H$\beta$ and H$\alpha$ lines at the redshift of NGC 3109 ($z=0.0013$). The spectral analysis revealed that the Balmer emission lines are intrinsic to the targeted object and are not coming from the ionized gas of NGC 3109. This implies that the source has indeed an extragalactic nature. However, the object would have a distance consistent with NGC 3109 and an off-nuclear origin. Using the redshift of NGC 3109 as the redshift of the object, we derive a luminosity of $\sim4\times10^{39}\rm\;erg\; s^{-1}$ in the 0.2--2.3 keV band, likely corresponding to a ULX-like object. The source is, therefore, discarded as a SMBHB candidate. We note that there is a possibility that the targeted source is not the real counterpart of the X-ray emission. However, the FoV surrounding eRASSt J1003-2607 is very crowded, and according to Fig. \ref{fig:lsdr10images}, there is only one evident optical counterpart at the position of the X-ray source. Other scenarios, such as an optically faint AGN in the background, are plausible but very unlikely.

\begin{figure}[h!]
    \centering
    \includegraphics[width=1.0\linewidth]{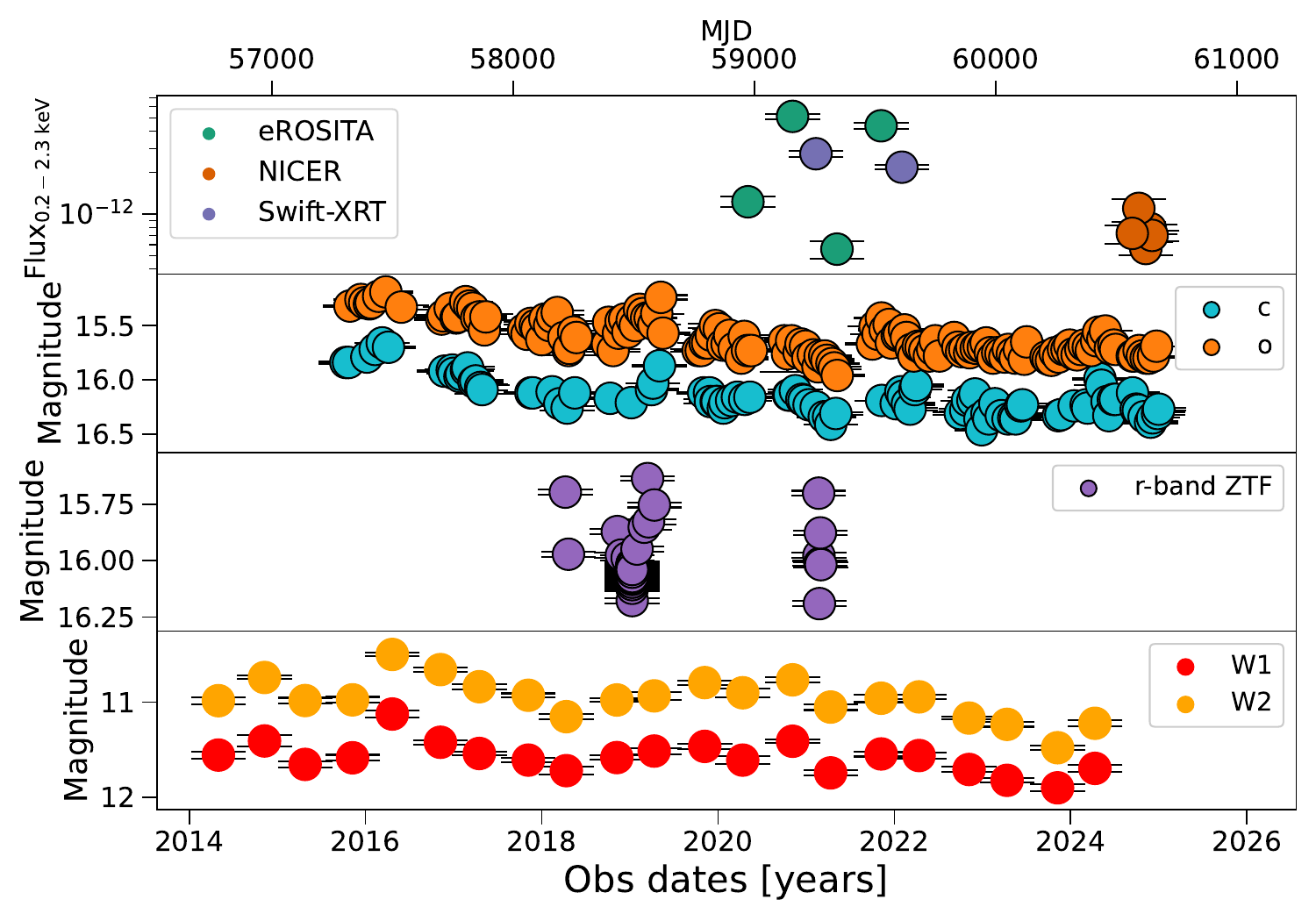}
    \caption{Multi-wavelength observations of eRASSt J0818-2252. The data is similar to the data of Fig. \ref{fig:erasst_j0344-3327} with additional ZTF r-band data in the third panel. The X-ray panel also displays NICER data with orange markers. }
    \label{fig:eRASSt_J0818-2252}
\end{figure}

\subsection{eRASSt J0818-2252}

eRASSt J0818-2252 (LEDA 80921, $z=0.034$) is a Seyfert I galaxy with strong AGN features such as broad Balmer lines and strong [\textsc{O iii}] lines. The X-ray from Fig. \ref{fig:allxraycandidates1} and Fig. \ref{fig:eRASSt_J0818-2252} seems to follow a quasi-periodic behavior with timescales of $\sim300$ days. More X-ray monitoring programs are needed to confirm periodicity in eRASSt J0818-2252. The IR and optical brightness increased by $\sim0.5$ mag in 2016 and 2019, respectively. The source was detected by the \ros all-sky survey and later by a pointed \ros HRI observation with fluxes of $\sim3\times10^{-12}\; \rm erg\; s^{-1}\;cm^{-2}$ and $\sim4\times10^{-13}\; \rm erg\; s^{-1}\;cm^{-2}$, respectively. Both \ros observations are consistent with the bright and faint eROSITA flux levels. 

\begin{figure}[h!]
    \centering
    \includegraphics[width=1.0\linewidth]{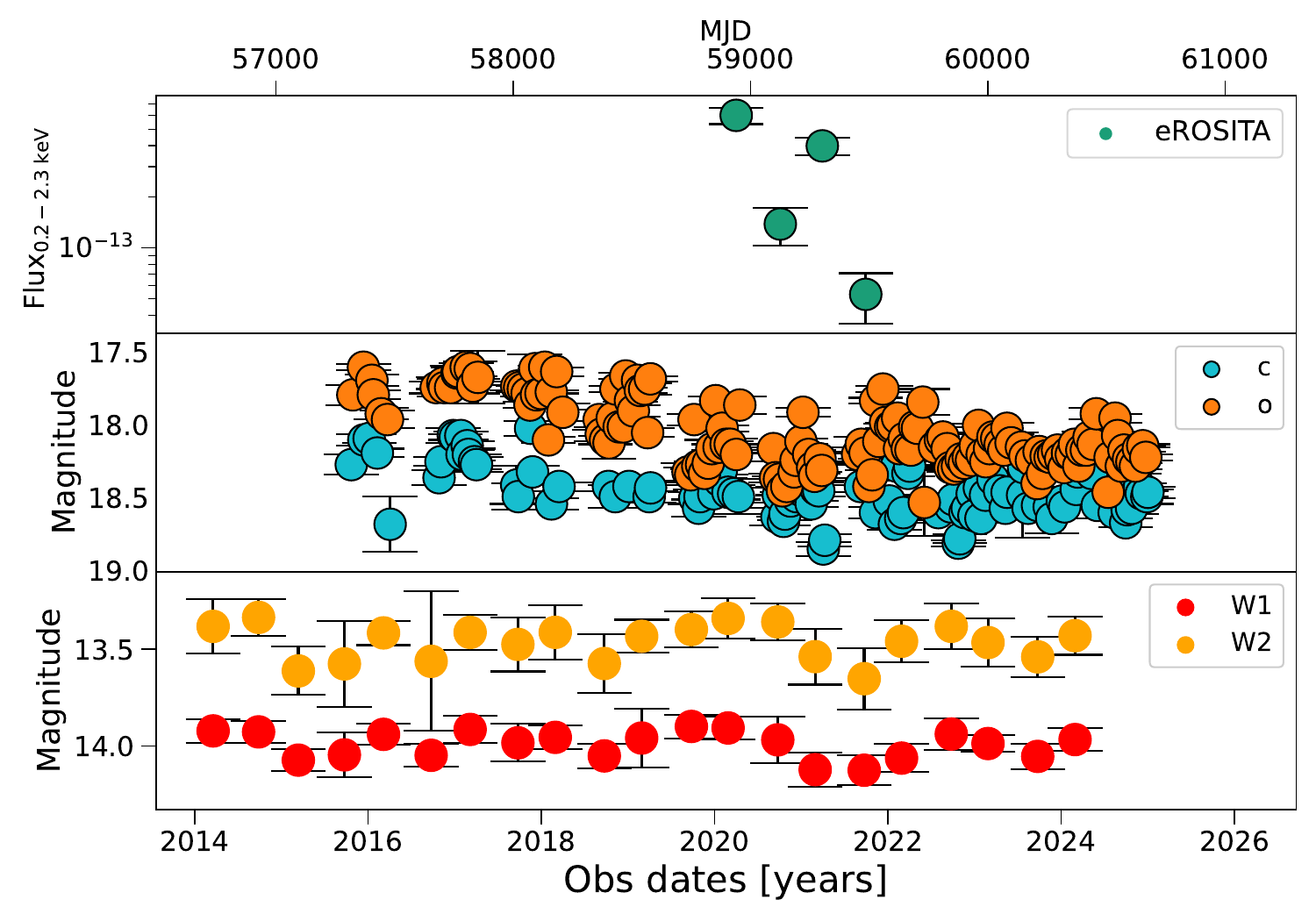}
    \caption{Multi-wavelength observations of eRASSt J0600-2939. The data is similar to the data of Fig. \ref{fig:erasst_j0344-3327}.}
    \label{fig:eRASSt_J0600-2939}
\end{figure}

\subsection{eRASSt J0600-2939}

eRASSt J0600-2939 is a galaxy at $z=0.104$ with an optical spectrum that shows weak H$\beta$+[\textsc{O iii}] emission lines. Only the broad H$\alpha$ line was used to estimate a single-epoch SMBH mass measurement. No clear trends are found in the multi-wavelength light curve (Fig. \ref{fig:eRASSt_J0600-2939}).

\begin{figure}[h!]
    \centering
    \includegraphics[width=1.0\linewidth]{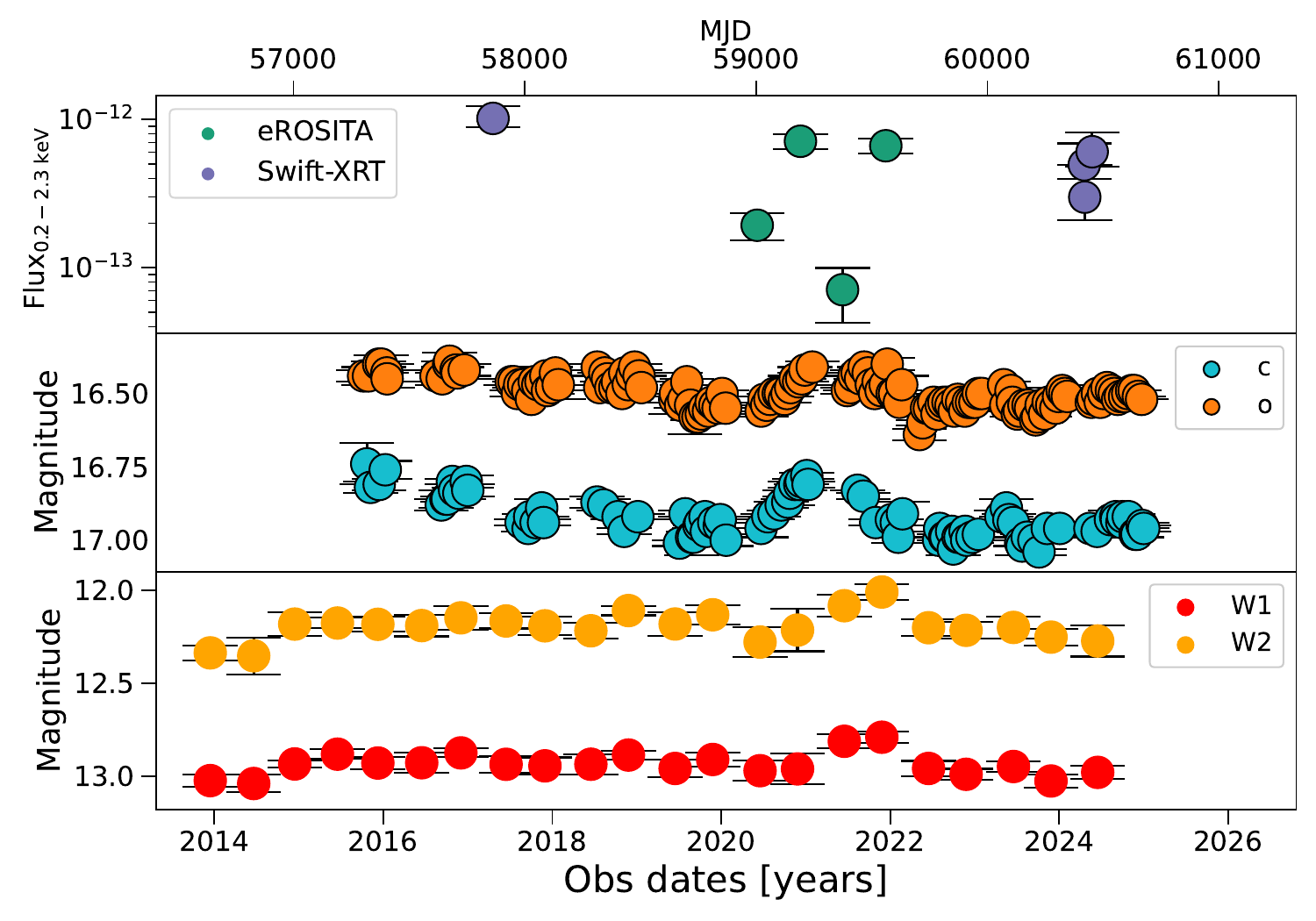}
    \caption{Multi-wavelength observations of eRASSt J0044-3313. The data is similar to the data of Fig. \ref{fig:erasst_j0344-3327}.}
    \label{fig:eRASSt_J0044-3313}
\end{figure}

\subsection{eRASSt J0044-3313}

eRASSt J0044-3313 shows evidence of AGN emission according to its broad Balmer lines and BPT classification. Similarly to eRASSt J1906-4850, the spectral profile of the broad H$\alpha$ component looks asymmetric and can be modeled by two broad Gaussian lines. This feature is only seen in H$\alpha$. A single broad component can model the H$\beta$ line due to the noise and low SNR in the blue part of the spectrum. The velocity shift between both components is $\Delta v\rm = 1800\pm500\; km\; s^{-1}$. Using the second H$\alpha$ component, we derive a SMBH mass of $\rm \log (M_{bh}/M_{\odot})=7.63\pm0.45$, which is consistent with the H$\beta$ measurement. The ATLAS data in Fig. \ref{fig:eRASSt_J0044-3313} shows an increase in the optical light curve that is temporally consistent with the X-ray data from eROSITA. The source was detected in X-ray by the \ros all-sky survey with a flux of $\sim4.5\times10^{-13}\; \rm erg\; s^{-1}\;cm^{-2}$ on the 0.2--2.0 keV band. 

\begin{figure}[h!]
    \centering
    \includegraphics[width=1.0\linewidth]{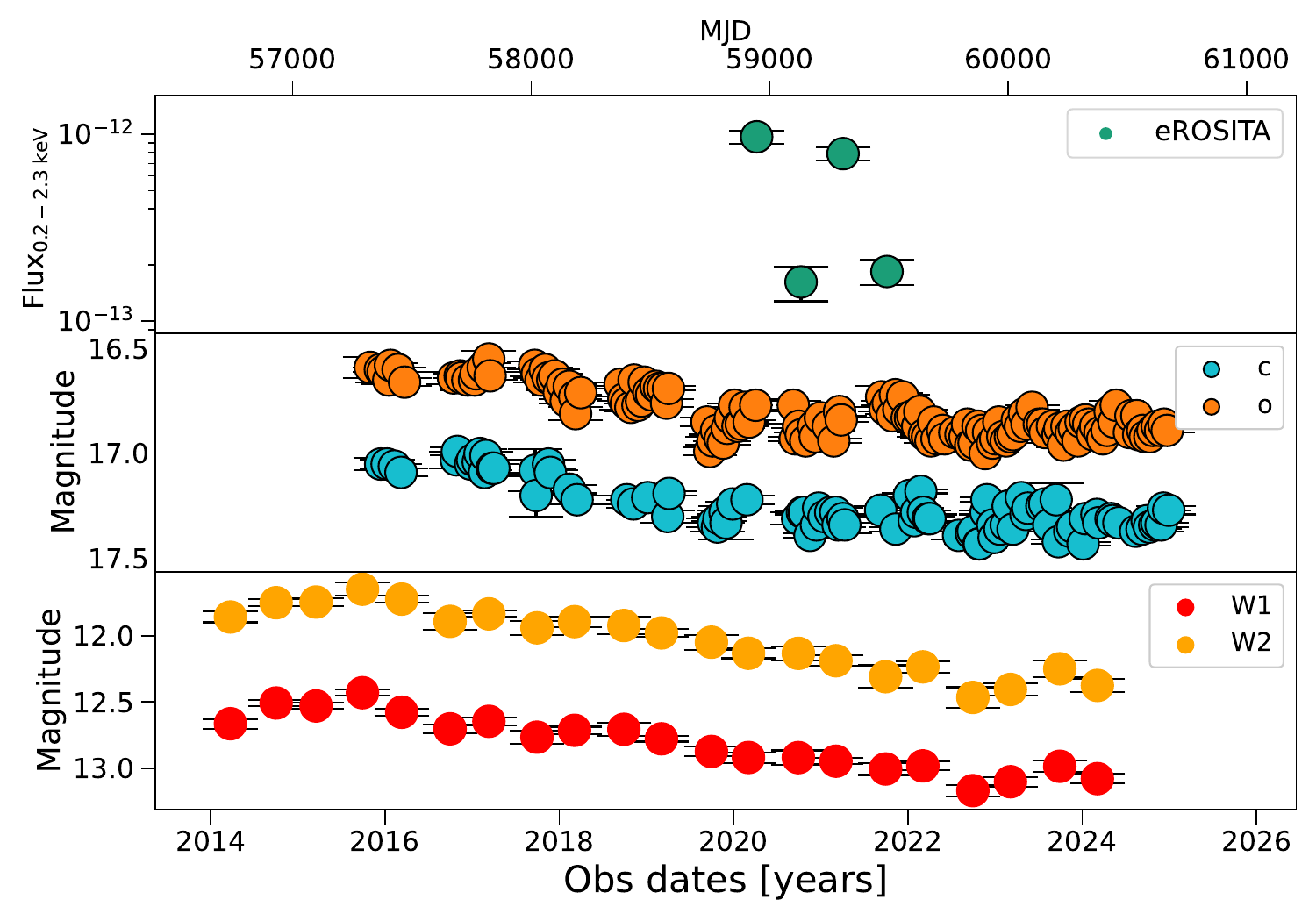}
    \caption{Multi-wavelength observations of eRASSt J0614-3835. The data is similar to the data of Fig. \ref{fig:erasst_j0344-3327}.}
    \label{fig:eRASSt_J0614-3835}
\end{figure}

\subsection{eRASSt J0614-3835}

eRASSt J0614-3835 shows a Seyfert II optical spectrum hosted by a galaxy at $z=0.054$. The source was detected in X-ray by the \ros all-sky survey with a flux of $\sim4\times10^{-13}\; \rm erg\; s^{-1}\;cm^{-2}$ on the 0.2--2.0 keV band, between the bright and faint eROSITA flux levels of Fig. \ref{fig:eRASSt_J0614-3835}.

\end{appendix}

\end{document}